\documentclass[preprint,showpacs]{revtex4}
\usepackage{ifthen}                 %%% package for conditionals in TeX
\ifx\pdfoutput\undefined %%%%%%%%%%%%%%%%%%%%%%%%%%%%%%%%%%%%%%%%% LATEX %%%
\usepackage[dvips]{graphicx}       %%% graphics for dvips
\DeclareGraphicsExtensions{.eps}   %%% standard extension for included graphics
\usepackage[ps2pdf]{thumbpdf}      %%% thumbnails for ps2pdf
\usepackage[ps2pdf,                %%% hyper-references for ps2pdf
bookmarks=true,%                   %%% generate bookmarks ...
bookmarksnumbered=true,%           %%% ... with numbers
hypertexnames=true,%              %%% needed for correct links to figures !!!
breaklinks=true,%                  %%% breaks lines, but links are very small
linkbordercolor={0 0 1},%          %%% blue frames around links
pdfborder={0 0 112.0}]{hyperref}%  %%% border-width of frames 
\else %%%%%%%%%%%%%%%%%%%%%%%%%%%%%%%%%%%%%%%%%%%%%%%%%%%%%%%%%% PDFLATEX %%%
\usepackage[pdftex]{graphicx}      %%% graphics for pdfLaTeX
\DeclareGraphicsExtensions{.pdf}   %%% standard extension for included graphics
\usepackage[pdftex]{thumbpdf}      %%% thumbnails for pdflatex
\usepackage[pdftex,                %%% hyper-references for pdflatex
bookmarks=true,%                   %%% generate bookmarks ...
bookmarksnumbered=true,%           %%% ... with numbers
hypertexnames=true,%              %%% needed for correct links to figures !!!
breaklinks=true,%                  %%% break links if exceeding a single line
linkbordercolor={0 0 1}]{hyperref} %%% blue frames around links
\fi %%%%%%%%%%%%%%%%%%%%%%%%%%%%%%%%%%%%%%%%%%%%%%%%%%% END OF CONDITION %%%

\newboolean{color}
\setboolean{color}{true}        %%% use color figures
\usepackage{dcolumn}% Align table columns on decimal point
\usepackage{bm}% bold math
\def\s#1{\slash\!\!\!{#1}}
\def\bee{\begin{eqnarray}}
\def\eee{\end{eqnarray}}
\def\nn{\nonumber\\}

\def\Tr{\makebox{Tr}}
\def\eq#1{Eq.~(\ref{#1})}
\def\fig#1{Fig.~\ref{#1}}
\def\sec#1{Section~\ref{#1}}

\def\myhref#1{\href{http://arXiv.org/abs/#1}{#1}}
\begin{document}

%Set an additional Bookmark to the first Page
%\pdfbookmark[1]{Preamble}{prea}
%Include your title-page here.
%\cleardoublepage                %%% start again on odd page
%
% Table of Contents:
%\pagenumbering{roman}           %%% Roman page numbers for ToC
%\pdfbookmark[1]{Contents}{toc}  %%% additional bookmark for ToC
%\thispagestyle{plain}           %%% uses the above defined fancy page-header
%\tableofcontents
%\markboth{Table of Contents}{Table of Contents} %%% for the page header
%\cleardoublepage                %%% start again on odd page
\pagenumbering{arabic}          %%% from now on Arabic page numbers
%\preprint{NCTU/HEP2001-2}
\title{QCD factorization at twist-3: the two parton contributions}
\author{Tsung-Wen Yeh}
\email{twyeh@ms3.ntcu.edu.tw}
\affiliation{Department of Science Application And Dissemination, National Taichung University,
Taichung 403, Taiwan}
%\maketitle
\begin{abstract}
In this paper, 
the twist-3 two parton corrections in charmless $B\to PP$ decays are shown to be factorizable 
under the QCD factorization approach.
The factorizability of the twist-3 two parton corrections is constructed on the following findings.
Under the energetic meson limit, 
the pseudoscalar distribution amplitude for a light pseudoscalar meson is allowed to be 
non-constant by the equations of motion for the quark.  
The non-constant pseudoscalar distribution amplitude is then used to regularize the end-point divergences 
in the hard spectator corrections at twist-3 order.
By retaining the momentum fraction variable of 
the spectator quark of the $B$ meson in the propagators,
the end-point divergence in the weak annihilation corrections at twist-3 order is resolved.
The factorization of the $O(\alpha_s)$ corrections under the two parton approximation is shown valid up-to $O(1/m_b)$ . 
The hard scattering kernels of order $O(\alpha_s)$ and $O(\Lambda_{QCD}/m_b)$ are explicitly given 
and found to be infrared finite.
The results are applied for making predictions for the branching ratios of $B\to \pi K $ decays.
\end{abstract}

\maketitle
%\narrowtext
\section{introduction}
The hadronic $B$ decays are good places for testing our understanding of standard model (SM)
and/or new physics (NP).
Until the present time, the B factories have obtained a lot of remarkable results.
For example, the $\sin{(2\beta)}$ has been determined precisely 
from the measurements of the mixing-induced CP-asymmetry in the $B\to J/\psi K_{s}$ decay 
\cite{Aubert:2002ic,Aubert:2004zt,Abe:2001xe,Chen:2006nk},
the branching ratios for $B\to \pi\pi,\pi K $ decays have been measured with only few percent errors 
\cite{Aubert:2002jb,Chao:2003ue,Bornheim:2003bv},
the CP asymmetry for $B\to\pi ^+ K^-$ has been measured precisely \cite{Abe:2005fz,Chao:2004mn,Aubert:2004qm},
and the $B\to\pi^0\pi^0$ decay mode has been confirmed experimentally with unexpectedly large branching ratio
\cite{Aubert:2004aq,Aubert:2003hf,Aubert:2003qj,Chao:2003ue}.
In contrast, the theory suffers from large uncertainties from the nonperturbative dynamics of QCD,
which are involved in the matrix elements contained in the decay amplitudes.
If the theoretical uncertainties can be largely reduced by effective methods, 
then a deterministic demonstration on the CP mechanism, 
an optimal extraction of CKM parameters from the wealth of experimental data,
and a clean separation of NP from SM could be derived from the investigations of hadronic $B$ decays 
\cite{Charles:2004jd}.

The major breakthrough in reducing the theoretical uncertainties 
comes from an observation \cite{Beneke:1999br,Bauer:2001cu}, which is based on a two loop analysis \cite{Beneke:2000ry}, 
that the decay amplitudes for nonleptonic B decays
can be factorized under the heavy quark mass infinity limit \cite{Beneke:1999br,Beneke:2000ry,Beneke:2001ev,Bauer:2001cu,Bauer:2004tj}.
The factorization of the decay amplitudes means 
that the amplitudes can be separated into perturbatively calculable short distance functions
and nonperturbatively incalculable long distance functions.
The finding leads to the QCD factorization (QCDF) \cite{Beneke:1999br,Beneke:2000ry,Beneke:2001ev}.
Similar facts have also been found under other approaches, 
the soft collinear effective theory (SCET) 
\cite{Bauer:2000ew,Bauer:2000yr,Bauer:2001yt,Beneke:2002ph,Beneke:2002ni,Chay:2002vy,Hill:2002vw,Becher:2003qh}
and the perturbative QCD (PQCD) factorization 
\cite{Li:1994cka,Li:1995jr,Li:1994iu,Keum:2000ph,Keum:2000wi,Keum:2000ms,Lu:2000em,Yeh:1997rq,Chang:1996dw}.

The important ingredient of QCDF is the factorization theorem \cite{Collins:1988ig}, 
which has been one successful method in the studies of hard scattering processes,
such as deep inelastic scattering, Drell-Yan processes, etc.
Unlike traditional hard scattering processes involving only one single energy scale,
the hadronic $B$ decays involve multiple energy scales:
the electroweak scale $\mu_{EW}\sim M_{W}$,
the hard scale $\mu_{H}\sim m_{b}$, 
the hard collinear scale $\mu_{HC}\sim \sqrt{m_{b}\Lambda_{QCD}}$,
and the soft scale $\mu_{S}\sim \Lambda_{QCD}$.  
This multi-scale characteristic feature of hadronic $B$ decays requires 
more efforts in the applications of factorization theorem.

The factorization theorem for $B\to M_{1} M_{2}$ processes
with $M_{1,2}$ being light mesons has been shown up-to leading order (LO) in the heavy mass expansion 
(or in the $1/m_b$ expansion)   
and next-to-leading order (NLO) in the loop expansion (or in the $\alpha_s$ expansion).
The $\alpha_s$ means the strong coupling constant and $m_b$ is the $b$ quark mass.
Throughout this paper, the final state mesons in $B\to M_{1} M_{2}$ processes are restricted to be
pseudoscalar light mesons, $\pi$ and $K$.
The extension of the results derived in this paper to other light mesons can be done similarly.  
Under QCDF, the matrix element of a four fermion effective operator $Q_{i}$ 
at $O(\alpha_s)$ and LO in $1/m_b$ expansion can be expressed as
\begin{eqnarray}\label{fact-1}
\langle M_{1} M_{2}|Q_{i}|B\rangle &=& 
F^{BM_1}_{j}(q^2)\int_{0}^1 du T^{I}_{ij}(u)\phi_{M_{2}}(u)+ (M_{1} \leftrightarrow M_{2})\nonumber\\ 
&&+ \int_0^1 d\xi \int_{0}^1 du \int_0^1 dv T^{II}_{i}(u)\phi_{B}(\xi)\phi_{M_1}(u)\phi_{M_{2}}(v)
\end{eqnarray}
where $F^{BM_1}_{j}(q^2)$ with $j=+, 0$ are the transition form factors , 
$T^{I,II}_{ij}$ are the short distance hard scattering functions
and $\phi_{B}, \phi_{M_{1}}, \phi_{M_{2}}$ denote the long distance light cone distribution amplitudes (LCDAs) 
for the external mesons, $B, M_{1}, M_{2}$, respectively.
The $T^{I}_{ij}$ function contains contributions of hard scale
and the $T^{II}_{i}$ function contains contributions of hard and hard-collinear scales.
The contributions below the soft scale are attributed to the meson LCDAs.
The contributions between any two energy scales are calculated by means of the renormalization group method.
The prediction power of factorization theorem given in \eq{fact-1} comes from the universality of the LCDAs.
Once the LCDAs are determined experimentally in some processes, 
they can be applied for making predictions for other processes.

However, the leading order predictions of QCDF based on Eq.~(\ref{fact-1}) 
can not consistently accommodate with experimental data for many decay processes.
For example, the theoretical predictions are about one half of the experimental data for 
the branching ratios of $B\to\pi K $ decays \cite{Beneke:2003zv,Du:2001hr}. 
For understanding the experiments, 
high order corrections to the factorization equation Eq.~(\ref{fact-1}) would be considered \cite{Beneke:2000wa}. 
The higher order corrections mean higher loop corrections
in the $\alpha_s$ expansion or higher twist corrections in the $1/m_b$ expansion.
As the higher order corrections are considered within QCDF formalism, 
it is an important issue to investigate whether the factorization of higher order corrections is still valid. 

So far, only partial results of order $O(\alpha_s^2)$ for $T^{I,II}$ have been calculated 
\cite{Li:2005wx,Beneke:2005gs,Beneke:2005yv,Beneke:2005vv,Kivel:2006xc,Bell:2006tz,Bell:2007tv, Jain:2007dy,Pilipp:2007wm}.
These calculations showed that the $O(\alpha_s^2)$ corrections 
seem preserving the factorization.
On the other hand, the factorization of higher twist corrections is still unclear.
It was found that some high twist hard spectator and weak annihilation corrections spoil the factorization. 
The twist-3 hard spectator corrections \cite{Beneke:2003zv} 
contain the divergent term 
\begin{eqnarray}\label{endpointdiv-1}
X_{H}=\int_{0}^{1} \frac{du}{u}\phi_{p}(u)\;,
\end{eqnarray}
where $\phi_{p}(u)$ is a twist-3 two particle pseudoscalar LCDA for a pseudoscalar light meson. 
Because $\phi_{p}(u)=1$ in the chiral limit \cite{Braun:1989iv, Ball:1998je}, 
$X_H$ diverges at $u= 0$.
Similar end-point divergences also happen in the weak annihilation corrections,
in which the divergent term is
\bee\label{endpointdiv-2}
X_{A}=\int_{0}^{1} \frac{du}{\bar{u}^2}\phi_{P}(u)\;,
\eee
where $\phi_{P}(u)$ is a twist-2 LCDA for a pseudoscalar light meson.
These end-point divergences spoil the QCD factorization at the order of $O(\alpha_s)$ and $O(1/m_b)$.
Since the $X_H$ term are related to chirally enhanced corrections,
which have a numerically large factor $r_{\chi}\sim O(1)$,
the power corrections can be equally important as the radiative corrections in charmless hadronic $B$ decays 
\cite{Beneke:2003zv}. 
From this respect, the extension of QCDF to subleading twist order is of some urgency
\cite{Charles:2004jd, Beneke:2003zv}. 

However, the systematic generalization of QCDF to subleading twist order is still not available.
For this reason,
we restrict ourselves in this paper to only investigate the physics 
related to the non-factorizability of the $X_{H,A}$
and search for a possible resolution to the related end-point divergences.
In QCDF, the factorization is based on the collinear factorization scheme in which
the partons participating in the hard scattering kernels are assumed to carry only collinear momenta.
The hard scattering kernels are calculated by means of a leading twist collinear expansion method
and perturbative QCD.  
The models for the LCDAs are submitted to nonperturbative theories, such as the QCD sum rules.
Since the $X_{H,A}$ terms are related to twist-3 contributions for $B\to M_1 M_2$ decays,
the calculation scheme for the leading twist hard scattering kernels needs to be generalized.
We will employ the collinear expansion (CE) proposed by Yeh \cite{Yeh:2007}
to calculate the relevant hard scattering functions up-to twist-3.
The CE calculation scheme is a generalization of the leading twist collinear expansion for hard scattering processes
\cite{Ellis:1982wd,Ellis:1982cd,Qiu:1988dn,Yeh:2001gu,Yeh:2001ys,Yeh:2002up,Yeh:2001ta,Yeh:2002rd}.
Similar extensions of the leading twist collinear expansion scheme 
for only two parton contributions have been proposed 
by Beneke and Neubert (the BN scheme) \cite{Beneke:2003zv} 
and by Du, Yang and Zhu (the DYZ scheme) \cite{Du:2001ns,Du:2001iu}.  

Not only the calculation scheme needs to be generalized,
but also the twist-3 LCDAs for the light mesons are required to be defined consistently
to derive a factorization theorem at twist-3 order for $B\to M_1 M_2$ decays.
For the $X_H$ term and other twist-3 two parton contributions for $B\to M_1 M_2$ decays, 
two twist-3 two particle LCDAs $\phi_{p}(u)$ and $\phi_{\sigma}(u)$ are involved.
For a pseudoscalar light meson, there are three twist-3 LCDAs ,
the pseudoscalar LCDA $\phi_{p}(u)$, the pseudotensor LCDA $\phi_{\sigma}(u)$, 
and the three particle LCDA $\phi_{3}(u,u^\prime)$.
These three twist-3 LCDAs are related to each other by equations of motion (EOM)\cite{Braun:1989iv, Ball:1998je}.
Due to its small normalization factor, 
the three particle LCDA $\phi_{3}(u,u^\prime)$ is usually neglected in literature.
However, this may not be a good approximation for some processes. 
For example, in the penguin dominated $B$ decays, 
the tree level contributions from the three particle LCDA could be as large as the tree level contributions
from the other two particle LCDAs \cite{Yeh:2007}.

In the approximation of neglecting the three particle contributions, 
$\phi_{p}(u)$ and $\phi_{\sigma}(u)$ are determined by the following equations
\begin{eqnarray}\label{equation-1}
\frac{\bar{u}}{2}(\phi_{p}(u)-\frac{1}{6}\frac{d\phi_{\sigma}(u)}{du}) &=& \frac{1}{6}\phi_{\sigma}(u)\;,\nn
\frac{u}{2}(\phi_{p}(u)+\frac{1}{6}\frac{d\phi_{\sigma}(u)}{d u}) &=& \frac{1}{6}\phi_{\sigma}(u)\;,
\end{eqnarray}
where $\bar{u}=1-u$.
The solutions are: $\phi_{\sigma}(u)=6u\bar{u}$ and $\phi_{p}(u)=1$.
We denote the solutions as the chiral limits of the $\phi_{p}(u)$ and $\phi_{\sigma}(u)$. 

As explained above, 
the substitution of $\phi_{p}(u)=1$ into the $X_H$ term results in an end-point divergence at $u=0$.
This end-point divergence may be due to the failure of the collinear factorization scheme, or
the incorrect use of a model for the $\phi_{p}(u)$.
In this paper, we intend to assume that the collinear factorization scheme is still applicable for the $X_H$ term 
and to study a consistent model for the $\phi_{p}(u)$.
This is different from the common viewpoint for this divergent problem as taken in the literature.
In this respect, we found that the energetic meson limit is 
an important condition for solving the divergent problem in 
the $X_{H,A}$ term.
The energetic meson limit is defined as the limit
at which the light meson's momentum becomes energetic.
The energetic momentum means that the momentum contains a large component 
which is much larger than the light meson's mass and other components of the momentum. 
One light meson carries an energetic momentum is defined as an energetic meson.
The energetic meson limit is the condition
under which the leading twist factorization theorem Eq.~(\ref{fact-1}) has been shown to be valid.
For example, the $O(\alpha_s)$ radiative corrections for the Feynman diagrams  
as depicted in Fig.~\ref{fig:fig1}(a)-(d) 
can only be  shown infrared finite
by requiring the emitted meson $M_2$ to carry an energetic momentum such that the partons inside the $M_2$ meson
can have large collinear momenta.
The collinear (infrared) divergences cancel out in pairs between these four diagrams under the condition that 
the external partons to the radiative loops are collinear to their parent meson $M_{2}$ 
\cite{Beneke:2000ry,Beneke:2001ev}.
Because no any evidence shows that the higher twist contributions require different kinematics,
we argue that, similar to the leading twist contributions,
the twist-3 contributions (including the $X_{H,A}$) should be derived by using the same energetic meson limit, too. 
This argument has been used to derive the $X_H$ term in \cite{Beneke:2003zv}.
Therefore, if the energetic meson limits for $\phi_{p}(u)$ and $\phi_{\sigma}(u)$ are different from
the chiral limits, then one can hope to find a resolution to the divergent problem in $X_H$. 

However, the energetic meson limits for $\phi_{p}(u)$ and $\phi_{\sigma}(u)$ have not been studied in literature.
We need to derive them in this paper.
The details for derivations of the energetic meson limits for $\phi_{p}(u)$ and $\phi_{\sigma}(u)$ will be given in 
Section \ref{sec:sec2}.
Here, we briefly describe how the energetic meson limits for $\phi_{p}(u)$ and $\phi_{\sigma}(u)$ can be different from
the chiral limits.  
If the pseudoscalar light meson is energetic,
then the $\phi_{\sigma}(u)$ at $u=O(1)$ 
is found to be of $O(\Lambda/E)$.
The $\phi^{\prime}_{\sigma}(u)=O(1)$ (the derivative of $\phi_{\sigma}(u)$) and the $\phi_{p}(u)=O(1)$ at $u=O(1)$.
The $E$ denotes the energy of the energetic meson and $\Lambda$ is of $O(\Lambda_{QCD})$.
It implies that the $\phi_{\sigma}(u)$ and the $\phi^{\prime}_{\sigma}(u)$ are of different
order under the energetic meson limit,
and they should be defined as different LCDAs.
To keep both sides of Eq.~(\ref{equation-1}) of the same order,
the $\phi_{\sigma}(u)$ should be dropped out
and Eq.~(\ref{equation-1}) is further reduced to contain only the $\phi_{p}(u)$ and the $\phi^{\prime}_{\sigma}(u)$.
The factor $1/6$ associated with $\phi^{\prime}_{\sigma}(u)$ in the Eq.~(\ref{equation-1}) is only for
normalization, 
it is then instructive to redefine the pseudotensor LCDA as that $\hat{\phi}_{\sigma}(u)$ defined in
Eq.~(\ref{definition-2}).
By using the $\hat{\phi}_{\sigma}(u)$, 
the appropriate EOM for $\phi_{p}(u)$ in the energetic meson limit becomes 
\begin{eqnarray}\label{equation-2}
\hat{\phi}_{p}(u) &=& \hat{\phi}_{\sigma}(u)\;,
\end{eqnarray}
where the $\hat{\phi}_{p}(u)$ and $\hat{\phi}_{\sigma}(u)$
are defined as the energetic meson limits of $\phi_{p}(u)$ and $\phi_{\sigma}(u)$, respectively. 

To have a better understanding of the above fact,
we can define $\Delta\phi_{\sigma}(u)=(\phi_{\sigma}(u)-\hat{\phi}_{\sigma}(u))$
as the difference between $\phi_{\sigma}(u)$ and $\hat{\phi}_{\sigma}(u)$
by referring to Eq.~(\ref{definition-1}) and Eq.~(\ref{definition-2}).
The $\phi_{\sigma}(u)$ in Eq.~(\ref{equation-1}) should be 
$\Delta\phi_{\sigma}(u)$, 
and $\Delta\phi_{\sigma}(u)$ is of $O(\Lambda/E)$ in comparison to $\phi^{\prime}_{\sigma}(u)$ at $u=O(1)$.
and should be identified as a twist-4 quantity in the energetic meson limit $E\gg \Lambda$.

Another interpretation is as following.
According to Eq.~(\ref{definition-1}), 
it is better to expand the coordinate variable $z_{\beta}$ in the spin projector for the $\phi_{\sigma}(u)$
into its collinear and transverse parts,
and to assign them by corresponding LCDAs, 
denoted by $\phi^\prime_{\sigma}(u)$ and $\phi_{\sigma}(u)$, respectively.
To have a more transparent notation,
we define $\phi_{\sigma}^{\parallel}(u)\equiv\phi^{\prime}_{\sigma}(u)$ 
and $\phi_{\sigma}^{\perp}(u)\equiv\phi_{\sigma}(u)$,
which respect the collinear degrees of freedom 
and the transverse degrees of freedom of the light meson state
$| M\rangle$, respectively. 
These two $\phi_{\sigma}^{\parallel}(u)$ and $\phi_{\sigma}^{\perp}(u)$ are equally important at the condition
$E\simeq \Lambda$, which is applicable for a soft light meson.
However, $\phi_{\sigma}^{\parallel}(u)$ and $\phi_{\sigma}^{\perp}(u)$ become of different order 
as we boost the reference frame
along the collinear direction of the light meson's momentum such that $E\gg \Lambda$, 
which is applicable for an energetic light meson.
In the $E\gg \Lambda$ reference frame, $\phi_{\sigma}^{\perp}(u)$ becomes suppressed by a factor $\Lambda/E$ than
$\phi_{\sigma}^{\parallel}(u)$.
This is consistent with the parton model picture
that the collinear degrees of freedom dominate over the transverse ones.
In summary, we arrive at two consistent explanations for the same thing by noting that
$\phi^{\parallel}_{\sigma}(u)\propto\hat{\phi}_{\sigma}(u)$ 
and $\phi^{\perp}_{\sigma}(u)\propto\Delta \phi_{\sigma}(u)$.

To clearly clarify the source for the divergence in the $X_{H}$ term, 
we also need to examine the calculation method related to the $\phi_{\sigma}(u)$.
The method proposed by Beneke and Neubert \cite{Beneke:2003zv} 
is to separate the spin projector for the $\phi_{\sigma}(u)$ 
into the collinear and transverse parts (the BN scheme).
The collinear part is transformed into a derivative over the momentum fraction $u$ 
and the (collinear) derivative is then defined to be applied on the $\phi_{\sigma}(u)$.
The transverse part is transformed into a momentum derivative in the transverse direction, 
and the (transverse) derivative is then defined to be applied on the hard scattering kernel. 
Alternatively, one can also let the collinear derivative applied on the hard scattering kernel
and the transverse derivative applied on the $\phi_{\sigma}(u)$.
These two approaches are equivalent mathematically, but may result in different physical results.
It is known as the projection ambiguity. 
To solve this ambiguity, Du, Yang and Zhu (DYZ) \cite{Du:2001ns,Du:2001iu} proposed 
that the whole momentum derivative should be 
applied on the hard scattering kernel (the DYZ scheme). 
As mentioned previously,
the $\phi^{\parallel}_{\sigma}(u)$ and the $\phi^{\perp}_{\sigma}(u)$
are of different magnitudes in the energetic meson limit,
and their associated hard scattering kernels should be calculated separately.
Since the above two calculation schemes did not consider the difference 
between $\phi^{\parallel}_{\sigma}(u)$ and $\phi^{\perp}_{\sigma}(u)$,
the results calculated by these two schemes require further examinations.
In this paper,  
we propose to employ the collinear expansion method to re-calculate the twist-3 two parton contributions.
The collinear expansion method will be described in details in Section \ref{sec:sec3}.

For the weak non-singlet annihilation corrections,
the end-point divergent problem is more severe
because it exists even for leading twist-2 LCDAs.
This can be seen from the $X_A$ term in \eq{endpointdiv-2}.
To solve this problem,
we propose to retain the momentum fraction variable of the spectator quark of the $B$ meson in the
denominators of the parton propagators.
In literature, the momentum fraction variable of the spectator quark of the $B$ meson is always neglected.
This is because the distribution function for the $B$ meson is highly asymmetric such that
the momentum fraction variable of the spectator quark of the $B$ meson is of order $\Lambda/m_b$.
The reason to retain the momentum fraction variable of the spectator quark of the $B$ meson is as following.
The divergence in $X_A$ term mainly arises as the parton propagators of the $B$ meson become on-shell.
To regularize the divergence in $X_A$ term, we let the propagator be slightly off-shell 
by adding a term of $O(\Lambda^2/m_b^2)$ into it.
After applying this for calculation, there are two same factors in the
numerator and the denominator of the propagator of the spectator particle, respectively.
As a result, these two factors cancel. 
The $X_{A}$ in \eq{endpointdiv-2} becomes
\bee
X_A\to\int_0^{1}d\xi \phi_{B}(\xi)\int_{0}^{1} \frac{du}{(\bar{u}-\xi)\bar{u}}\phi_{P}(u)\;.
\eee
It is obvious that the original end-point divergence is regularized by the momentum fraction $\xi$
carried by the spectator particle.
The only price we need to pay is to retain the $\phi_{B}(\xi)$ without integrating it out.
From the resultant expression, 
the spectator particle is interpreted to carry a collinear momentum,
although the spectator particle's momentum is soft.  
The above argument is valid for $(V-A)(V\pm A)$ and $-2(S-P)(S + P)$ operators.
Therefore, the divergent problem associated with the $X_{A}$ term is resolved.
Similar fact for the factorizability of the annihilation contributions has been observed in \cite{Arnesen:2006vb,Arnesen:2006dc}

The organization of this paper is as follows.
In Section \ref{sec:sec2}, the energetic meson limits
for $\phi_{p}(u)$ and $\phi_{\sigma}(u)$ and related physics will be studied in details.
In Section \ref{sec:sec3}, a simple introduction to the CE expansion scheme will be given first.
The CE scheme is then applied to analyze the next-to-leading order radiative corrections.
The factorization for the next-to-leading order radiative corrections is shown to be valid at $O(1/E)$ or $O(1/m_b)$.
In Section \ref{sec:sec4}, the CE scheme is compared to the BN and DYZ schemes for 
the contributions related to the $\phi_{\sigma}(u)$.
In Section \ref{sec:sec5},
the decay amplitudes at twist-3 order and at $\alpha_s$ order
will be recalculated under the CE scheme.
The explicit expressions for the amplitudes for $B\to PP$ decays, in which the final state $P P$ means
pseudoscalar light mesons, will be given in this Section, too.
The predictions for the branching ratios of $B\to\pi K $ decays will be also present.
The last Section is devoted for conclusions. 
The calculation details for the twist-3 $O(\alpha_s)$ vertex and penguin corrections are given in Appendix A and B.

\section{The energetic meson limit and the light cone distribution amplitudes for light pseudoscalar mesons}
\label{sec:sec2}
In the $B\to M_{1} M_{2}$ decays with $M_{1,2}$ being light mesons, 
the momenta $P_{1}$ and $P_{2}$ of the $M_{1}$ and $M_2$ mesons can have a component 
being much larger than the other components and the meson mass. 
In the rest mass frame of the decaying $B$ meson, 
the momentum conservation $P_{B}=P_{1} + P_{2}$ 
and the smallness of the meson masses $m_{M_1}, m_{M_2}\ll M_{B}$ 
lead to $P_{1}\cdot P_2 = (M_{B}^2-m_{M_1}^2- m_{M_2}^2)/2 \simeq M_{B}^2/2$ 
and $P_{1}^{\mu}=(P_{1}^+, P_{2}^-, \vec{P}_{1\perp})=(M_{B}/\sqrt{2},0,\vec{0})$ 
and $P_{2}^{\mu}=(P_{2}^+, P_{2}^-, \vec{P}_{2\perp})=(0,M_{B}/\sqrt{2},\vec{0})$.
The light mesons in this situation are identified as energetic mesons.
The energetic mesons are not limited in the charmless hadronic $B$ decays.
The light mesons in the hadronic $B\to D M$ decays with heavy-light final state mesons, 
or the exclusive hard scattering processes,e.g., $\pi\gamma^{*}\to \gamma$, $\pi\gamma^{*}\to\pi$, 
can all be considered as energetic mesons. 
For convenience, we parameterize the momentum $P^{\mu}$ of an energetic meson as
\begin{eqnarray}\label{meson_momentum}
P^{\mu}=E\bar{n}^{\mu}+\frac{m_{M}^2}{2E}n^{\mu}\;,
\end{eqnarray} 
in which $E=M_{B}/\sqrt{2}$ and
two light-like vectors 
$\bar{n}^{\mu}=(\bar{n}^{+},\bar{n}^{-},\vec{\bar{n}}_{\perp})=(1,0,\vec{0}_{\perp})$ 
and $n^{\mu}=(n^{+},n^{-},\vec{n}_{\perp})=(0,1,\vec{0}_{\perp})$ 
are introduced.
The vectors $n^\mu$ and $\bar{n}^\mu$ satisfy $\bar{n}^2=n^{2}=0$ and $\bar{n}\cdot n=1$. 
For later discussions, we define the chiral limit as $E\sim m_{M}\sim \Lambda_{QCD}$, 
and the energetic meson limit as $E\gg m_{M}, \Lambda_{QCD}$.

The meson distribution amplitudes with a given twist order are defined as 
the momentum fraction distributions of partons in a particular Fock state of a meson.
Up to twist-3, the distribution amplitudes of a pseudoscalar meson $M$ are defined 
by the matrix element of nonlocal operators \cite{Braun:1989iv}
\bee\label{definition-1}
&&\left.\langle M(P)|\bar{q}(0)[0;z]q(z)|0\rangle\right|_{z^2=0}\nn
&=&-i\frac{f_{M}}{4}\int_0^1 du e^{i\bar{u}P \cdot z}
[\gamma_5\gamma^{\mu}P_{\mu}\phi_P(u) + \mu^{M}_{\chi}\gamma_5 
\left(\phi_p(u) 
- \sigma^{\mu\nu}P_{\mu}z_{\nu}\frac{\phi_{\sigma}(u)}{6}\right) ]
\eee
where $[0;z]$ denotes the Wilson line for preserving the gauge invariance of the matrix element.
$P$ is the meson momentum, 
$f_M$ the decay constant of the meson $M$, 
and $\mu^{M}_{\chi}=m_M^2/(m_{q}+m_{\bar{q}})$ with $m_M$, $m_{q}$ 
and $m_{\bar{q}}$ being the meson's mass and the quark and anti-quark current masses.
The twist-2 distribution amplitude $\phi_{P}(u)$
and the twist-3 two particle distribution amplitudes $\phi_{p}(u)$ and $\phi_{\sigma}(u)$ 
have been studied, in the chiral limit,
based on nonlocal product expansion and conformal expansion. 
In addition, the equations of motion of on-shell quarks in the meson
were used to obtain two differential-integral relations between 
the twist-3 two particle distribution amplitudes $\phi_{p}(u)$ and $\phi_{\sigma}(u)$
and the twist-3 three particle distribution amplitude $\phi_{3}(u,u^\prime)$.
The differential-integral relations can be solved by means of moment.
$\phi_{p}(u)$ and $\phi_{\sigma}(u)$ are determined by $\phi_{3}(u,u^\prime)$ \cite{Braun:1989iv}.

Some simplifications can be obtained by neglecting 
the twist-3 three particle distribution amplitude $\phi_{3}(u,u^\prime)$ 
due to its small normalization constant. 
The two differential-integral relations are then reduced to the relations as shown in Eq.~(\ref{equation-1}).
We identify the solutions to Eq.~(\ref{equation-1}) 
as the chiral limits $\phi^c_{p}(u)$ and $\phi^c_{\sigma}(u)$ of $\phi_{p}(u)$ and $\phi_{\sigma}(u)$,
in which $\phi^c_{p}(u)=1$ and $\phi^c_{\sigma}(u)=6u\bar{u}$ with $\bar{u}=1-u$.

Referring to \eq{endpointdiv-1}, the $X_H$ term from the hard spectator diagrams contains the partonic part
$1/u$ and the hadronic part $\phi_{p}(u)$.
As the chiral limit solution $\phi^c_{p}(u)$ is substituted into the hadronic part $\phi_{p}(u)$ of the $X_H$ term,
an end-point divergence arises as $u\to 0$.
The end-point divergence spoils the factorization at the twist-3 order for $B\to M_1 M_2$ processes.
In literature, the common viewpoint is to identify that the breakdown of the factorization for the $X_H$ term 
is due to the failure of the factorization scheme, the QCD factorization.
In this paper, we propose to take another viewpoint that the source for the breakdown of 
the factorization for the $X_H$ term   
may be due to the use of the chiral limit solution for the $\phi_{p}(u)$. 
Our consideration is the following.
The partonic part of the $X_H$ term is derived under the assumption 
that the external meson is taken in its energetic limit \cite{Beneke:2003zv},
while the hadronic part is using the chiral limit solution for the
relevant distribution amplitude.
If the energetic limit solution for $\phi_{p}(u)$ can be different 
from the chiral limit solution,
then we can expect to find a resolution for the end-point divergence. 

%The inconsistency in the application of the factorization theorem for the $X_H$ term 
%also exist in physical quantities for other hard exclusive processes.
%For example, in the $\pi\gamma^{*}\to \gamma$ process,
%the leading twist factorization formula for the transition form factor $F^{\pi\gamma}(Q^2)$ 
%are expressed in terms of the twist-2 pion distribution amplitude, 
%$\phi_{\pi}(u)$ and the corresponding hard scattering functions.
%The factorization formula was derived by assuming that the external pion carries an energetic momentum
%and the $\phi_{\pi}(u)$ was used its chiral limit solution.
%The prediction of the leading twist factorization formula for $F^{\pi\gamma}(Q^2)$ 
%can only accomodate with the experimental data at very large $Q^2$.
%The deviation between the theoretical prediction and the experimental data become larger for smaller $Q^2$.  
%This deviation at low $Q^2$ can be identified as the inconsistency in the application of leading twist factorization
%theorem by using a chiral limit solution for the relevant $\phi_{\pi}(u)$.
%The deviation at low $Q^2$ can be reduced by adding twist-4 contributions 
%to the leading twist factorization formula for $F^{\pi\gamma}(Q^2)$ [Yeh].
%Since the difference between the chiral limit and the energetic limit 
%of the twist-2 pion distrbution amplitude starts from twist-4 order,
%this shows one evidence that the inconsistency indeed exists in the application of leading twist factorization
%theorem .

To see whether the above argument is correct,
it is necessary to find out the energetic limits,
$\hat{\phi}_{p}(u)$ and $\hat{\phi}_{\sigma}(u)$, 
of the $\phi_{p}(u)$ and $\phi_{\sigma}(u)$.
Since the energetic limits have not been studied, we will derive them in this paper.
The first main result of this paper is to show that the energetic limits,
$\hat{\phi}_{p}(u)$ and $\hat{\phi}_{\sigma}(u)$, indeed exist and are different from the chiral limits.
We will employ a simplified method to derive similar relations to those in \eq{equation-1} 
for $\hat{\phi}_{p}(u)$ and $\hat{\phi}_{\sigma}(u)$.
According to the relations for $\hat{\phi}_{p}(u)$ and $\hat{\phi}_{\sigma}(u)$, 
a non-constant solution for the $\hat{\phi}_{p}(u)$ will be obtained.
We will also develop a expansion scheme consistent with assumption for taking the energetic limit for the light mesons.
The expansion scheme is the second main result of this paper and will be given in \sec{sec:sec3}.
By using the expansion scheme to re-derive the $X_H$ term 
and substituting the non-constant solution for the $\hat{\phi}_{p}(u)$ into the hadronic part of the $X_H$ term,
the end-point divergence is then resolved. 

In the following derivation, 
only the asymptotic solutions for $\hat{\phi}_{p}(u)$ and $\hat{\phi}_{\sigma}(u)$ are considered.
The complete solutions for the $\hat{\phi}_{p}(u)$ and $\hat{\phi}_{\sigma}(u)$ by following
the traditional approaches \cite{Braun:1989iv,Ball:1998sk,Ball:1998je} will be given in another place.
%The twist-2 $\phi_{\pi}(u)$ for the hard scattering exclusive processes, 
%such as $\pi\gamma^{*}\to \gamma$, $\pi\gamma^{*}\to\pi$,
%is required to be defined on the light-cone $z^2\simeq 0$ 
%and small transverse distances $ z_{\perp}\sim 1/E$ with $E$ being the relevant energy scale for the processes.
Usually, the twist-2 LCDA $\phi_{P}(u)$ is defined as the probability of the transition of the meson $M$ 
into the $q(u)\bar{q}(1-u)$ pair at zero transverse distance.
We assume that the same conditions are also applicable 
for the $\hat{\phi}_{p}(u)$ and $\hat{\phi}_{\sigma}(u)$.
To define the $\hat{\phi}_{p}(u)$ and $\hat{\phi}_{\sigma}(u)$, 
we start from \eq{definition-1}.
For convenience, the coordinate $z$ in the matrix element can be parameterized under the energetic limit 
as the following 
\begin{eqnarray}\label{z-expansion}
z_{\beta}=\frac{z^2_{\perp} E}{2\lambda }\bar{n}_{\beta}+\frac{\lambda}{E} n_{\beta}+z_{\perp\beta}\;,
\end{eqnarray}
where the variable $\lambda$ is for boost invariance in the collinear direction
and $z_{\perp}$ is assumed to be of order $O(1/E)$. 
It is noted that the $\lambda$ is required to be large 
to insure that the collinear component $\bar{z}^{\mu}\equiv\lambda/E n^{\mu}$  
dominates.
This requirement for $\lambda$ is consistent with the condition for an energetic meson in the
highly boost frame.
The quark field $q(z)$ in Eq.~(\ref{definition-1}) can be expanded with respect to 
$\bar{z}^{\mu}\equiv\lambda/E n^{\mu}$ as
\begin{eqnarray}\label{qfield-expansion}
q(z)=q(\bar{z})+\left.\frac{\partial q(z)}{\partial z^{\mu}}\right|_{z=\bar{z}}(z-\bar{z})^{\mu}+\cdots\;,
\end{eqnarray}
where dots means terms of order $O(z_\perp^n)$ with $n\ge 2$.
By using Eqs.~(\ref{meson_momentum}) and (\ref{z-expansion}), 
the spin projector $[P_{\alpha},z_{\beta}]$ associated with $\phi_{\sigma}(u)$ defined in \eq{definition-1} 
can be written as
\begin{eqnarray}\label{spin_proj_1}
[P_{\alpha},z_{\beta}]
&=&\lambda[\bar{n}_{\alpha},n_{\beta}]-\frac{m_{M}^2 z_{\perp}^2}{4\lambda}[\bar{n}_{\alpha},n_{\beta}]\nonumber\\
&&+E[\bar{n}_{\alpha},z_{\perp\beta}]+\frac{m_{M}^2}{2 E}[n_{\alpha},z_{\perp\beta}]\;.
\end{eqnarray}
By substituting Eqs.~(\ref{qfield-expansion}) and (\ref{spin_proj_1}), 
into Eq.~(\ref{definition-1}), 
we arrive at the following identity up-to $O(z_{\perp})$ 
\begin{eqnarray}\label{definition-1-2} 
&&\langle M(P)|\bar{q}(0)\sigma_{\alpha\beta}\gamma_5 q(\lambda/E n)|0\rangle
+ \left.\langle M(P)|\bar{q}(0)\sigma_{\alpha\beta}\gamma_5 \frac{\partial q(z)}{\partial z^{\mu}}|0\rangle
\right|_{z=\bar{z}}z^{\mu}_{\perp}\nonumber\\
&=& -if_M \mu^{M}_{\chi} \left\{-i[\bar{n}_{\alpha},n_{\beta}]\int_{0}^{1} d u e^{i \bar{u} \lambda}
\frac{1}{6}\frac{d\phi_{\sigma}(u)}{du} 
 + E[\bar{n}_{\alpha},z_{\perp\beta}]\int_{0}^{1} d u e^{i \bar{u}  \lambda}\frac{\phi_{\sigma}(u)}{6}\right.\nonumber\\
&& + i\left.[\bar{n}_{\alpha},n_{\beta}] \left(e^{i \bar{u}  \lambda}\frac{\phi_{\sigma}(u)}{6}\right)_{u=0}^{u=1}
\right\}\;,
\end{eqnarray}
where the $P\cdot z$ in the phase factor $\exp(i \bar{u} P\cdot z)$ has been approximated to be $\lambda$, 
and the terms proportional to $m_{M}$ or $z_{\perp}^2$ have been neglected.
An integration by parts has been used to obtain the first 
and third terms in the right hand side of Eq.~(\ref{definition-1-2}).
According to the convention of \cite{Beneke:2001ev,Beneke:2003zv}, 
we let $\phi^{\prime}_{\sigma}(u)$ and $\phi_{\sigma}(u)$ correspond to 
the collinear part and the transverse part of the spin projector, respectively.
Comparing both sides of Eq.~(\ref{definition-1-2}), 
we choose the following identities according to the order of the $z_{\perp}$ factor
\begin{eqnarray}
&&\langle M(P)|\bar{q}(0)\sigma_{\alpha\beta}\gamma_5 q(\lambda/E n)|0\rangle
= -f_M \mu^{M}_{\chi} [\bar{n}_{\alpha},n_{\beta}]\frac{1}{6}\left\{
\int_{0}^{1} d u e^{i \bar{u} \lambda}\frac{d\phi_{\sigma}(u)}{du}
-\left(e^{i \bar{u} \lambda}\phi_{\sigma}(u)\right)_{u=0}^{u=1}\right\}\;,\label{definition-1-3}\nn\\
&&\langle M(P)|\bar{q}(0)\sigma_{\alpha\beta}\gamma_5\partial_{\mu} q(\lambda/E n)|0\rangle z^{\mu}_{\perp}
= -if_M \mu^{M}_{\chi} E[\bar{n}_{\alpha},z_{\perp\beta}]\int_{0}^{1} d u e^{i \bar{u} \lambda}
\frac{\phi_{\sigma}(u)}{6}\;.
\label{definition-1-4}
\end{eqnarray} 

From Eqs.~(\ref{definition-1-3}), (\ref{definition-1-4}) and (\ref{definition-1}), 
we observe the following issues needed for further examinations in the energetic limit:
\begin{itemize}
\item  The meaning of the momentum fraction $u$: 
according to the parton model, the momentum fraction $u$ is defined 
to be the ratio of the collinear parts of the parton's momentum and the meson's momentum.
That is, if we let $k$ denote the parton's momentum, 
then the momentum fraction is defined as $\hat{u}=k^+/P^+$, 
where we have used the $\hat{u}$ to distinguish from the $u$.  
However, the momentum fraction $u$ in Eq.~(\ref{definition-1}) is not defined as the $\hat{u}$.
On the other hand, the $u$ in Eq.~(\ref{definition-1}) can be interpreted as the fraction of the whole momentum $P$.
This can be seen from the $\exp(i\bar{u} P\cdot z)$ in Eq.~(\ref{definition-1}). 
If the parton momentum is denoted as $k$, then $u=k\cdot z/P\cdot z$.
By using the parameterizations for $P$ and $z$, the $u$ is expanded as
\begin{eqnarray}\label{faction-1}
u=\frac{k\cdot z}{P\cdot z}=\frac{\frac{z_{\perp}^2 k_{\perp}^2 E}{4\lambda k\cdot n}
+ k\cdot n\frac{\lambda}{E}-\vec{k}_{\perp}\cdot \vec{z}_{\perp}}
{\lambda + \frac{m_{M}^2 z_{\perp}^2}{4\lambda}}\;,
\end{eqnarray}
where the parton momentum $k^{\mu}$ has been parameterized as
\begin{eqnarray}
k^{\mu} &=& k\cdot n \bar{n}^{\mu}+\frac{k_{\perp}^2}{2 k\cdot n}n^{\mu}+\vec{k}_{\perp}\;.
\end{eqnarray}
The momentum fraction $u$ carried by the collinear partons should be defined under the limit $|z_{\perp}|\to 0$, $\hat{u}=\lim_{|z_{\perp}|\to 0} u(z_{\perp})\equiv k\cdot n/P\cdot n$.
The fraction $u$ in Eqs.~(\ref{definition-1-3}) and (\ref{definition-1-4}) should be interpreted as $\hat{u}$.
In the following text, the $u$ is always interpreted as $\hat{u}$ to simplify the notations.

\item The boundary condition: 
the boundary terms in Eq.~(\ref{definition-1-3}) implies the following equation
\begin{eqnarray}\label{bc-1}
\left(\phi_{\sigma}(1)-e^{i\lambda} \phi_{\sigma}(0)\right)= 0\;.
\end{eqnarray}
Although the solution $\phi_{\sigma}(1)=\phi_{\sigma}(0)=0$ can satisfy the above equation,
but the end-point $u\to 1$ behavior of $\phi_{\sigma}(u)$ (, or, by translation invariance, 
the end-point $u\to 0$ behavior of $\phi_{\sigma}(u)$) is related to $\lambda$.
Since the $\lambda$ depends on the reference frame, 
this is in contradiction with the universality assumption for the $\phi_{\sigma}(u)$.

\item The definition for the $\phi_{\sigma}(u)$:
take a differentiation on both sides of Eq.~(\ref{definition-1-4}) in $z_{\perp}$ as
\begin{eqnarray}\label{definition-1-5}
&&\langle M(P)|\bar{q}(0)\sigma_{\alpha\beta}\gamma_5\partial_{\mu}(\lambda/E n) q(\lambda/E n)|0\rangle
= -if_M \mu_M E[\bar{n}_{\alpha},d_{\perp\mu\beta}]\int_{0}^{1} d u e^{i \bar{u} \lambda}\frac{\phi_{\sigma}(u)}{6}\;.
\end{eqnarray}
We observe from Eqs.~(\ref{definition-1-3}) and (\ref{definition-1-5}) 
that the same $\phi_{\sigma}(u)$ corresponds to two different matrix elements in the energetic limit.
There arise confusions which definition, \eq{definition-1-3} or \eq{definition-1-5}
should be used for $\phi_{\sigma}(u)$.
\item The equations of motion:
to avoid the above confusions in the definition for $\phi_{\sigma}(u)$,
we suggest to re-define $\phi^{\parallel}_{\sigma}(u)=\phi^{\prime}_{\sigma}(u)$ and
$\phi^{\perp}_{\sigma}(u)=\phi_{\sigma}(u)$ according to \eq{definition-1-3} and \eq{definition-1-5}, respectively.
Because the boundary term in Eq.~(\ref{definition-1-3}) may not vanish, 
the boundary term is defined to be absorbed by $\phi^{\parallel}_{\sigma}(u)$.
By using $\phi^{\parallel}_{\sigma}(u)$ and $\phi^{\perp}_{\sigma}(u)$,
the first equation in Eq.~(\ref{equation-1}) can be written as
\begin{eqnarray}\label{equation-1-2}
\frac{\bar{u}}{2}(\phi_{p}(u)-\frac{1}{6}\phi^{\parallel}_{\sigma}(u)) &=& \frac{1}{6}\phi^{\perp}_{\sigma}(u)\;.
\end{eqnarray} 
To find out the energetic limit of the above equation,
it is instructive to rewrite
$\phi^{\parallel}_{\sigma}(u)$ and $\phi^{\perp}_{\sigma}(u)$ as the following expressions by taking
Fourier transformations for Eqs.~(\ref{definition-1-3}) and (\ref{definition-1-5})
\bee
\phi^{\parallel}_{\sigma}(u)&=& 
\frac{1}{f_M \mu_{\chi}^M}\int_0^\infty\frac{d\lambda}{2\pi}e^{-i\bar{u}\lambda}
\langle M(P)|\bar{q}(0)\sigma_{\alpha\beta}\gamma_5 n^{\alpha}\bar{n}^{\beta}q(\lambda/E n)|0\rangle\;,
\label{sigma-parall}\\
\phi^{\perp}_{\sigma}(u)&=& 
 \frac{i}{f_M \mu_{\chi}^M}\int_0^\infty\frac{d\lambda}{2\pi}\frac{e^{-i\bar{u}\lambda}}{E}
\langle M(P)|\bar{q}(0)\sigma_{\alpha\beta}\gamma_5 n^{\alpha}\partial^{\beta}_{\perp}(\lambda/E n) q(\lambda/E n)
|0\rangle\;.\label{sigma-perp}
\eee 
It is seen that, in \eq{sigma-perp}, there is a large factor $1/E$ for $\phi^{\perp}_{\sigma}(u)$, 
which is of short distance.
The transverse derivative $\partial^\beta_{\perp}$ in the matrix element
$\langle M(P)|\bar{q}(0)\sigma_{\alpha\beta}\gamma_5 n^{\alpha}\partial^{\beta}_{\perp}(\lambda/E n) 
q(\lambda/E n)|0\rangle$ corresponds to the transverse momentum $k^{\beta}_{\perp}$ for the quarks in the meson,
which is of order $O(\Lambda)$.
Therefore, $\phi^{\perp}_{\sigma}(u)$ is of order $O(\Lambda/E)$ for $u=O(1)$ in the energetic limit. 
On the other hand, $\phi^{\parallel}_{\sigma}(u)$ and $\phi_{p}(u)$ are of order $O(1)$ for $u=O(1)$ 
in the energetic limit.
The power counting for $\phi^{\perp}_{\sigma}(u)$, $\phi^{\parallel}_{\sigma}(u)$ and $\phi_{p}(u)$ implies that
\eq{equation-1-2} 
is reduced to $\phi_{p}(u)=\phi^{\parallel}_{\sigma}(u)/6$ in the energetic limit. 
\end{itemize}

Based on the above discussions,
we arrive at the following definitions for the $\hat{\phi}_{p}(u)$ and $\hat{\phi}_{\sigma}(u)$
\begin{eqnarray}
\langle M(P)|\bar{q}(0)i\gamma_5 q(\lambda/E n)|0\rangle
&=& f_M \mu^{M}_{\chi} \int_{0}^{1} d u e^{i \bar{u} \lambda}\hat{\phi}_{p}(u)\label{definition-p-1}\;,\\
\langle M(P)|\bar{q}(0)\sigma_{\alpha\beta}\gamma_5 q(\lambda n/E )|0\rangle
&=& -f_M \mu^{M}_{\chi} [\bar{n}_{\alpha},n_{\beta}]
\int_{0}^{1} d u e^{i \bar{u} \lambda}\hat{\phi}_{\sigma}(u)\;.\label{definition-2}
\end{eqnarray}

We now show that the equations of motion for $\phi_{p}(u)$ and $\phi_{\sigma}(u)$ in the energetic meson limit 
is  
\begin{eqnarray}
\hat{\phi}_{p}(u) &=& \hat{\phi}_{\sigma}(u)\;.
\end{eqnarray}
Let's start from the following equation
\bee
if_M \mu^{M}_{\chi} (\bar{u}\hat{\phi}_{p}(u))=\int_{0}^{\infty}\frac{d\lambda}{2\pi}\frac{e^{-i\bar{u}\lambda }}{E}
\langle M|\bar{q}(0)\gamma_5 i n\cdot\partial(\lambda n/E )q(\lambda n/E )|0\rangle -f_M \mu^{M}_{\chi}\;.
\eee
To arrive at the above equation, we have assumed the boundary conditions 
\bee
\lim_{\lambda\to\infty} e^{-i\bar{u}\lambda }\langle M|\bar{q}(0)\gamma_5 q(\lambda n/E)|0\rangle=0\;,
\lim_{\lambda\to 0} e^{-i\bar{u}\lambda }\langle M|\bar{q}(0)\gamma_5 q(\lambda n/E)|0\rangle=-if_M \mu^{M}_{\chi}\;,
\eee
where the first holds due to the large fluctuation in $e^{-i \bar{u}\lambda}$ under $\lambda\to \infty$,
and the second is the normalization condition. 
By using the identity 
$n\cdot (i\partial)=\s{n}i\s{\partial}-\sigma_{\alpha\beta}n^{\alpha}(\partial^{\beta})$
and the fact that 
\[
\int_{0}^{\infty}\frac{d\lambda}{2\pi}\frac{e^{-i\bar{u}\lambda }}{E}
\langle M|\bar{q}(0)\gamma_5 \sigma_{\alpha\beta}n^{\alpha}\partial^{\beta}(\lambda n/E )q(\lambda n/E )|0\rangle
=-i f_M \mu^{M}_{\chi} (\bar{u}\hat{\phi}_{\sigma}(u)) -f_M\mu^{M}_{\chi}\;,
\]
we then obtain
\begin{eqnarray}
if_M \mu_M (\bar{u}\hat{\phi}_{p}(u))=\int_{0}^{\infty}\frac{d\lambda}{2\pi}e^{-i\bar{u}\lambda }
\langle M|\bar{q}(0)\gamma_5 \frac{\s{n}}{E}i\s{\partial}(\lambda/E n)q(\lambda n)|0\rangle
+i f_M \mu^{M}_{\chi}(\bar{u}\hat{\phi}_{\sigma}(u))\;,
\end{eqnarray}
where $\hat{\phi}_{\sigma}(u)$ has been defined in Eq.~(\ref{definition-2}).

For latter uses, we use the identity
\bee
\gamma_5 \sigma_{\alpha\beta}=\frac{i}{2}\epsilon_{\alpha\beta\eta\lambda}\sigma^{\eta\lambda}
\eee
to re-express $\hat{\phi}_{\sigma}(u)$ in a form as
\bee
f_M \mu^{M}_{\chi}\hat{\phi}_{\sigma}(u)=
-\frac{i}{2}
\int_0^{\infty}\frac{d\lambda}{2\pi}e^{-i\bar{u}\lambda}
\langle M|\bar{q}(0)\epsilon_{\perp}^{\alpha\beta}\sigma_{\alpha\beta}q(\lambda n/E)|0\rangle\;,
\eee
in which $\epsilon_{\perp}^{\alpha\beta}=\epsilon^{\alpha\beta\gamma\lambda}n_{\gamma}\bar{n}_{\lambda}$.
This form for $\hat{\phi}_{\sigma}(u)$ is convenient for our following calculations of
the hard scattering kernels.  
The equations of motion for the $q$ in the energetic limit is 
\[ i\s{\partial}(\lambda n/E)q(\lambda n/E)=0 + O(1/E^2)\;, 
\]
which is valid up-to $O(1/E^2)$.
We finally arrive at Eq.~(\ref{equation-2}).

By using $\hat{\phi}_{\sigma}(u)$, 
there is no the so-called projection ambiguity in the calculations
of the hard scattering kernels,
because there are no coordinate variables in the spin projector of $\hat{\phi}_{\sigma}(u)$.
In the energetic meson limit, 
the ambiguity that the same $\phi_{\sigma}(u)$ is defined for different components with different magnitudes 
of the spin projector has been avoided.
We conclude that
$\hat{\phi}_{p}(u)$ and $\hat{\phi}_{\sigma}(u)$ are more appropriate than
$\phi^c_{p}(u)$ and $\phi^c_{\sigma}(u)$ for uses in the calculation of the $X_H$ term 
and similar twist-3 contributions.

The model for $\hat{\phi}_{p}(u)$ can be obtained by solving  Eq.~(\ref{equation-2}).
Since the partons in the two parton Fock state of a light pseudoscalar meson can share the meson's momentum equally,
we can assume that the parameterization models for $\hat{\phi}_{p}(u)$ and $\hat{\phi}_{\sigma}(u)$ 
can have the asymptotic form $6u(1-u)$, for simplicity.
Because now the $\hat{\phi}_{p}(u)$ is no longer a constant, 
if the $\phi_{p}(u)$ in $X_{H}$ can be replaced by $\hat{\phi}_{p}(u)$,
then the end-point divergent problem can be resolved.
However, this requires us to examine the contributions related to $X_H$ term whether they are completely
coming from the $\hat{\phi}_{p}(u)$ under the energetic meson limit.
It is necessary to develop an appropriate calculation scheme for the twist-3 contributions.

\section{The collinear expansion}\label{sec:sec3}
As mentioned previously,
the spin projector for $\phi_{\sigma}(u)$ contains a coordinate variable $z$.
This leads to a projection ambiguity problem.
For comparison, we will first review what this problem is. 
Next, we will introduce a collinear expansion method for calculations of the hard scattering kernels of the one loop
corrections to the matrix element.

\subsection{The spin projection ambiguity and infrared divergences}
As mentioned in previous sections,
the spin projector $[P_{\alpha}, z_{\beta}]$ for $\phi_{\sigma}(u)$ contains a coordinate variable $z$.
There are two methods for performing the calculations of the contributions from the $\phi_{\sigma}(u)$.
The first method proposed by Beneke and Neubert (BN)\cite{Beneke:2003zv} is 
to separate the $z$ in $[P_{\alpha}, z_{\beta}]$ into its collinear and transverse parts as
\begin{eqnarray}
z_{\beta}\to (-i)\frac{\partial}{\partial k_{\beta}}=(-i)\left(\frac{n^{\beta}}{E}\frac{\partial}{\partial u}
+\frac{\partial}{\partial k_{\perp\beta}}+\cdots\right)\;,
\end{eqnarray}
where $k$ is assumed to be the momentum carried by the quark.
The $\partial/\partial u$ is defined to be applied on $\phi_{\sigma}(u)$,
and the $\partial/\partial k_{\perp\beta}$ is interpreted to act
on the associated hard scattering kernel for $\phi_{\sigma}(u)$.
However, it is also legal to let the whole momentum derivative $\partial/\partial k_{\beta}$ 
applied on the hard scattering kernel.
The latter method was proposed by Du, Yang, and Zhu (DYZ) \cite{Du:2001ns,Du:2001iu}.
However, both methods cannot avoid the infrared divergence in the $X_H$ term.

\subsection{A preface to collinear expansion method}
In this section, the CE calculation scheme proposed by Yeh in \cite{Yeh:2007} will be used for calculations of
$O(\alpha_s)$ contributions.
The CE scheme is only applicable 
to calculate the decay amplitudes in the energetic meson limit.
Under the CE scheme, the calculated twist-3 contributions are interpreted to be composed of collinear partons. 
A higher twist, which is composed of collinear partons, 
is usually called a dynamical power correction \cite{Qiu:1988dn}.
There exist other types of power corrections, such as the power corrections from soft gluon or renormalon contributions.
We identify these as non-partonic power corrections.
For these non-partonic power corrections, the CE may not be applicable.
To include these non-partonic power corrections within QCDF requires further assumptions.
For example, the soft gluon contributions are better determined by 
the QCD sum rules or the lattice QCD. 
In this work, we only investigate how the dynamical power corrections can be calculated by the CE method.

The idea of CE was first made by Polizer \cite{Politzer:1980me}. 
The systematical method was latter developed by Ellis, Furmanski and Petronzio (EFP) \cite{Ellis:1982wd,Ellis:1982cd}. 
Using CE, Ellis {\it el at} showed that, for DIS processes, the twist-4 power corrections (corrections suppressed by $Q^{-2}$ with $Q$ the relevant hard scale in the processes) can be factorized into its short distance and long distance parts as the factorization of the leading twist contributions.
However, a parton interpretation for the twist-4 power corrections is lost. 
To recover the parton model picture for the twist-4 power corrections, 
Qiu \cite{Qiu:1988dn} introduced a Feynman-diagram approach  to re-formula the EFP's method.
In this Feynman-diagram language, a parton model interpretation for the twist-4 power corrections becomes trivial.

\subsection{Preliminary}
The organization of this section is as following.
We will first describe how CE can be applied for tree level diagrams.
We then apply CE to calculate one loop corrections to the matrix elements under the two parton approximation.
The factorization of the amplitudes of the one loop corrections will be shown up-to twist-3 order.
In the following, we will use $1/E$ instead of using $1/m_b$ to represent the twist-3 order. 
To be specific, we shall consider the decay processes of a $B$ meson into two pseudoscalar light mesons.
The decay processes involve three restrictedly ordered energy scales: 
the $W$ boson mass $\mu_W\sim M_W$ scale, the factorization scale $\mu_F\sim m_b$, 
and the characteristic energy scale of nonperturbative QCD $\Lambda_{QCD}$.   
With the help of operator product expansion (OPE), the relevant $|\Delta B|=1$ effective Hamiltonian
is given by 
\bee
H&=&\frac{G_F}{\sqrt{2}}[\sum_{q=u,c}v_q(C_1(\mu)Q_1^q(\mu)+C_2(\mu)Q_2^q(\mu))\nn
&&+\sum_{k=3}^{10}C_k(\mu)Q_k(\mu))-v_t(C_{7\gamma}(\mu)Q_{7\gamma}(\mu)\nn
&&+C_{8G}(\mu)Q_{8G}(\mu))]+H.C.\ ,
\eee  
where $v_q=V_{q b}V^*_{q d}$(for $b\to d$ transition) or $v_q=V_{q b}V^*_{qt}$ 
(for $b\to s$ transition) and $C_i(\mu)$ are the Wilson coefficients 
which have been evaluated to next-to-leading order approximation by means of perturbative QCD 
and renormalization group \cite{Buras:1992zv,Ciuchini:1992tj,Ciuchini:1993vr,Buchalla:1995vs}. 
The four quark operators $Q_i$ are given by
\bee
Q_1^u&=&(\bar{u}_\alpha b_\alpha)_{V-A}(\bar{q}_\beta u_\beta)_{V-A}\ ,
Q_1^c=(\bar{c}_\alpha b_\alpha)_{V-A}(\bar{q}_\beta c_\beta)_{V-A}\ ,\nn
Q_2^u&=&(\bar{u}_\alpha b_\beta)_{V-A}(\bar{q}_\beta u_\alpha)_{V-A}\ ,
Q_2^c=(\bar{c}_\alpha b_\beta)_{V-A}(\bar{q}_\beta c_\alpha)_{V-A}\ ,\nn
Q_3&=&(\bar{q}_\alpha b_\alpha)_{V-A}\sum_{q^\prime}(\bar{q}^\prime_\beta q^\prime_\beta)_{V-A}\ ,
Q_4=(\bar{q}_\beta b_\alpha)_{V-A}(\sum_{q^\prime}(\bar{q}^\prime_\alpha q^\prime_\beta)_{V-A}\ ,\nn
Q_5&=&(\bar{q}_\alpha b_\alpha)_{V-A}\sum_{q^\prime}(\bar{q}^\prime_\beta q^\prime_\beta)_{V+A}\ ,
Q_6=(\bar{q}_\beta b_\alpha)_{V-A}\sum_{q^\prime}(\bar{q}^\prime_\alpha q^\prime_\beta)_{V+A}\ ,\nn
Q_7&=&\frac{3}{2}(\bar{q}_\alpha b_\alpha)_{V-A}\sum_{q^\prime}e_{q^\prime}
(\bar{q}^\prime_\beta q^\prime_\beta)_{V+A}\ ,
Q_8=\frac{3}{2}(\bar{q}_\beta b_\alpha)_{V-A}
(\sum_{q^\prime}e_{q^\prime}\bar{q}^\prime_\alpha q^\prime_\beta)_{V+A}\ ,\nn
Q_9&=&\frac{3}{2}(\bar{q}_\alpha b_\alpha)_{V-A}
\sum_{q^\prime}e_{q^\prime}(\bar{q}^\prime_\beta q^\prime_\beta)_{V-A}\ ,
Q_{10}=\frac{3}{2}(\bar{q}_\beta b_\alpha)_{V-A}
\sum_{q^\prime}e_{q^\prime}(\bar{q}^\prime_\alpha q^\prime_\beta)_{V-A}\ ,\nn
\eee
and
\bee
Q_{7\gamma}&=&-\frac{e}{8\pi^2}m_b(\bar{q}_\alpha \sigma^{\mu\nu}(1+\gamma_5)b_\alpha)F_{\mu\nu}\ ,\nn
Q_{8G}&=&-\frac{g}{8\pi^2}m_b(\bar{q}_\alpha \sigma^{\mu\nu}(1+\gamma_5)t^a_{\alpha\beta}b_\beta)G^a_{\mu\nu}\ , 
\eee
where $Q_1^q$ and $Q_2^q$ are the tree operators, 
$Q_3-Q_{6}$ the QCD penguin operators, 
$Q_7-Q_{10}$ the electroweak penguin operators, 
and $Q_{7\gamma}$ and $Q_{8,G}$ the magnetic and chromo-magnetic penguin operators.

The contributions between the scales of $\mu_F$ 
and $\mu_W$ are attributed to the Wilson coefficients, 
and the contributions between the scales of $\mu_F$ and $\Lambda_{QCD}$ 
are then included into the matrix elements of the operators.
By choosing an appropriate regularization method for the infrared singularities,
the Wilson coefficients can be calculated to be independent of the external states 
and can be factorized from the matrix element $\langle M_1(p)M_2(q)|Q_i|\bar{B}(P_B)\rangle$. 
For an energetic $M_2$ meson in the $\bar{B}\to M_1 M_2$ processes,
the color transparency leads to the naive factorization \cite{Bauer:1986bm} for the matrix element such that
the matrix element $\langle M_1(p)M_2(q)|Q_i|\bar{B}(P_B)\rangle$ can be written as a product of 
a transition form factor and a decay constant in the following way
\bee\label{naive-fact}
\langle M_1(p)M_2(q)|Q_i|\bar{B}(P_B)\rangle\approx \langle M_1(p)|J_1^{i}|B(P_B)\rangle \langle M_2(q)|J_2^{i}|0\rangle
+ (M_1 \leftrightarrow M_2)\;.
\eee
The form factors are defined as
\bee
&&\langle M_1(p)|\bar{q}(0)\gamma^{\mu}b(0)|\bar{B}\rangle \nn
&=& F_+^{B\to M_1}(q^2)(P_{B}^{\mu} + p^{\mu})
+[F_0^{B\to M_1}(q^2)-F_+^{B\to M_1}(q^2)]\frac{m_B^2-m_{M_1}^2}{q^2}q^{\mu}\;.
\eee
The form factors coincide as $q^2=0$, $F^{B\to M_1}_+ (0)=F_0^{B\to M_1}(0)$.
Due to the conservation of the currents $J_{1,2}^{i}$,  
the scale invariance of the decay amplitudes is broken under the naive factorization.
The radiative corrections to the matrix elements are necessary to compensate the scale dependence
of the matrix element \cite{Beneke:1999br,Beneke:2000ry, Beneke:2001ev}.

The bottom meson momentum $P_{B}$ is defined in a light-cone coordinate frame such that it can be written as 
$P_{B}^{\mu}=(p^{\mu}+q^{\mu})$ with two light-like vectors $q^{\mu}=(q^+,q^-,q^{i}_{\perp})=E\bar{n}^{\mu}$
and $p^\mu=(p^+,p^-,p^i_{\perp})=E n^{\mu}$ in the plus and minus directions, respectively.
The $M_1$ meson is defined to receive the spectator quark of the bottom meson as its quark component.
The $M_2$ meson is defined as the emitted meson from the weak interaction vertex.
We associate the momentum $p^{\mu}$ for the $M_1$ meson and the momentum $q^{\mu}$ for the $M_2$ meson.
The following metric and antisymmetric tensors are useful in our calculations
\bee
&&w^{\mu}_{\mu^\prime}=g^{\mu}_{\mu^\prime}-\bar{n}^{\mu}n_{\mu^\prime}\;,\;
\bar{w}^{\mu}_{\mu^\prime}=g^{\mu}_{\mu^\prime}-n^{\mu}\bar{n}_{\mu^\prime}\;,\nn
&&d^{\mu}_{\mu^\prime}=g^{\mu}_{\mu^\prime}-\bar{n}^{\mu}n_{\mu^\prime}-n^{\mu}\bar{n}_{\mu^\prime}\;,\;
\bar{d}^{\mu}_{\mu^\prime}=g^{\mu}_{\mu^\prime}-n^{\mu}\bar{n}_{\mu^\prime}-\bar{n}^{\mu}n_{\mu^\prime}\;,\nn
&&\epsilon^{\mu\nu}_{\perp}=\epsilon^{\mu\nu\alpha\beta}n_{\alpha}\bar{n}_{\beta}\;,\;
\bar{\epsilon}^{\mu\nu}_{\perp}=\epsilon^{\mu\nu\alpha\beta}\bar{n}_{\alpha}n_{\beta}\;.
\eee
These tensors satisfy the following identities
\bee
q_{\mu}w^{\mu}_{\mu^\prime}=0\;,\;
p_{\mu}\bar{w}^{\mu}_{\mu^\prime}=0\;,\;
d^{\mu}_{\mu}=\bar{d}^{\mu}_{\mu}=2\;,\;
\epsilon^{\mu\nu}_{\perp}\epsilon_{\perp\mu\nu}=\bar{\epsilon}^{\mu\nu}_{\perp}\bar{\epsilon}_{\perp\mu\nu}= 2\;.
\eee
For a collinear loop parton of the $M_2$ meson, 
it is convenient to parametrize its momentum $l^{\mu}$ into its components proportional to 
the meson momentum $q^{\mu}$,
the light-cone vector $n^{\nu}$, and the transversal directions
\bee\label{eq3}
l^{\mu} &=& E \bar{n}^{\mu} + \frac{l^2+ l_{\perp}^2}{2 E}n^{\mu}+l^{\mu}_{\perp}\;.
\eee 
For convenience, we further define the collinear component, $\hat{l}^{\mu}$, 
the on shell component, $l_{L}^{\mu}$, and the off shell component, $l_{S}^{\mu}$, of the momentum $l^{\mu}$ as
\bee
\hat{l}^{\mu}&=& n\cdot l \bar{n}^{\mu}\;,\nn
l^{\mu}_{L}&=& \hat{l}^{\mu}+\frac{l_{\perp}^2}{2n\cdot l}n^{\mu}+l_{\perp}^{\mu}\;,\nn
l_{S}^{\mu}&=&\frac{l^2}{2n\cdot l}n^{\mu}\;.
\eee
For the $M_1$ meson, the momenta of the collinear loop partons flowing through the $M_1$ meson 
can be parametrized in a similar way.
We parametrize the collinear momentum in terms of $p^{\mu}$ and $\bar{n}^{\mu}$.
Because the mass effects from the mesons' masses and the light quarks' masses are very small in the decay processes
considered in this work, we shall neglect them completely in this paper.
In this approximation, we let $m_{M_{i}}=0$, $i=1,2$, and $q^2=0$.
And the loop partons are assumed massless.
 
According to Eq.~(\ref{eq3}),
a parton propagator can be separated into its long distance part and short distance part.
If we write the loop parton propagator as
\bee\label{lp1}
F(y,z)&=&\int\frac{d^4 l}{(2\pi)^4}e^{i l\cdot (y-z)}[F_L(l)+F_S(l)]\nn
      &=& F_L(y,z)+ F_S(y,z)\;,
\eee
where
\bee
F_L(l)=\frac{i\s{l}_{L}}{l^2-i\epsilon}\;,\;\;
F_S(l)=\frac{i \s{n}}{2n\cdot l-i\epsilon}\;.
\eee
The  $F_{L}(l)$ propagator corresponds to the long distance part of the propagator,
since $F_L(y,z)\propto \theta(y-z)$. 
The $F_{S}(l)$ propagator represents the short distance part because $F_S(y,z)\propto \delta(y-z)$.
We now describe one important property of the long distance propagator $F_{L}(l)$.
As $F_{L}(l)$ contacts with a $\s{q}n^{\mu}$ component of a vertex $\gamma^{\mu}$ in the parton amplitudes,
the $\s{q}n^{\mu}$ vertex will extract one short distance propagator $F_{S}(l)$ 
and one interaction vertex $i\gamma_{\nu}$ from the hadron amplitude in the following ways
\bee\label{long-prop-1}
\frac{i\s{l}_{L}}{l^2}\s{q}=\frac{i\s{l}_L}{l^2}(i\gamma_{\nu})\frac{i\s{n}}{2n\cdot l}\s{q}(l-\hat{l})^{\nu}\;.
\eee
The momentum factor $(l-\hat{l})^{\nu}$ is then absorbed by the hadron amplitude due to the Ward identity 
\cite{Ellis:1982wd,Ellis:1982cd,Qiu:1988dn}.
The above identity is obtained by a simple manipulation \cite{Qiu:1988dn}.
We first insert an identity $1=(\s{l}^2)/l^2$ into the left hand side of Eq.~(\ref{long-prop-1})
and expand the $\s{l}$ as $\s{l}_L+\s{l}_S$ as the following 
\begin{eqnarray}\label{long-prop-2}
\frac{i\s{l}_{L}}{l^2}\frac{\s{l}\s{l}}{l^2}\s{q}
=\frac{i\s{l}_{L}}{l^2}\frac{(\s{l}_L +\s{l}_S)(\s{l}_L +\s{l}_S)}{l^2}\s{q}
\end{eqnarray}
Since $(\s{l}_L)^2=0=(\s{l}_S)^2$,
we then obtain
\begin{eqnarray}
\frac{i\s{l}_{L}}{l^2}\frac{\s{l}\s{l}}{l^2}\s{q}
=\frac{i\s{l}_{L}}{l^2}\frac{(\s{l}_L \s{l}_S +\s{l}_S \s{l}_L)}{l^2}\s{q}
\end{eqnarray} 
where the first term $\s{l}_L \s{l}_S$ in the right hand side of the above equation 
leads to a vanishing result as it contacts with its
left hand side ${i\s{l}_{L}}/{l^2}$ term.
The only contribution can only come from the second term $\s{l}_S \s{l}_L$.
We further expand the $\s{l}_L$ in the following way
\[
\s{l}_S \s{l}_L\s{q}=l^2\frac{\s{n}}{2n\cdot l}(n\cdot l \s{q} + \frac{l_{\perp}^2\s{n}}{2n\cdot l}+\s{l}_{\perp})\s{q}
\]
Due to $\s{q}^2=\s{n}^2=0$, the remaining result becomes
\[
l^2\frac{\s{n}}{2n\cdot l}(\s{l}_{\perp})\s{q}\;.
\]
By substituting the above back into Eq.~(\ref{long-prop-2}), 
Eq.~(\ref{long-prop-1}) is derived by noting that 
\[
\s{l}_{L}(i\gamma_{\alpha})(i\s{n})\s{q}(l-\hat{l})^{\alpha}=\s{l}_{L}\s{n}\s{l}_{\perp}\s{q}\;.
\]
Using Eq.~(\ref{long-prop-1}), one can systematically include the effects from the non-collinearity and off-shell-ness
of collinear partons external to the hard scattering function of a Feynman diagram.
This property of the long distance propagator plays an important role in our following analysis.

\begin{figure}
%{\centerline{\includegraphics{diagrams/fourparton_tree}}}
{\centerline{\includegraphics*[scale=0.5,viewport=0 580 700 800]{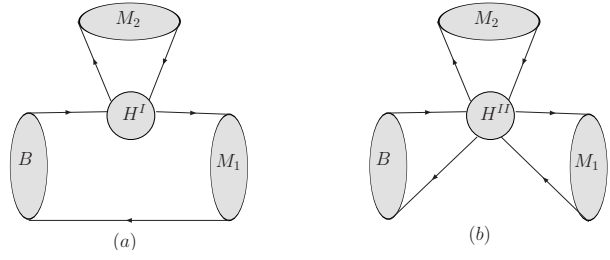}}}
\caption{(a)The type $I$ diagram for $B\to M_1 M_2$ decays. 
(b) The type $II$ diagram for $B\to M_1 M_2$ decays. 
The central blobs with symbols $H^{I,II}$ represent the parton scattering functions.
The initial state $B$ meson is represented by the circle with a symbol $B$ in the left hand side of each diagram.
The final state $M_1$ meson is represented by the circle with a symbol $M_1$ in the right hand side of each diagram.
The final state $M_2$ meson is represented by the circle with a symbol $M_2$ in the upper part of each diagram.
The lines with a arrow are the fermion partons.}
  \label{fig:fig1}                 
\end{figure}

Let's first describe how the collinear expansion method can be applied for tree level amplitudes to
recover the NF amplitudes.
After the tree level analysis, we will investigate how CE method can be applied for the one loop amplitudes
and show that the factorization theorem \eq{fact-1} up-to $O(1/E)$.
We begin from the diagrams of \fig{fig:fig1} for the $\bar{B}\to M_1 M_2$ process.
The parton interactions in the diagrams of \fig{fig:fig1}~(a) and (b) are assumed by 
an effective four quark operator 
$Q_{i}=(\bar{q}_{a}\Gamma_{i}b)(\bar{q}_{b}\Gamma_{i}q_{c})$.
In order to discuss the collinear expansion,
we propose to formally write 
the transition matrix element $\langle M_1 M_2|Q_{i}|\bar{B}\rangle$ involved in the amplitude 
for the diagrams of \fig{fig:fig1}~(a) and (b) as the following expression
\begin{eqnarray}\label{CE-tree-1}
\langle M_1 M_2|Q_{i}|\bar{B}\rangle
&=&F_{j}^{B\to M_{1}}(q^2)\int\frac{d^4 l_{M_{2}}}{(2\pi)^4}Tr[H_{ij}^{I}(l_{M_{2}})\phi^{M_{2}}(l_{M_{2}})]\nn
&&+\int\frac{d^4 l_{B}}{(2\pi)^4}
\int\frac{d^4 l_{M_{1}}}{(2\pi)^4}
\int\frac{d^4 l_{M_{2}}}{(2\pi)^4}
Tr[H_{i}^{II}(l_{M_{2}})\phi_B(l_B)\phi^{M_{1}}(l_{M_{1}})\phi^{M_{2}}(l_{M_{2}})]\nn
&&+\cdots\;,
\end{eqnarray}
where $F^{B\to M_1}_{j}(q^2)$ represents the transition form factor denoted by the bottom part of the diagram 
\fig{fig:fig1}~(a).
The scattering functions $H^{I,II}$ contain the parton interactions.
The $H^{I}$ function represents four parton interactions with multiple radiative gluons.
The $H^{II}$ function represents six parton interactions with multiple radiative gluons.
The dots denote higher order terms which could contain interactions between the three parton Fock state of $M_2$ 
and the other partons from the $M_1$ or $B$ mesons. 
The contributions from the three parton Fock state of $M_1$ or $M_2$ are neglected completely in this paper.
The dots' terms are not shown explicitly.

The diagrams in \fig{fig:fig1}~(a) and (b) represent the processes that 
the initial state $\bar{B}$ undertakes a transition into the final state $M_1$ by means of
the $b\to q_{1}q_{2}\bar{q}_{3}$ decays accompanying multiple radiative interactions.
The partonic radiative interactions associated with $b\to q_{1}q_{2}\bar{q}_{3}$ decays are collected
into $H^{I,II}$.
The probability for the transition of $\bar{B}$ into $b\bar{q}_s$ pair is denoted by $\phi_{B}(l_B)$.
The $q_{2}\bar{q}_{3}$ pair produced from the interaction center, i.e. $H^{I,II}$,
then combine to form the $M_{2}$ meson after a long distance travel away from the interaction center.
The probability for the transition of the $q_{2}\bar{q}_{3}$ pair into the $M_{2}$ 
is represented by $\phi^{M_{2}}(l_{M_{2}})$.
In the diagram of \fig{fig:fig1}~(a), 
the spectator quark $\bar{q}_s$ of $B$ combines with the $q_1$ produced from $H^{I}$ to form the $M_1$.
The transition from $\bar{B}$ into $M_1$ is represented by $F^{B\to M_1}$.
In the diagram of \fig{fig:fig1}~(b), the $\bar{q}_s$ gets involved in $H^{II}$ 
and then combines with $q_1$ to form $M_1$.
The probability for the transition of $\bar{q}_s q_1$ into $M_1$ is represented by $\phi^{M_{1}}(l_{M_{1}})$.

We first discuss the diagram in \fig{fig:fig1}~(a), which is represented by the first term of \eq{CE-tree-1}.
In the expression of the first term of \eq{CE-tree-1},
the scattering kernel $H_{ij}^{I}(l_{M_{2}})$ and the meson amplitude $\phi^{M_{2}}(l_{M_2})$
are correlated by the loop parton momentum $l_{M_2}$, the color indices, and the spin indices.
The loop parton momentum $l_{M_2}$ is defined to flow from the antiquark line to the quark line of $M_2$.
The expression for $\phi^{M_{2}}(l_{M_{2}})$ is defined as
\begin{eqnarray}
\phi^{M_{2}}(l_{M_{2}})=\int d^4z e^{-i\bar{l}_{M_{2}}\cdot z}
\langle M_{2}(q)|\bar{q}_{2}(0)q_{3}(z)|0\rangle\;,
\end{eqnarray} 
where $\bar{l}_{M_{2}}=q-l_{M_{2}}$ and the color and spin indices are not shown explicitly.
The $Tr$ denote traces over the color and spin indices.
To complete the factorization of the first term of \eq{CE-tree-1} into the short distance and long distance parts,  
we need to disentangle the correlations in the integration over $l_{M_2}$, the color indices, and the spin indices.

In order to derive a factorization theorem similar to \eq{fact-1},
we propose to employ the expansion scheme made in \cite{Yeh:2007}.
The $H^{I}_{ij}(l_{M_{2}})$  is expanded in $\alpha_s$ and $l_{M_{2}\perp}$.
Similarly, the $H^{II}_{i}$ is expanded in  $\alpha_s$ and $l_{M_{i}\perp}$, $i=1,2$.
Namely, we first expand $H^{I}_{ij}(l_{M_{2}})$ in $\alpha_s$ as
\bee\label{H-I-expansion}
H^{I}_{ij}(l_{M_{2}})&=&H^{I(0)}_{ij}+H^{I(1)}_{ij}(l_{M_{2}})+\cdots\;,\\
H^{II}_{i}(l_{M_1},l_{M_{2}})&=& H^{II(1)}_{i}(l_{M_1},l_{M_{2}})+\cdots\;
\eee
where $H^{I(0)}_{ij}$ is of order $O(\alpha_s^0)$ 
and $H^{I(1)}_{ij}$ and $H^{II(1)}_i$ are of order $O(\alpha^{1}_s)$.
The dots are of order $O(\alpha_s^n)$ with $n\ge 2$.
Each term in \eq{H-I-expansion} is then expanded in $l_{M_{1}\perp}$ or $l_{M_{2}\perp}$.
The expansion of $H^{I(1)}_{ij}$ and $H^{II(1)}_{ij}$ in $l_{M_{1}\perp}$ or $l_{M_{2}\perp}$ are left to 
latter discussions for one loop corrections.

\subsection{Tree level expansion}
The $H^{I(0)}_{ij}$ is independent of $l_{M_2}$ and $H^{I(0)}_{ij}=\Gamma_{i}\delta_{ij}\delta^{ab}$.
The color factor $\delta_{a b}$ is then absorbed by $\phi^{M_{2}}$.
By substituting $H^{I(0)}_{ij}$ back into the first term of \eq{CE-tree-1}, we arrive at
\bee
A_1= F_{j}^{B\to M_{1}}(q^2)\int\frac{d^4 l_{M_{2}}}{(2\pi)^4}Tr[H_{ij}^{I(0)}\phi^{M_{2}}(l_{M_{2}})]\;.
\eee 
We further use the integration transformation \cite{Ellis:1982wd,Qiu:1988dn} 
\begin{eqnarray}\label{Int-Tranf-1}
1=\int_{0}^{1} dv \delta(v- l_{M_2}\cdot n /q\cdot n)
=\int_{0}^{1} dv \int_{0}^{\infty}\frac{d\lambda_{2}}{2\pi} 
e^{i\lambda_{2}(v- l_{M_2}\cdot n /E)}
\end{eqnarray}
to rewrite $A_{1}$ as
\begin{eqnarray}
A_{1}=F_{j}^{B\to M_{1}}(q^2)\int_0^1 dv Tr[H_{ij}^{I(0)}\phi^{M_{2}}(v)]\;.
\end{eqnarray}
The distribution amplitude $\phi^{M_{2}}(v)$ is given by
\begin{eqnarray}
\phi^{M_{2}}(v)
=\int_{0}^{\infty}\frac{d\lambda_{2}}{2\pi} e^{iv\lambda_{2}}\phi^{M_{2}}(\lambda_{2})
\end{eqnarray}
where 
\begin{eqnarray}
\phi^{M_{2}}(\lambda_{2})
=\int d^4 z e^{-iq\cdot z}\int\frac{d^4 l_{M_{2}}}{(2\pi)^4} e^{i l_{M_2}\cdot ( z - \lambda_{2}n/E)}
\langle M_{2}(P_{2})|\bar{q}_{b}(0)q_{c}(z)|0\rangle\;.
\end{eqnarray}
We further use the fact that the integrations over $l_{M_2}$ and $z$ result in a delta function
\begin{eqnarray}
\int d^4 z\int\frac{d^4 l_{M_{2}}}{(2\pi )^4} e^{i l_{M_2}\cdot ( z -\lambda_{2}n/E )}
=\int d^4 z\delta^{(4)}(z-\lambda_{2}n/E)\;.
\end{eqnarray}
By this, we write $\phi^{M_{2}}(v)$ as
\begin{eqnarray}
\phi^{M_{2}}(v)
=\int_{0}^{\infty}\frac{d\lambda_{2}}{2\pi} e^{-i\bar{v}\lambda_{2}}
\langle M_{2}(q)|\bar{q}_{2}(0)q_{3}(\lambda_{2} \frac{n}{E})|0\rangle\;,
\end{eqnarray} 
where $\bar{v}=1-v$.
To factorize the remaining spin indices in the trace, we make use of the Fierz identity
\begin{eqnarray}
\delta_{ij}\delta_{kl}=\frac{1}{4}((\gamma^{\mu})_{ij}(\gamma_{\mu})_{kl}
+(\gamma^{\mu}\gamma_{5})_{ij}(\gamma_5\gamma_{\mu})_{kl}
+(\gamma_{5})_{ij}(\gamma_5)_{kl}
+\frac{1}{2}(\sigma^{\mu\nu}\gamma_5)_{ij}(\sigma_{\mu\nu}\gamma_5)_{kl})
\end{eqnarray} 
to obtain
\begin{eqnarray}
A_{1}\simeq F_{j}^{B\to M_{1}}(q^2)\sum_{k}\int_{0}^{1} dv \frac{1}{4}Tr[H^{(0)}_{ij}\Gamma^{\prime}_{k}]
Tr[\bar{\Gamma}^{\prime}_{k}\phi^{M_{2}}(v)]\;,
\end{eqnarray}
where $\Gamma^{\prime}_{k} \bar{\Gamma}^{\prime}_{k}=(\gamma^{\mu})(\gamma_{\mu}), 
(\gamma^{\mu}\gamma_{5})(\gamma_5\gamma_{\mu}),
(\gamma_{5})(\gamma_5),
\frac{1}{2}(\sigma^{\mu\nu}\gamma_5)(\sigma_{\mu\nu}\gamma_5)$ for $k=1,\cdots, 4$.
To project onto the specific spin state of $M_2$,
we introduce the following definitions
\begin{eqnarray}
Tr[\gamma^{\mu}\phi^{M_{2}}(v)] &=& 0 \;,\nonumber\\
Tr[\gamma^{\mu}\gamma_5\phi^{M_{2}}(v)] &=& -if_{M_{2}} E \bar{n}^{\mu} \phi^{M_2}_{P}(v) \;,\nonumber\\
Tr[\gamma_5\phi_{M_{2}}(v)] &=&-i f_{M_{2}}\mu_{M_2} \hat{\phi}^{M_2}_{p}(v) \;,\nonumber\\
Tr[\sigma^{\mu\nu}\gamma_{5}\phi^{M_{2}}(v)] 
&=& -f_{M_{2}}\mu_{M_2}[\bar{n}^{\mu},n^{\nu}] \hat{\phi}^{M_2}_{\sigma}(v) \;,
\end{eqnarray}
where  $\phi^{M_2}_{P}(v)$ is the twist-2 LCDA,
$\hat{\phi}^{M_2}_{p}(v)$ and $\hat{\phi}^{M_2}_{\sigma}(v)$ are twist-3 LCDAs,
and the LCDAs of twist order higher than three have been omitted.
The final result depends on Dirac structure of the effective operator $Q_{i}$.
The Dirac structure can be 
$\Gamma_{i}\otimes\Gamma_{i}=(\gamma_{\mu}(1-\gamma_5))\otimes (\gamma^{\mu}(1\pm \gamma_5))$ 
for $(V-A)(V\pm A)$ operators, 
and $\Gamma_{i}\otimes\Gamma_{i}=-2((1-\gamma_5))\otimes ((1+\gamma_5))$ for
$-2(S-P)(S+P)$ operators.
By substituting the Dirac matrices for the operators and using the definitions for the spin state of $M_2$,
the amplitude becomes
\begin{eqnarray}
A_{1}^{(V-A)(V\pm A)}=\pm if_{M_{2}}F_{+}^{B\to M_{1}}(q^2)m_{B}^2\int_{0}^1 dv \phi^{M_{2}}_{P}(v)\;, \nonumber\\
A_{1}^{-2(S-P)(S+P)}=if_{M_{2}}F_{+}^{B\to M_{1}}(q^2)m_{B}^2 r_{\chi}^{M_2}\int_{0}^1 dv \hat{\phi}^{M_{2}}_{p}(v)\;, 
\end{eqnarray}
where $r_{\chi}^{M_2} = 2 m_{M_2}^2/(\bar{m}_b (\bar{m}_{q_{2}} + \bar{m}_{\bar{q}_3}))$ 
with $\bar{m}_b, \bar{m}_{q_2}, \bar{m}_{q_{3}}$ being current quark masses.
In the above example,
we have present how the collinear expansion method can be applied for tree level diagrams.
As a necessary condition, it can recover the result obtained by naive factorization (NF).

\subsection{Factorizable one loop diagrams}
\begin{figure}
%{\centerline{\includegraphics{diagrams/fourparton_tree}}}
{\centerline{\includegraphics*[viewport=80 530 480 620]{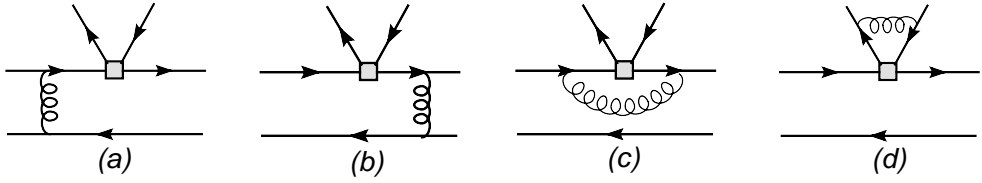}}}
\caption{The one loop factorizable diagrams. The external meson states are not shown.}
  \label{fig:fig2}                 
\end{figure}
The $O(\alpha_s)$ scattering function $H^{I(1)}_{ij}(l_{M_2})$ are classified into factorizable and
non-factorizable parts
\bee
H^{I(1)}_{ij}(l_{M_2})=H^{I(1),F}_{ij}(l_{M_2})+H^{I(1),NF}_{ij}(l_{M_2})\;.
\eee
The $H^{I(1),F}_{ij}(l_{M_2})$ corresponds to the the diagram in \fig{fig:fig2}~(d).
However, there are factorizable one loop diagrams as depicted in \fig{fig:fig2}~(a)-(c),
which are attributed to the $B\to M_1$ form factors $F_{j}^{B\to M_1}$.
The sum of the first two diagrams \fig{fig:fig2}~(a)-(b) for the form factor are dominated by soft gluons.
Under QCD factorization approach, 
these leading soft contributions are defined to be absorbed by the physical form factors 
$F_{j}^{B\to M_1}$ \cite{Beneke:2000ry}.
The third factorizable diagram is to re-normalize the $V-A$ current 
associated with the $b\to q_{1}$ process.
The fourth factorizable diagram is to re-normalize the $V-A$ current 
associated with the $M_2$ meson and the equations of motion.
Since the $V-A$ current is conserved, 
the $V-A$ current and the equations of motion receive no renormalization.
As a result, 
the factorizable diagrams lead to finite contributions under QCDF approach. 

\subsection{Non-factorizable one loop diagrams and collinear expansion}
\begin{figure}
%{\centerline{\includegraphics{diagrams/fourparton}}}
{\centerline{\includegraphics*[scale=0.7,viewport=80 410 540 600]{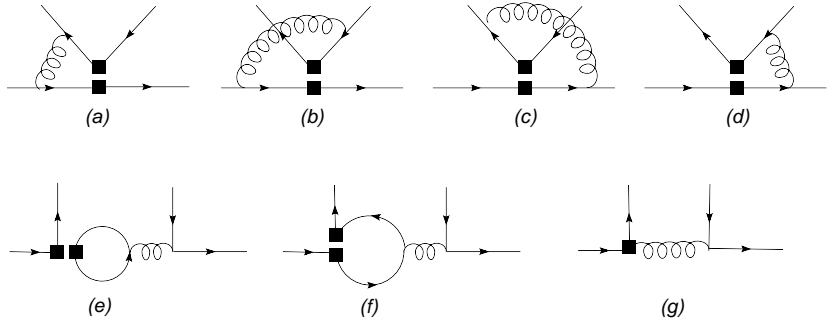}}}
\caption{(a)-(d) The vertex diagrams. (e)-(g) The penguin diagrams.}
  \label{fig:fig3}                 
\end{figure}
The $H^{I(1),NF}_{ij}(l_{M_2})$ corresponds to the non-factorizable one loop diagrams in \fig{fig:fig3}~(a)-(g).
According to the types of the diagrams, we write $H^{I(1),NF}_{ij}(l_{M_2})$ as
\bee
H^{I(1),NF}_{ij}(l_{M_2})=V_{ij}(l_{M_2})+P_{ij}(l_{M_2})+P_{8g,ij}(l_{M_2})\;,
\eee
where $V_{ij}$ denotes the vertex corrections from the diagrams in \fig{fig:fig3}~(a)-(d),
$P_{ij}$ denotes the penguin corrections from the diagrams in \fig{fig:fig3}~(e)-(f),
and $P_{8g,ij}$ denotes the magnetic dipole moment corrections from the diagrams in \fig{fig:fig3}~(g).
Each type of corrections will be analyzed in following each subsection, respectively.

We take the vertex correction $V_{ij}(l_{M_2})$ as an example 
to show the main feature of the application of the CE method to the one loop corrections.
In the following, we omit the subscript $M_2$ in $l_{M_2}$ to simplify the notation.
We write the amplitude for the diagrams in \fig{fig:fig3}~(a)-(d) as
\begin{eqnarray}\label{vertex-example}
A_{V}\sim F^{B\to M_1}(q^2)\int\frac{d^4 l}{(2\pi)^4}\Tr[V(l)\phi^{M}(l)]\;.
\end{eqnarray}
In the above expression, the color and spin indices and the subscription of the form factor have been omitted 
for simplification.
The irrelevant factors associated with the amplitude are also omitted for simplicity.
To separate the collinear limiting $V(\hat{l})$ of $V(l)$ from the others, 
we make a Taylor expansion for $V(l)$ with respect to $\hat{l}$ as
\begin{eqnarray}\label{Taylor-expansion-Vertex-1}
V(l)=V(\hat{l})
+\left.{\frac{\partial V(l)}{\partial l^{\alpha}}}\right|_{l=\hat{l}}
(l-\hat{l})^{\alpha}+\cdots\;,
\end{eqnarray}
where dots are higher derivative terms. 
The first derivative term is shown for our latter comparison with the BN and DYZ schemes
and its related contribution will not be considered in this work.
By using 
\[
w^{\alpha}_{\alpha^{\prime}}=g^{\alpha}_{\alpha^{\prime}}-\bar{n}^{\alpha}n_{\alpha^{\prime}}\;,
\]
we can write
\[
(l-\hat{l})^{\alpha}=w^{\alpha}_{\alpha^{\prime}}l^{\alpha^{\prime}}\;.
\]
$l^{\alpha^{\prime}}$ is then absorbed into $\phi^{M_2}(l)$ by Ward identity.
With substitution of the first two terms of \eq{Taylor-expansion-Vertex-1} back into \eq{vertex-example}, 
the result appears
\begin{eqnarray}\label{vertex-d-1}
A_{V}\sim
\int\frac{d^4 l}{(2\pi)^4}\Tr[V(v)\phi^{M_{2}}(l)]
+ \int\frac{d^4 l}{(2\pi)^4}\Tr[V_{\alpha}(v,v)w^{\alpha}_{\alpha^{\prime}}
\phi^{M_{2}\alpha^{\prime}}_{\partial}(l,l)]+\cdots\;,
\end{eqnarray} 
where we have used the low energy theorems
\begin{eqnarray}
V(v)&\equiv& V(\hat{l})\;,\\
V_{\alpha}(v,v) &\equiv&
\left.{\frac{\partial V(l)}{\partial l^{\alpha}}}\right|_{l=\hat{l}}\;,
\end{eqnarray}
and have defined
\begin{eqnarray}\label{twist-4-DA-1}
\phi^{M_{2}\alpha^{\prime}}_{\partial}(l,l)=
\int d^4 z e^{-i\bar{l}\cdot z}\langle M_{2}(q)|\bar{q}(0)i\partial^{\alpha^{\prime}}(z)q(z)|0\rangle\;.
\end{eqnarray}
The parameterization $\hat{l}^{\mu}=v E \bar{n}^{\mu}$ has been used.
The $v$ has the meaning of momentum fraction carried by the partons. 
The collinear momenta $\hat{l}$ of the partons are defined to be parallel to the $M_2$'s momentum $q$ 
in the $\bar{n}^{\mu}$ direction.  
The functions $V(v)$ and $V_{\alpha}(v,v)$ have been introduced to emphasize the fact that, 
under the collinear limiting, $V(\hat{l})$ and $V_{\alpha}(\hat{l},\hat{l})$ and $\phi^{M_2}(l)$ are only
correlated by $v$.
Similar integral transformations to \eq{Int-Tranf-1} can be used to rewrite \eq{vertex-d-1} as
\bee
A_{V}\sim
\int_0^1 dv\Tr[V(v)\phi^{M_{2}}(v)]
+ \int_0^1 dv \Tr[V_{\alpha}(v,v)w^{\alpha}_{\alpha^{\prime}}
\phi^{M_{2}\alpha^{\prime}}_{\partial}(v,v)]+\cdots\;,
\eee
where $\phi^{M_{2}}(v)$ and
\bee
\phi^{M_{2}\alpha^{\prime}}_{\partial}(v,v)=
\int_{0}^{\infty}\frac{d\lambda}{2\pi} e^{-i\lambda\bar{v}}
\langle M_{2}(q)|\bar{q}_{2}(0)i\partial^{\alpha^\prime}(\lambda \frac{n}{E})q_{3}(\lambda \frac{n}{E})|0\rangle
\eee
have been used.

\begin{figure}
%{\centerline{\includegraphics{diagrams/fourparton}}}
{\centerline{\includegraphics*[scale=0.4,viewport=60 280 800 800]{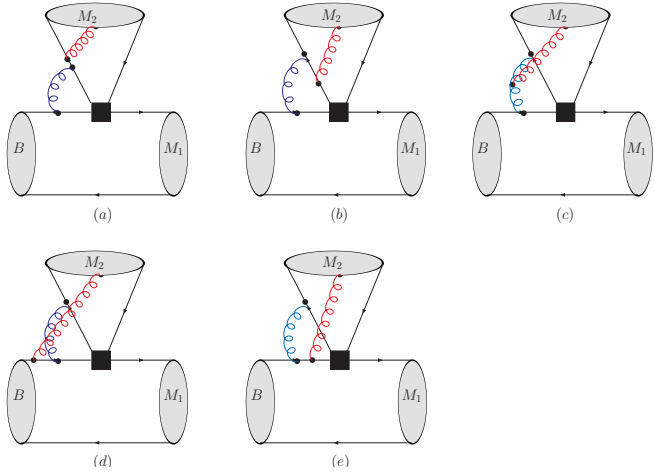}}}
\caption{The diagrams for $V^{(1)}_{\alpha}(\hat{l},\hat{l}^{\prime})$. 
For each diagram, there are similar diagrams with different gluon attachment.}
  \label{fig:fig4}                 
\end{figure} 
For gauge invariance, we need to consider the diagrams as depicted in \fig{fig:fig4}.
We write the amplitude for the diagrams in \fig{fig:fig4} as
\begin{eqnarray}\label{vertex-A-0}
A^{\prime}_{V}\sim F^{B\to M_1}(q^2)\int\frac{d^4 l}{(2\pi)^4}\int\frac{d^4 l^{\prime}}{(2\pi)^4}
\Tr[V_{\alpha}(l,l^{\prime})\phi^{M_{2}\alpha}_{A}(l,l^{\prime})]\;,
\end{eqnarray}
where the function $V_{\alpha}(l,l^{\prime})$ represents 
\begin{eqnarray}\label{vertex-A-00}
V_{\alpha}(l,l^{\prime})=V^{(a)}_{\alpha}(l,l^{\prime})
+V^{(b)}_{\alpha}(l,l^{\prime})
+V^{(c)}_{\alpha}(l,l^{\prime})
+V^{(d)}_{\alpha}(l,l^{\prime})
+V^{(e)}_{\alpha}(l,l^{\prime})\;.
\end{eqnarray}

Each term in the right hand side of Eq.~(\ref{vertex-A-00}) corresponds to the diagrams as depicted in 
\fig{fig:fig4}~(a)-(e), respectively.
For each diagram in \fig{fig:fig4}, there are other similar diagrams with different gluons' attachments.
They are not shown explicitly. 
The amplitude $\phi^{M_{2}\alpha}_{A}(l,l^{\prime})$ is defined as
\begin{eqnarray}\label{twist-4-DA-2}
\phi^{M_{2}\alpha}_{A}(l,l^{\prime})=
\int d^4 y \int d^4 z e^{-i(\bar{l}^{\prime}-\bar{l})\cdot y}e^{-i\bar{l}\cdot z}
\langle M_{2}(q)|\bar{q}(0)(-)g A^{\alpha,a}(y)T^{a}q(z)|0\rangle\;.
\end{eqnarray}
In the following, we only consider the contributions related to $V^{(b)}_{\alpha}(l,l^{\prime})$,
which results in contributions related to $\langle M_{2}|\bar{q}(0)iD^{\mu}(z^{\prime})q(z)|0\rangle$.
The other contributions, $V^{(i)}_{\alpha}$, $i\ne b$, are related to
the $\langle M_{2}|\bar{q}(0)G^{\mu\nu}(z^{\prime})q(z)|0\rangle$.
The twist-3 contributions related to $\langle M_{2}|\bar{q}(0)G^{\mu\nu}(z^{\prime})q(z)|0\rangle$ 
are left to our another preparing paper.

The collinear limiting part of $V^{(b)}_{\alpha}(l,l^{\prime})$ is derived by using CE
\begin{eqnarray}\label{vertex-A-1}
V^{(b)}_{\alpha}(l,l^{\prime})=V^{(b)}_{\alpha}(\hat{l},\hat{l}^{\prime})+
\sum_{k=l,l^{\prime}}
\left.\frac{\partial V^{(b)}_{\alpha}(l,l^{\prime})}{\partial k^{\beta}}\right
|_{k=\hat{k}}(k-\hat{k})^{\beta}+\cdots\;,
\end{eqnarray}

By substituting Eq.~(\ref{vertex-A-1}) into Eq.~(\ref{vertex-A-0})
and neglecting the other terms except of $V^{(b)}_{\alpha}(\hat{l},\hat{l}^{\prime})$,
we arrive at
\begin{eqnarray}\label{vertex-A-2}
A^{\prime}_V\sim F^{B\to M_1}(q^2)\int\frac{d^4 l}{(2\pi)^4}\int\frac{d^4 l^{\prime}}{(2\pi)^4}
\Tr[V^{(b)}_{\alpha}(\hat{l},\hat{l}^{\prime})\phi^{M_{2}\alpha}_{A}(l,l^{\prime})]\;.
\end{eqnarray} 
Further more, 
it is convenient to rewrite $A^{\alpha}(z^\prime)$ in  $\phi^{M_{2}\alpha}_{A}(l,l^{\prime})$ as
\begin{eqnarray}\label{vertex-A-3}
A^{\alpha}(z^\prime)=\bar{n}^{\alpha}n\cdot A (z^\prime)+w^{\alpha}_{\alpha^{\prime}}A^{\alpha^{\prime}}(z^\prime)\;.
\end{eqnarray} 
By substituting the above expansion for $A^{\alpha}(z^\prime)$ into
Eq.~(\ref{vertex-A-2}), we arrive at
\begin{eqnarray}\label{vertex-A-4}
A^{\prime}_V\sim F^{B\to M_1}(q^2)
&&\int\frac{d^4 l}{(2\pi)^4}\int\frac{d^4 l^{\prime}}{(2\pi)^4}\nn
&&\times\left\{\right.
\Tr[V^{(b)}_{\alpha}(\hat{l},\hat{l}^{\prime})\bar{n}^{\alpha}n_{\alpha^\prime}
\phi^{M_{2}\alpha^{\prime}}_{A}(l,l^{\prime})]\nn
&&\hspace{1cm}+
\Tr[V^{(b)}_{\alpha}(\hat{l},\hat{l}^{\prime})w^{\alpha}_{\alpha^\prime}
\phi^{M_{2}\alpha^{\prime}}_{A}(l,l^{\prime})]\;.
\end{eqnarray} 

For covariant gauge, the first term of Eq.~(\ref{vertex-A-4}) can be transferred in the gauge phase factor
and the second term of Eq.~(\ref{vertex-A-4}) can be combined 
with the second term of Eq.~(\ref{vertex-d-1}) (See below further explanations.).
For physical gauge, such as $n\cdot A=0$, it is automatically vanishing.
In this work, we choose to use the covariant gauge in our following analysis.
Under the covariant gauge, there are other contributions from diagrams with more partonic gluons of $M_2$. 
The collinear limitings of these contributions are equally important and should be considered.
However, they can be shown to contribute to the gauge phase factor.
In the following, we exhibit this fact by only considering one partonic gluon case, 
the first term of Eq.~(\ref{vertex-A-4}). 
We now explain how the first term of Eq.~(\ref{vertex-A-4}) can be transferred into a gauge phase factor.
By using the identity
\begin{eqnarray}
V^{(b)}_{\alpha}(\hat{l},\hat{l}^{\prime})\bar{n}^{\alpha}
=V^{(b)}_{\alpha}(\hat{l},\hat{l}^{\prime})
\left.\frac{k^{\alpha}}{n\cdot k}\right|_{k=(\hat{l}^{\prime} -\hat{l})}
\end{eqnarray}
and the Feynman identity
\begin{eqnarray}
k^{\alpha}V^{(b)}_{\alpha}(\hat{l},\hat{l}^{\prime})
= V(\hat{l})-V(\hat{l}^{\prime})\;, 
\end{eqnarray}
we can rewrite the first term of Eq.~(\ref{vertex-A-4}) in the following form
\begin{eqnarray}
\int\frac{d^4 l}{(2\pi)^4}\int\frac{d^4 l^{\prime}}{(2\pi)^4}
\left(\Tr[V(\hat{l})\phi^{M_{2}}_{n\cdot A}(l,l^{\prime})]
-\Tr[V(\hat{l}^{\prime})\phi^{M_{2}}_{n\cdot A}(l,l^{\prime})]\right)\;,
\end{eqnarray}
where
\begin{eqnarray}
\phi^{M_{2}\alpha}_{n\cdot A}(l,l^{\prime})=
\int d^4 y \int d^4 z e^{ik \cdot( y-n\eta)}e^{-i\bar{l}\cdot z}
\langle M_{2}(q)|\bar{q}(0)(-ig) \int_0^{\infty} d\eta n\cdot A^a(y)T^{a}q(z)|0\rangle\;.
\end{eqnarray}
in which the identity
\begin{eqnarray}
\frac{i}{n\cdot k-i\epsilon}=\int_{0}^{\infty}d\eta e^{ -i\eta n\cdot k}
\end{eqnarray}
has been used.
Completing the momentum and coordinate integrals, we obtain
\begin{eqnarray}
\int dv^{\prime}
\Tr[V^{(1)}(v)\phi^{M_{2}}_{n\cdot A}(v)]
-\int du \Tr[V^{(1)}(v^\prime)\phi^{M_{2}}_{n\cdot A}(v^\prime)]\;,
\end{eqnarray}
where
\bee
\phi^{M_{2}}_{n\cdot A}(v)=
\int_0^{\infty}\frac{d\lambda}{2\pi}e^{-i\bar{v}\lambda}
\langle M_{2}(q)|\bar{q}(0)(-ig) \int_0^{\infty} d\eta n\cdot A^a(\eta n/E)T^{a}q(\lambda n/E)|0\rangle\;.
\eee

Since $V_{\alpha}(\hat{l},\hat{l})$ 
and $V^{(b)}_{\alpha}(\hat{l},\hat{l}^{\prime})$ 
have similar structures, 
this enable us to add up $A_V$ and $A^{\prime}_V$ to obtain
\begin{eqnarray}
&&A_V+A^{\prime}_V\sim F^{B\to M_1}(q^2)\nn
&&\times\left\{\right.
\int\frac{d^4 l}{(2\pi )^4}
\Tr[V(\hat{l})\phi^{M_{2}}(l)]\nn
&&\hspace{1cm}+
\int\frac{d^4 l}{(2\pi )^4}
\int\frac{d^4 l^{\prime}}{(2\pi )^4}
\Tr[V^{(b)}_{\alpha}(\hat{l},\hat{l}^{\prime})w^{\alpha}_{\alpha^{\prime}}
\phi^{M_{2}\alpha^{\prime}}_{D}(l,l^{\prime})]\left.\right\}
\end{eqnarray}
where 
\begin{eqnarray}\label{twist-4-DA-3}
\phi^{M_{2}\alpha}_{D}(l,l^{\prime})=
\int d^4 y \int d^4 z e^{-i(\bar{l}^{\prime}-\bar{l})\cdot y}e^{-i\bar{l}\cdot z}
\langle M_{2}(q)|\bar{q}(0)iD^{\alpha}(y)q(z)|0\rangle\;.
\end{eqnarray}
with $iD^{\alpha}=i\partial^{\alpha} - g A^{\alpha}$.
By using the integral transformations for $l$ and $l^\prime$,
we arrive at
\begin{eqnarray}
A+A^{\prime}\sim && F^{B\to M_1}(q^2)
\times\left(
\int_{0}^{1} dv 
\Tr[V(v)
\phi^{M_{2}}(v)] \right. \nn
&& \left. + \int_{0}^{1} dv \int_{0}^{1} dv^{\prime}
\Tr[V^{(b)}_{\alpha}(v,v^\prime)w^{\alpha}_{\alpha^{\prime}}
\phi^{M_{2}\alpha^{\prime}}_{D}(v,v^\prime)]\right)\;,
\end{eqnarray} 
where
\bee
V(v) &\equiv& V(\hat{l})\;,\nn
V^{(b)}_{\alpha}(v,v^\prime) &\equiv& V^{(b)}_{\alpha}(\hat{l},\hat{l}^{\prime})\;,\nn
\phi^{M_{2}}(v) &\equiv &  
\int_0^{\infty}\frac{d\lambda}{2\pi}
e^{-i\bar{v}\lambda}
\langle M_{2}(q)|\bar{q}(0)q(\lambda n/E)|0\rangle\;,\nn
\phi^{M_{2}\alpha^{\prime}}_{D}(v,v^\prime)
&\equiv &  
\int_0^{\infty}\frac{d\eta}{2\pi}\int_0^{\infty}\frac{d\lambda}{2\pi}
e^{-i(\bar{v}^\prime-\bar{v})\eta}
e^{-i\bar{v}\lambda}
\langle M_{2}(q)|\bar{q}(0)iD^{\alpha^{\prime}}(\eta n/E) q(\lambda n/E)|0\rangle\;.
\eee
The contributions associated with the $V^{(b)}_{\alpha}(v,v^\prime)$ are at least of $O(1/E^2)$
and will be neglected.
In the above, we have written the term related to $V^{(b)}_{\alpha}(v,v^\prime)$.
It is given here for comparison of the CE scheme with the BN and DYZ schemes.
We have introduced how the CE is applied to derive the contributions related to the collinear limiting part
$V(v)$ and how the CE can separate different contributions related to different number of collinear partons 
of $M_2$.
Because we are only concerning the twist-3 contributions,
our remaining task is to show that $V(v)$ is infrared finite up-to $O(1/E)$.
This is given below.

\subsection{Collinear expansion for vertex corrections}
The relevant term in the amplitude for the four diagrams depicted in \fig{fig:fig3}~(a)-(d) is written as
\begin{eqnarray}\label{vertex}
A_{V}^{Q_{i}}\sim F_{j}^{B\to M_{1}}(q^2)\int\frac{d^4l_{2}}{(2\pi)^4}\Tr[V_{i j}(l)\phi^{M_{2}}(l)]
\end{eqnarray}
where 
\bee\label{vertex-1}
\text{Tr}[V_{i j}(l)\phi^{M_2}(l)]
&=& 2i \frac{ \pi\alpha_s}{N_c}\int\frac{d^4 k}{(2\pi)^4}
\text{Tr}[T^{a}T^{b}\bar{\Gamma}_{j}(\Gamma_i \frac{(2P_{b,\alpha}-\gamma_{\alpha}\s{k})}{2P_b\cdot k+k^2}
- \frac{(2p_{\alpha}+\s{k}\gamma_{\alpha})}{2p\cdot k -k^2}\Gamma_i)]\nn
&\times&\text{Tr}[T^{a}T^{b}( \gamma^{\alpha} \frac{(\s{l}+\s{k})}{(l+k)^2}\bar{\Gamma}_{i}
-\bar{\Gamma}_{i}\frac{(\bar{\s{l}}+\s{k})}{(\bar{l}+k)^2}\gamma^{\alpha} )\phi^{M_2}(l)]\frac{1}{k^2}\;,
\eee 
where we have employed equations of motion for the $b$ and $q_{1}$ quarks 
for the terms inside the first trace bracket.
The virtual gluon's momentum is represented by $k$.
We first perform the following expansion for $V_{i j}(l)$ to derive its collinear limiting part $V(\hat{l})$
\bee
V(l)&=& V(\hat{l})+\left.\frac{\partial V}{\partial l^{\alpha}}\right|_{l=\hat{l}}
(l-\hat{l})^{\alpha}+\cdots\;.
\eee
The virtual gluons could be hard $k_H^{\mu}\sim (E,E,E)$, soft $k_S^{\mu}\sim (\lambda,\lambda,\lambda)$
, or collinear to $q^{\mu}$ as $k_C^{\mu}\sim (E,\lambda^2/E,\lambda)$ or collinear to $p^{\mu}$ as 
$k_{C^\prime}^{\mu}\sim (\lambda^2/E,E,\lambda)$.
Therefore, we divide the $k$ integral into three regions corresponding to the soft $k_S$, 
the collinear $k_C$ or $k_{C^\prime}$, 
and the hard $k_H$.

To analyze the infrared structure of $V^{(1)}(\hat{l})$,
we define the soft and collinear limiting parts of $V^{(1)}_{i j}(\hat{l})$, 
in which $k$ are set as soft $k_S$ or collinear $k_{C}$ or $k_{C^\prime}$, 
as $V^{(1)}_{i j,S}(\hat{l})$ and $V^{(1)}_{i j,C}(\hat{l})$ and $V^{(1)}_{i j,C^\prime}(\hat{l})$, respectively.

\subsubsection{Soft finiteness}
We first write $\text{Tr}[V^{(1)}_{i j,S}(\hat{l})\phi^{M_2}(l)]$ in its explicit form as 
\begin{eqnarray}\label{vertex-soft-1}
\text{Tr}[V^{(1)}_{i j,S}(\hat{l})\phi^{M_2}(l)]
&=&i\frac{C_F\pi\alpha_s}{N_c}\int\frac{d^4 k_S}{(2\pi)^4}
\text{Tr}[\bar{\Gamma}_{j}\Gamma_i(\frac{2P_{b,\alpha}}{2 P_b\cdot k_S}
- \frac{2 p_{\alpha}}{2p\cdot k_S}) ]\nn
&\times&\text{Tr}[( \gamma^{\alpha} \frac{\s{\hat{l}}+\s{k}_S}{(\hat{l}+ k_S)^2}\bar{\Gamma}_{i}
-\bar{\Gamma}_{i}\frac{\bar{\s{\hat{l}}}+\s{k}_{S}}{(\bar{l}+ k_S)^2}
\gamma^{\alpha} )\phi^{M_2}(l)]\frac{1}{k^2_S}\;.
\end{eqnarray}
In order to find the $O(1/E)$ contributions in the above expression,
we have written the full part (i.e. $\hat{l}+k_{s}$ or $\hat{\bar{l}}+k_{s}$ ) of
the internal parton propagators in the upper part of the diagrams in \fig{fig:fig3}~(a)-(d).
Take an example for explanation,
we consider the propagator for the left internal parton propagator in the upper part of the first diagram in
\fig{fig:fig3}~(a).
Because $k_S^2\ll \hat{l}\cdot k_{S}$, 
the propagator appears as 
\begin{eqnarray}\label{soft-propagator-1}
\frac{\s{\hat{l}}+\s{k}_S}{(\hat{l}+ k_S)^2}=\frac{\s{\hat{l}}}{2\hat{l}\cdot k_{S}}
+\frac{\s{n}}{2 n\cdot\hat{l} }
+\frac{\s{t}_{\perp}}{2 n\cdot\hat{l} }
\end{eqnarray}
where $t^{\mu}_{\perp}$ is an unit vector in the transverse directions, $t^{2}_{\perp}=1$.
Since the denominator of the first term in Eq.~(\ref{soft-propagator-1}) is of $O(E\Lambda)$, 
and the denominators of the last two terms are of $O(E)$,
we obtain the first term is of $O(1/\Lambda)$ 
and the last two terms are of $O(1/E)$.
Since $E\gg\Lambda$, the leading contribution comes from the first term 
and the last two terms are power suppressed as $O(\Lambda/E)$ in comparison with the leading first term.
Since the last two terms are independent of $k_{S}$,
they can be decomposed from the loop momentum integrations over $k_{S}$.

After we have separated the leading and subleading terms,
where the latter are power and soft suppressed as $O(\Lambda/E)$ than the former,
we now show that the leading part of $V_{S}$ gives a vanishing result.
By observing Eq.~(\ref{soft-propagator-1}), we can see that the relevant terms of $V_{S}$ are
\bee\label{vertex-soft-3}
&&\text{Tr}[\bar{\Gamma}_{j}\Gamma_i(\frac{2P_{b,\alpha}}{2 P_b\cdot k_S}
- \frac{2 p_{\alpha}}{2p\cdot k_S}) ]
\text{Tr}[( \gamma^{\alpha} \frac{\s{\hat{l}}}{2\hat{l}\cdot k_S}\bar{\Gamma}_{i}
-\bar{\Gamma}_{i}\frac{\s{\hat{\bar{l}}}}{2\hat{\bar{l}}\cdot k_S}
\gamma^{\alpha} )\phi^{M_2}(l)]\frac{1}{k^2_S}\;.
\eee
The $\gamma^{\alpha}$ in the second trace term can be 
$\gamma^{\alpha}=\s{n}\bar{n}^{\alpha}$, $\s{\bar{n}}n^{\alpha}$, or,
 $d_{\alpha}^{\alpha^{\prime}}\gamma^{\alpha^{\prime}}$.
If $\gamma^{\alpha}=\bar{\s{n}}n^{\alpha}$ or $\s{n}\bar{n}^{\alpha}$,
the second trace vanishes as 
\begin{eqnarray}\label{soft-vertex-2}
&&\text{Tr}[( \s{\bar{n}}n^{\alpha} \frac{\s{\hat{l}}}{2\hat{l}\cdot k_S}\bar{\Gamma}_{i}
-\bar{\Gamma}_{i}\frac{\s{\hat{\bar{l}}}}{2\bar{l}\cdot k_S}
\s{\bar{n}}n^{\alpha} )\phi^{M_2}(l)]\nn
&=&\text{Tr}[( \s{n}\bar{n}^{\alpha} \frac{\s{\hat{l}}}{2\hat{l}\cdot k_S}\bar{\Gamma}_{i}
-\bar{\Gamma}_{i}\frac{\s{\hat{\bar{l}}}}{2\hat{\bar{l}}\cdot k_S}
\s{n}\bar{n}^{\alpha})\phi^{M_2}(l)]\nn
&=&0\;,
\end{eqnarray}
where the first line vanishes due to $\s{\bar{n}}\hat{l}\propto \bar{n}^2=0$,
and the second line vanishes since 
$\s{n}\s{\hat{l}}=v E$, $\s{n}\s{\hat{\bar{l}}}=\bar{v} E$ and
\begin{eqnarray}
&&(\frac{\s{n}\s{\hat{l}}}{2 n \cdot \hat{l}}\bar{\Gamma}_{i}
-\bar{\Gamma}_{i}\frac{\s{\hat{\bar{l}}}\s{n}}{2 n\cdot \hat{\bar{l}}}
)\bar{n}^{\alpha}\nn
&=&\frac{\bar{n}^{\alpha}}{2}(\bar{\Gamma}_{i}-\bar{\Gamma}_{i})=0\;.
\end{eqnarray}
If $\gamma^{\alpha}=d_{\alpha}^{\alpha^{\prime}}\gamma^{\alpha^{\prime}}$, 
the contraction of $d^{\alpha}_{\alpha^{\prime}}$ with
the first trace gives a vanishing result as
\[
d^{\alpha}_{\alpha^{\prime}}
\left(\frac{2P_{b,\alpha}}{2 P_b\cdot k_S}
- \frac{2 p_{\alpha}}{2p\cdot k_S}
\right)=0\;.
\]

The $O(\Lambda/E)$ contributions from Eq.~(\ref{soft-propagator-1}) are related to the spin state of $M_{2}$.
Because up-to twist-3 order, 
$q_{2}\bar{q}_{3}$ pair for the $M_{2}$ can be proportional to $\s{\bar{n}}\gamma_5$, $\gamma_5$, or 
$\epsilon^{\alpha\beta}_{\perp}\sigma_{\alpha\beta}$,
the substitution of these spin terms into the trace over those $O(\Lambda/E)$ terms in the internal propagators,
such as those last two terms in the right hand side of Eq.~(\ref{soft-propagator-1}),
results in vanishing results by an explicit manipulation.
This concludes that the $O(1/E)$ soft contributions vanish and the uncertainties are of $O(1/E^2)$.
In summary, we have shown that the $V^{(1)}_{i j,S}(\hat{l})$ vanishes up-to $O(1/E)$ explicitly.
The uncertainties are estimated to be of $O(1/E^2)$ which has beyond our precision used in this paper.

\subsubsection{Collinear finiteness}
We now consider the collinear part $V^{(1)}_{i j,C}(\hat{l})$.
It is convenient to combine $\hat{l}^{\mu} + k_C^{\mu}=l^{\prime,\mu}$ 
\bee\label{vertex-collinear-1}
\Tr[V^{(1)}_{i j,C}(\hat{l})\phi^{M_2}(l_2)]&=& i\frac{C_F\pi\alpha_s}{N_c}\int\frac{d^4 k_C}{(2\pi)^4}
\text{Tr}[\bar{\Gamma}_{j}(\frac{\Gamma_i (2P_{b,\alpha}-\gamma_{\alpha}\s{k}_C)}{2P_b\cdot k_C}
- \frac{(2p_{\alpha}+\s{k}_C\gamma_{\alpha})}{2p\cdot k_C }\Gamma_i)]\nn
&\times&\Tr[( \gamma^{\alpha} \frac{\s{l}^{\prime}}{(l^{\prime})^2}\bar{\Gamma}_{i}
-\bar{\Gamma}_{i}\frac{\bar{\s{l}^{\prime}}}{(\bar{l}^{\prime})^2}\gamma^{\alpha} \phi^{M_2}(l_2)]\frac{1}{k_C^2}\;.
\eee
We separate the internal parton propagators into the long distance and
short distance parts, and analyze their contributions, respectively.
Let's first consider the term
\bee
\text{Tr}[( \gamma^{\alpha}( \frac{\s{l^{\prime}}_L}{{l^\prime}^2}
+\frac{\s{n}}{2 n\cdot \hat{l}^{\prime}})\bar{\Gamma}_{i}
-\bar{\Gamma}_{i}(\frac{\bar{\s{l}^{\prime}_L}}{(\bar{l}^{\prime})^2}
+\frac{\s{n}}{2 n\cdot \bar{\hat{l}}^{\prime}})\gamma^{\alpha} )\phi^{M_2}(l)]\;,
\eee
where we have explicitly written the long distance term
\bee\label{vertex-collinear-long}
\text{Tr}[( \gamma^{\alpha} \frac{\s{l^{\prime}_L}}{{l^\prime}^2}\bar{\Gamma}_{i}
-\bar{\Gamma}_{i}\frac{\bar{\s{l}_L^{\prime}}}{(\bar{l}^{\prime})^2}\gamma^{\alpha} )\phi^{M_2}(l)]\;,
\eee
and
the short distance term
\bee
\text{Tr}[( \gamma^{\alpha}\frac{\s{n}}{2 n\cdot \hat{l}^{\prime}}\bar{\Gamma}_{i}
-\bar{\Gamma}_{i}\frac{\s{ n}}{2 n\cdot \hat{\bar{l}}^{\prime}}\gamma^{\alpha} )\phi^{M_2}(l)]\;.
\eee
Because the short distance part is suppressed than the leading part of the long distance term by an 
$O(\Lambda^2/E^2)$ factor,  
we may safely ignore the contributions associated with the short distance term.

In \eq{vertex-collinear-long}, 
the $\gamma^{\alpha}$ can be 
$\gamma^{\alpha}=\s{n}\bar{n}^{\alpha}$, $\s{\bar{n}}n^{\alpha}$, or,
 $d^{\alpha}_{\alpha^{\prime}}\gamma^{\alpha^{\prime}}$. 
The final result also depends on the spin state of the $M_{2}$.
Because up-to twist-3 order, $q_{2}\bar{q}_{3}$ pair for the $M_{2}$ 
can be proportional to $\s{\bar{n}}\gamma_5$, $\gamma_5$, or 
$\epsilon^{\alpha\beta}_{\perp}\sigma_{\alpha\beta}$,
the substitution of these spin terms into the second trace term \eq{vertex-collinear-1}
leads to the following nine results.
We explain them term by term:
\begin{itemize}
\item{$\gamma^{\alpha}=\s{\bar{n}}n^{\alpha}$ and $[q_{2}\bar{q}_{3}]\propto\s{\bar{n}}\gamma_5$:} 
the trace term in \eq{vertex-collinear-long} vanishes due to $\bar{n}^2=0$. 

\item{$\gamma^{\alpha}=\s{n}\bar{n}^{\alpha}$ and $[q_{2}\bar{q}_{3}]\propto\s{\bar{n}}\gamma_5$:} 
the trace term in \eq{vertex-collinear-long} is proportional to $\bar{n}^{\alpha}$.
The contraction of $\bar{n}^{\alpha}$ with the first trace term in Eq.~(\ref{vertex-collinear-1}) gives
\begin{eqnarray}
\bar{n}^{\alpha}\text{Tr}[\bar{\Gamma}_{j}(\Gamma_i\frac{ (2P_{b,\alpha}-\gamma_{\alpha}\s{k}_C)}{2P_b\cdot k_C}
- \frac{(2p_{\alpha}+\s{k}_C\gamma_{\alpha})}{2p\cdot k_C }\Gamma_i)]\simeq 0+O(\frac{\Lambda^2}{E^3})\;. 
\end{eqnarray}
This is because $\bar{n}\cdot P_{b}=\bar{n}\cdot p$ and $P_{b}\cdot k_{C}\simeq p\cdot k_{C}+ O(\Lambda^2)$.
The errors are from the $\s{n}$ component of $\s{k}_{C}$.

\item{$\gamma^{\alpha}=d_{\alpha}^{\alpha^{\prime}}\gamma^{\alpha^{\prime}}$ and 
$[q_{2}\bar{q}_{3}]\propto\s{\bar{n}}\gamma_5$:}
the trace term in \eq{vertex-collinear-long} 
\begin{eqnarray}
&&d^{\alpha}_{\alpha^{\prime}}\text{Tr}[( \gamma^{\alpha^{\prime}} \frac{\s{l^{\prime}_L}}{{l^\prime}^2}\bar{\Gamma}_{i}
-\bar{\Gamma}_{i}\frac{\bar{\s{l}_L^{\prime}}}{(\bar{l}^{\prime})^2}\gamma^{\alpha^{\prime}}\s{\bar{n}}\gamma_5)]
\propto
d^{\alpha}_{\alpha^{\prime}}
\end{eqnarray}
is proportional to $d^{\alpha}_{\alpha^{\prime}}$.
The leading contributions of the trace term in \eq{vertex-collinear-long} are of $O(E/\Lambda^2)$.
The contraction of $d^{\alpha}_{\alpha^{\prime}}$ with the first trace term in \eq{vertex-collinear-1} 
results in 
\begin{eqnarray}
&&d^{\alpha}_{\alpha^{\prime}}\text{Tr}[\bar{\Gamma}_{j}(\Gamma_i\frac{ (2P_{b,\alpha}-\gamma_{\alpha}\s{k}_C)}{2P_b\cdot k_C}
- \frac{(2p_{\alpha}+\s{k}_C\gamma_{\alpha})}{2p\cdot k_C }\Gamma_i)]\nn
&\propto& \frac{d^{\alpha}_{\alpha^{\prime}}k_{C,\alpha}}{p\cdot k_C } \;,
\end{eqnarray} 
which is of order $O(\Lambda/E^2)$.
Due to the loop integration over $k_C$, 
the single $k_{C,\perp}$ factor selects another $k_{C,\perp}$ factor 
from the second trace term in \eq{vertex-collinear-1}.
The product of these two trace terms are of $O(\Lambda^2/E^3)$.

\item{$\gamma^{\alpha}=\s{\bar{n}}n^{\alpha}$ and $[q_{2}\bar{q}_{3}]\propto \gamma_5$:} 
the trace term in \eq{vertex-collinear-long} is then proportional to
\begin{eqnarray}
(\frac{1}{n\cdot \hat{l}^{\prime}}-\frac{1}{n\cdot \hat{\bar{l}}^{\prime}})n^{\alpha}\;.
\end{eqnarray}
The contraction of $n^{\alpha}$ with the first trace term in \eq{vertex-collinear-1} gives 
\begin{eqnarray}
&& n^{\alpha}\text{Tr}[\bar{\Gamma}_{j}(\Gamma_i\frac{ (2P_{b,\alpha}-\gamma_{\alpha}\s{k}_C)}{2P_b\cdot k_C}
- \frac{(2p_{\alpha}+\s{k}_C\gamma_{\alpha})}{2p\cdot k_C }\Gamma_i)]\nn
&\propto&\frac{1}{p\cdot k_{C}}(P_{b}\cdot n-k_{c}\cdot n)=\frac{\bar{\alpha}}{\alpha E}\;, 
\end{eqnarray}
where we have used the fact that $\Gamma_{i}\otimes \bar{\Gamma}_{i}$ should be $-2(S-P)\otimes(S+P)$
and the parameterization $k^{\mu}_{C}=\alpha E \bar{n}^{\mu}+\cdots$.
The combination of these two trace terms are of $O(1/E^2)$.

\item{$\gamma^{\alpha}=\s{n}\bar{n}^{\alpha}$ and $[q_{2}\bar{q}_{3}]\propto\gamma_5$:} 
the trace term in \eq{vertex-collinear-long} vanishes similar to \eq{soft-vertex-2}.

\item{$\gamma^{\alpha}=d_{\alpha}^{\alpha^{\prime}}\gamma^{\alpha^{\prime}}$ and $[q_{2}\bar{q}_{3}]\propto\gamma_5$:}
the trace term in \eq{vertex-collinear-long} is proportional to $k^{\alpha}_{C,\perp}$ and is of order $O(\Lambda/E)$.
The contraction of $k^{\alpha}_{C,\perp}$ with the first trace term in \eq{vertex-collinear-1}
 gives contributions of order $O(\Lambda/E^2)$.
The product of these two trace terms is of order $O(\Lambda^2/E^3)$. 
We neglect it.

\item{$\gamma^{\alpha}=\s{\bar{n}}n^{\alpha}$ 
and $[q_{2}\bar{q}_{3}]\propto\epsilon^{\mu\nu}_{\perp}\sigma_{\mu\nu}$:}
only $\Gamma_{i}\otimes \bar{\Gamma}_{i}=-2(S-P)\otimes(S+P)$ can contribute.
The second trace term in \eq{vertex-collinear-long} gives a result proportional to 
\begin{eqnarray}
(\frac{1}{n\cdot l^{\prime}}+\frac{1}{n\cdot \bar{l}^{\prime}})\bar{n}^{\alpha}
\end{eqnarray}
which is of order $O(\Lambda/E)$.
The contraction of $\bar{n}^{\alpha}$ with the first trace term in \eq{vertex-collinear-1}  
gives a contribution of $O(1/E)$.
The combination of these two trace terms is of order $O(\Lambda/E^2)$. 

\item{$\gamma^{\alpha}=\s{n}\bar{n}^{\alpha}$ 
and $[q_{2}\bar{q}_{3}]\propto\epsilon^{\mu\nu}_{\perp}\sigma_{\mu\nu}$:}
only $\Gamma_{i}\otimes \bar{\Gamma}_{i}=-2(S-P)\otimes(S+P)$ can contribute.
The trace term in \eq{vertex-collinear-long} is proportional to $\bar{n}^{\alpha}$.
The contraction of $\bar{n}^{\alpha}$ with the first trace term in \eq{vertex-collinear-1} 
gives a vanishing result with errors of $O(\Lambda^2/E^3)$.
\end{itemize}
In summary, the $\Tr[V_{C}(\hat{l}_2)\phi^{M_2}(l)]$ vanishes up-to $O(1/E)$.
Similarly, we can show that the term $\Tr[V_{ij,C^\prime}(\hat{l})\phi^{M_2}(l)]$ vanishes up-to $O(1/E)$.
Therefore, up-to $O(1/E)$, we can neglect collinear divergences.
Note that the above estimated errors should be multiplied by a factor $\Lambda^2 E$ for the 
$[q_{2}\bar{q}_3]_{M_2}$  spin state being proportional to $\s{\bar{n}}\gamma_5$, 
 and $\Lambda^2 \mu_{M_{2}}$ for the $\gamma_5$ and 
$\epsilon_{\perp}\cdot\sigma$ spin states, respectively. 

\subsubsection{Infrared finite one loop vertex corrections}
Based on the above analysis for the soft and collinear limitings of $V(\hat{l})$, 
we now describe how to use the subtraction method to show the infrared finiteness of $\Tr[V(\hat{l})\phi^{M_2}(l)]$.
The proof is given by the following series of identities 
\bee\label{vertex-expansion-1}
A_{V}^{Q_{i}}
&=&F_{j}^{B\to M_{1}}(q^2)\int\frac{d^4l}{(2\pi)^4}\Tr[V_{i j}(l)\phi^{M_{2}}(l)]\nn
&=&F_{j}^{B\to M_{1}}(q^2)\int\frac{d^4l}{(2\pi)^4}(\Tr[(V_{i j}(\hat{l})\phi^{M_{2}}(l)]+\cdots)\nn
&=&F_{j}^{B\to M_{1}}(q^2)\int\frac{d^4l}{(2\pi)^4}
[(\Tr[V_{i j}(\hat{l})\phi^{M_{2}}(l)]-
\Tr[V_{i j;S}(\hat{l})\phi^{M_{2}}(l)]-
\Tr[V_{i j;C}(\hat{l})\phi^{M_{2}}(l)]\nn
&&\hspace{3.5cm}
-\Tr[V_{i j;C^\prime}(\hat{l})\phi^{M_{2}}(l)])
+\Tr[V_{i j;S}(\hat{l})\phi^{M_{2}}(l)]+\Tr[V_{i j;C}(\hat{l})\phi^{M_{2}}(l)]\nn
&&\hspace{3.5cm}+\Tr[V_{i j;C^\prime}(\hat{l})\phi^{M_{2}}(l)]+\cdots]\nn
&=&F_{j}^{B\to M_{1}}(q^2)\int_{0}^{1} dv
(\Tr[V_{i j;H}(v)\phi^{M_{2}}(v)]+\cdots)\;,
\eee
where the $V_{i j;H}(v)$ is defined as
\bee
V_{i j;H}(v)=V_{i j}(v)-V_{i j;S}(v)-V_{i j;C}(v)-V_{i j;C^\prime}(v)\;,
\eee 
and the dots represent higher order terms which are of order $O(1/E^2)$.
In the last identity, the integration transformation 
\[
\int\frac{d^4l}{(2\pi)^4}\phi^{M_2}(l)\to \int_{0}^{1} dv\phi^{M_2}(v)
\]
has been performed.
Up-to twist-3 order, the spin factorization can be obtained by using the identity
\bee
\phi^{M_2}(v)
= -\frac{if_{M_2}}{4N_c}[\gamma_5 \s{q}\phi^{M_2}_{P}(v)
+\mu_{\chi}^{M_2}
(\gamma_5\hat{\phi}^{M_2}_{p}(v)-\frac{1}{2}\epsilon_{\perp}\cdot\sigma\hat{\phi}^{M_2}_{\sigma}(v)]\;,\nn
\eee
where $\epsilon_{\perp}\cdot\sigma=\epsilon_{\perp}^{\alpha\beta}\sigma_{\alpha\beta}$, and 
$\epsilon_{\perp}^{\alpha\beta}=\epsilon^{\alpha\beta\eta\gamma}n_{\eta}\bar{n}_{\gamma}$.
The remaining task is to complete the one loop momentum integration over $k$ 
by using the naive dimensional regularization.
The final expression will be given in \sec{sec:sec5}.

\subsection{Collinear expansion for corrections from penguin contractions}
The summation of the contributions from the two diagrams in \fig{fig:fig3}~(e) and (f) 
 contain the following expression
\bee
F_{j}^{B\to M_{1}}(q^2)\int\frac{d^4 l}{(2\pi)^4}\text{Tr}[P_{i j}(l)\phi^{M_2}(l)]
\eee
where the function $P_{i j}(l)$ is defined to contain the quark loop
\bee 
&&\text{Tr}[P_{i j}(l)\phi^{M_2}(l)]\nn
&=& i\frac{C_F\pi\alpha_s}{N_c}\int\frac{d^4 k}{(2\pi)^4}
\times\left\{
\text{Tr}[\bar{\Gamma}_j \Gamma_i \frac{\s{k}+m_q}{k^2-m_q^2}\gamma_{\alpha}
\frac{\s{k}+\s{p}+\s{l}+m_q}{(k+p+l)^2-m_q^2}\bar{\Gamma}_i \phi^{M_2}(l)\gamma^{\alpha}]
\right.\nn
&+&
\left.
\text{Tr}[\bar{\Gamma}_j  \Gamma_i \phi^{M_2}(l) \gamma^{\alpha}]
\text{Tr}[\frac{\s{k}+m_q}{k^2-m_q^2}\gamma_{\alpha}
\frac{\s{k}+\s{p}+\s{l}+m_q}{(k+p+l)^2-m_q^2}\bar{\Gamma}_i ]
\right\}\frac{1}{(p+l)^2}\;.
\eee
The internal quark loop momentum $k$ and the internal quark mass $m_{q}$ has been used in the above expression. 
To extract the leading contributions, we perform the collinear expansion for $P_{i j}(l)$ 
in the following way
\bee
P_{i j}(l)=P_{i j}(\hat{l})
+\frac{\partial P_{i j}}{\partial l^{\beta}}|_{l=\hat{l}}(l-\hat{l})^{\beta}
+\cdots\;.
\eee
Due the momentum derivative over $l$, 
the second term in the right hand side of the above equation is suppressed by a factor of $O(1/E)$ than the first term.
We first concentrate on the first term $P_{i j}(\hat{l}_{2})$. 
Because both the recoil and emitted mesons are energetic,
the radiative gluon's momentum scales as $(p+l_{2})^{\mu}\sim (E,E,0)$.
The infrared structure of $P_{i j}(\hat{l})$ then depends on the quark loop integration over $k$,
in which $k$ could be soft or hard.
If $k$ is soft, the numerator of the integrand of the $k$ integration decreases one power of $k^2$.
With respect to the loop integration over hard $k$,
the loop integration over soft $k$ is suppressed as $O(1/E^{2})$.
The hard $k$ region is dominated.
The integral transformation is made to transform $\phi^{M_2}(l)$ into $\phi^{M_2}(v)$.
%Up-to twist-3, the spin factorization leads to 
%\bee
%&&\int d v \text{Tr}[P_{i j}(v)\phi^{M_2}(v)]\nn
%&=&\int d v 
%\left(\right.
%-\frac{1}{4}\text{Tr}[P_{i j}(v)\s{q}\gamma_5]\text{Tr}[\s{n}\gamma_5 \phi^{M_2}(v)] 
%+\frac{1}{4}\text{Tr}[P_{i j}(v)\gamma_5]\text{Tr}[\gamma_5 \phi^{M_2}(v)] \nn
%&&-\frac{1}{32}\text{Tr}[P_{i j}(v)\epsilon_{\perp}\cdot\sigma]
%\text{Tr}[\epsilon_{\perp}\cdot\sigma \phi^{M_2}(v)] 
%\left.\right)\;.
%\eee
The $k$ integration inside $P_{i j}(v)$ is calculated by using NDR. 
The explicit expression for the above penguin corrections is given in \sec{sec:sec5}.

\subsection{Collinear expansion for hard spectator corrections} 
\begin{figure}
%{\centerline{\includegraphics{diagrams/fourparton}}}
{\centerline{\includegraphics*[scale=0.4,viewport=60 470 550 800]{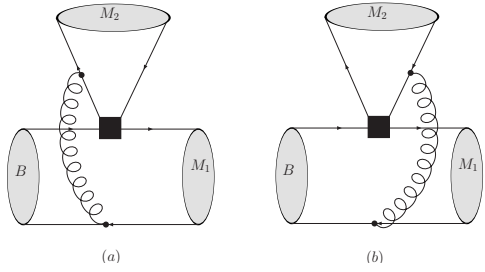}}}
\caption{The hard spectator diagrams.}
  \label{fig:fig5}                 
\end{figure}
The relevant expression  for the spectator diagrams depicted in \fig{fig:fig5} 
is written as
\bee
\int\frac{d^4 l_B}{(2\pi)^4}\int\frac{d^4 l_1}{(2\pi)^4}\int\frac{d^4 l_2}{(2\pi)^4}
\text{Tr}[H(l_B, l_1, l_2)\phi_{B}(l_B)\phi^{M_1}(l_1)\phi^{M_2}(l_2)]\;,
\eee
where the spectator scattering function $H(l_B, l_1, l_2)$ is expressed as
\bee\label{eq:spect-1}
&&\text{Tr}[H(l_B, l_1, l_2)\phi_{B}(l_B)\phi^{M_1}(l_1)\phi^{M_2}(l_2)]\nn
&=&-4\frac{\pi\alpha_s C_F}{N_c^2}\text{Tr}[\Gamma_i\phi_{B}(l_B)\gamma_{\alpha}\phi^{M_1}(l_1)]\nn
&&\times \text{Tr}[\left(\gamma^{\alpha}\frac{\s{l}_2+\s{k}}{(l_2+k)^2}\bar{\Gamma}_i 
-\bar{\Gamma}_i \frac{\bar{\s{l}}_2+\s{k}}{(\bar{l}_2+k)^2}\gamma^{\alpha}\right)\phi^{M_2}(l_2)]\frac{1}{k^2}\nn 
\eee
where $k=l_B - l_1$.
We first perform the collinear expansion of $H(l_B, l_1, l_2)$ 
with respect to the collinear momenta $\hat{l}_i, i=B, 1, 2$,
\bee
H(l_B, l_1, l_2)
=H(\hat{l}_B, \hat{l}_1, \hat{l}_2)
+\sum_{i=1,2}\frac{\partial H^{(1)}}{\partial l^\nu_i}|_{l_i=\hat{l}_i}(l_i -\hat{l}_i)^{\nu}+\cdots\;.
\eee
Since the $M_1$ and $M_2$ mesons are energetic and the spectator quark of the $B$ meson can only carry soft momentum, 
the virtual gluon momentum $k$ can be soft or hard-collinear.
For soft $k$ ,
the second trace term vanishes 
\bee
\text{Tr}[(\frac{q^{\alpha}}{2q\cdot k}\bar{\Gamma}_i 
-\bar{\Gamma}_i \frac{q^{\alpha}}{q\cdot k})\phi^{M_2}(l_2)]\simeq 0 +O(1/E^2)\;,
\eee
where the errors are estimated to be of $O(1/E^2)$.
This shows that the hard spectator diagram contributions are free from infrared divergences up-to $O(1/E)$.
However, it has been noted in the literature that there exist end point divergences $X_H$ at $O(1/E)$ as mentioned in Introduction.
The term $X_{H}$ becomes divergent because the pseudoscalar LCDA $\phi_{p}^{M_1}(u)$ is a constant.
Since the constant model for $\phi_{p}^{M_1}(u)$ is determined by the equation of motion \eq{equation-1},   
as we have shown in \sec{sec:sec2}, 
the end point divergences can be identified as a mixing effect between the twist-3 and twist-4 
LCDAs for the $M_1$ meson.
In addition, we also showed in \sec{sec:sec2} that, in the energetic meson limit, 
the pseudoscalar LCDA $\phi_{p}^{M_1}(u)$ is reduced to be
$\hat{\phi}^{M_1}_p(u)$, which is no longer a constant according to the reduced equation of motion \eq{equation-2}.
The divergence in $X_H$ as $u\to 0$ is then regularized by the $\hat{\phi}^{M_1}_p(u)$.
The explicit expression for this fact will be given in \sec{sec:sec5}.

We now show that the $O(1/E)$ contributions can only come from the two parton twist-3 LCDAs for the $M_{1}$ meson,
and the similar contributions from the two parton twist-3 LCDAs for the $M_{2}$ meson 
are vanishing at $O(1/E)$.
The $\gamma^{\alpha}$ in the second trace term in Eq.~(\ref{eq:spect-1}) can be 
$\gamma^{\alpha}=\s{n}\bar{n}^{\alpha}$, $\s{\bar{n}}n^{\alpha}$, or,
$d_{\alpha}^{\alpha^{\prime}}\gamma^{\alpha^{\prime}}$.
For the spin state of $M_{2}$ being $\gamma_5$, the $\Gamma_{i}\otimes \bar{\Gamma}_{i}$ can only be $-2(S-P)(S+P)$.
%The magnitude of $\hat{l}_B$ is very small as compared to $\hat{l}_1$, $\hat{l}_2$, or $\hat{\bar{l}_2}$,
%since the spectator can carry soft momentum.
For $H^{(1)}(\hat{l}_B, \hat{l}_1, \hat{l}_2)$, 
the dominant contributions in the second trace term can only come from $\s{\hat{l}}_2-\s{\hat{l}}_1$, or, 
$\s{\hat{\bar{l}}}_2-\s{\hat{l}}_1$, which are proportional to $\s{n}$ or $\s{\bar{n}}$.
This selects $\gamma^{\alpha}=\s{n}\bar{n}^{\alpha}$, or $\s{\bar{n}}n^{\alpha}$.
For $\gamma^{\alpha}=\s{\bar{n}}n^{\alpha}$, 
the contraction of $n^{\alpha}$ with the first trace term in Eq.~(\ref{eq:spect-1}) gives
\begin{eqnarray}
n^{\alpha}\text{Tr}[\Gamma_i\phi_{B}(l_B)\gamma_{\alpha}\phi^{M_1}(l_1)]\;.
\end{eqnarray}
The contact of $\s{n}$ with $\phi^{M_1}(l_1)$ extracts one short distance propagator 
\[
\frac{-i\s{\bar{n}}}{2\bar{n}\cdot l_{1}}
\]
and a vertex $-i\gamma^{\beta}$. 
The result appears as
\begin{eqnarray}
n^{\alpha}\text{Tr}[\Gamma_i\phi_{B}(l_B)\gamma_{\alpha}\frac{-i\s{\bar{n}}}{2\bar{n}\cdot l_{1}}
(-i\gamma_{\beta})\bar{w}^{\beta}_{\beta^{\prime}}\phi^{M_1\beta}_{\partial}(l_1,l_1)]\;,
\end{eqnarray}  
where $\bar{w}^{\beta}_{\beta^{\prime}}=g^{\beta}_{\beta^{\prime}}-n^{\beta}\bar{n}_{\beta^{\prime}}$. 
Because the short distance propagator is of $O(1/E)$, 
the dimension of hard scattering function is then decreased by one order.
The related contributions are of next twist than what we have considered.
The other possibility is that $\gamma^{\alpha}=\s{n}\bar{n}^{\alpha}$,
which then selects the $\s{\hat{l}}_2$, or $\s{\hat{\bar{l}}}_2$ parts of the propagators in the second trace term.
The result is proportional to
\begin{eqnarray}
\left(\frac{1}{2 n\cdot q}
-\frac{1}{2 n \cdot q}\right)\bar{n}^{\alpha}=0
\end{eqnarray}
which is obviously vanishing. 

For the spin state of $M_{2}$ being $\epsilon_{\perp}\cdot\sigma$, 
the $\Gamma_{i}\otimes \bar{\Gamma}_{i}$ can only be $-2(S-P)(S+P)$.
Similar to the situation for the $\gamma_{5}$ spin state of $M_{2}$,
the possible twist-3 contribution can only come from $\gamma^{\alpha}=\s{\bar{n}}n^{\alpha}$,
since the $\mu$, $\nu$ indices of $\sigma^{\mu\nu}$ of $\epsilon_{\perp}\cdot\sigma$ are transversal.
The second trace term is then proportional to
\begin{eqnarray}
\left(\frac{1}{2 n\cdot q}
-\frac{1}{2 n \cdot q}\right)
\epsilon_{\perp}^{\mu\nu}
\epsilon_{\perp,{\mu\nu}}=0
\end{eqnarray}
which is also vanishing.

\subsection{Collinear expansion for annihilation corrections from final state emission gluons}
\begin{figure}
%{\centerline{\includegraphics{diagrams/fourparton}}}
{\centerline{\includegraphics*[scale=0.4,viewport=60 460 580 800]{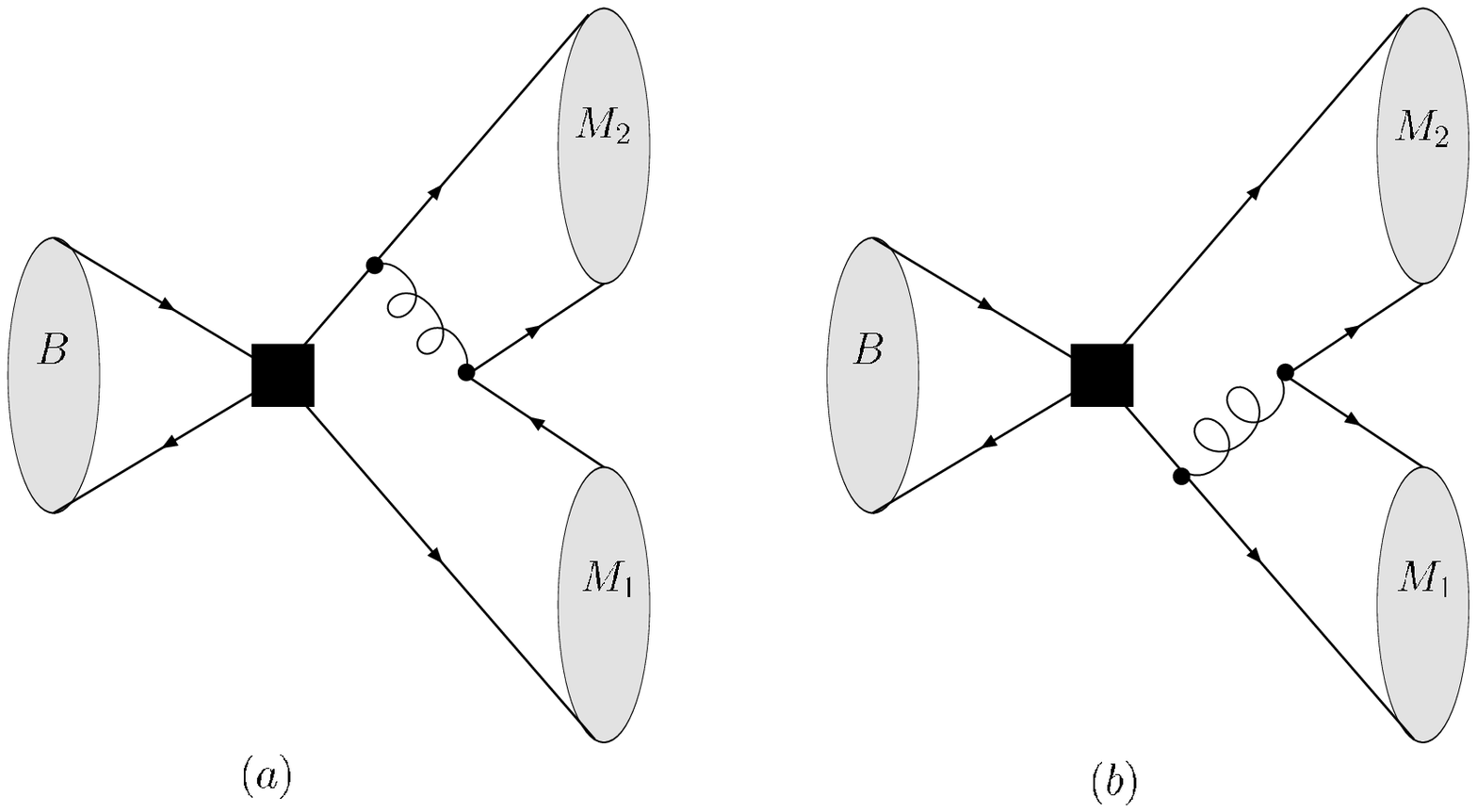}}}
\caption{The diagrams for annihilation corrections from final state emission gluons.}
  \label{fig:fig6}                 
\end{figure}
The relevant term for the annihilation diagrams depicted in \fig{fig:fig6} , 
in which the gluons are emitted from the final state mesons,
is written as,
\bee
\text{Tr}[\phi_{B}(l_B)\Gamma_i][A^{f}_{j}(l_1,l_2)\phi^{M_1}(l_1)\phi^{M_2}(l_2)]
\eee
where
\bee
&&\text{Tr}[A^{f}_{j}(l_1,l_2)\phi^{M_1}(l_1)\phi^{M_2}(l_2)]\nn
&=&4\frac{\pi\alpha_s C_F}{N_c^2}
\text{Tr}[(\gamma_{\alpha}\frac{(\s{l}_2-\s{k})}{(l_2-k)^2}\bar{\Gamma}_i-
\bar{\Gamma}_i\frac{(\s{l}_1-\s{k})}{(l_1-k)^2}\gamma_{\alpha})\phi^{M_1}(l_1)
\gamma^{\alpha}\phi^{M_2}(l_2)]\frac{1}{k^2}
\eee
with $k=\bar{l}_1-l_2$.
The collinear expansion of $A^{f}_{j}(l_1,l_2)$ is
\bee
A^{f}_{j}(l_1,l_2)=A^{f}_{j}(\hat{l}_1,\hat{l}_2)+\cdots\;,
\eee
where dots denote the terms of higher than twist-3.
Since $M_1$ and $M_2$ are assumed to move in opposite directions and carry energetic momenta,
the gluon momentum $k$ can be hard or soft. 
For soft $k$,
the following vanishes 
\bee
\text{Tr}[(\frac{q_{\alpha}}{q\cdot k}\bar{\Gamma}_i-
\bar{\Gamma}_i\frac{p_{\alpha}}{p\cdot k})\phi^{M_1}(l_1)
\gamma^{\alpha}\phi^{M_2}(l_2)]\simeq 0 + O(1/E^2)\;.
\eee
This is because 
\begin{eqnarray}
\phi^{M_1}(l_2)\s{q}\phi^{M_2}(l_2)=\phi^{M_1}(l_1)\s{p}\phi^{M_2}(l_2)\simeq 0 + O(1/E^2)\;,
\end{eqnarray}
where we have used the property of the long distance part of the parton propagators.

\subsection{Collinear expansion for annihilation corrections from initial state emission gluons}
\begin{figure}
%{\centerline{\includegraphics{diagrams/fourparton}}}
{\centerline{\includegraphics*[scale=0.4,viewport=60 460 580 800]{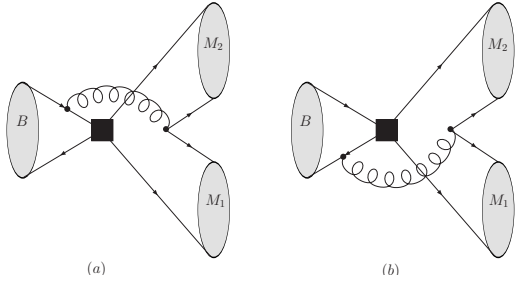}}}
\caption{The diagrams for annihilation corrections from initial state emission gluons.}
  \label{fig:fig7}                 
\end{figure}
The relevant expression for the annihilation diagrams as depicted in \fig{fig:fig7} is given by
\bee\label{eq:ann-0}
&&\text{Tr}[ A^{i}_{j}(l_B,l_1,l_2)\phi_{B}(l_B)\phi^{M_1}(l_1)\phi^{M_2}(l_2)]\nn
&=&-2\frac{\pi\alpha_s C_F}{N_c^2}
\text{Tr}[(\gamma_{\alpha}\frac{(\bar{\s{l}}_B+\s{k}+m_b)}{(\bar{l}_B+k)^2-m_b^2}\Gamma_j-
\Gamma_j\frac{(\s{l}_B+\s{k})}{(l_B+k)^2}\gamma_{\alpha})\phi_{B}(l_B)]\nn
&&\text{Tr}[\bar{\Gamma}_j\phi^{M_1}(l_1)\gamma^{\alpha}\phi^{M_2}(l_2)]\frac{1}{k^2}
\eee
with $k=\bar{l}_1-l_2$.
The collinear expansion for $A^{i}_{j}(l_B, l_1,l_2)$ with respect to $\hat{l}_{B,1,2}$ is given by
\bee
A^{i}_{j}(l_B,l_1,l_2)=A^{i}_{j}(\hat{l}_B,\hat{l}_1,\hat{l}_2)+\cdots\;,
\eee
where dots denote terms of higher than twist-3.
The $\hat{l}_B$ is defined as $\hat{l}_B=\xi P_B$ and the $\xi$ is identified as the momentum fraction carried by
the spectator anti-quark of the $B$ meson. 
Some of order $O(\Lambda^2/E^2)$ terms have been added into the propagator of the spectator anti-quark 
in $A^{i}_{j}(\hat{l}_B,\hat{l}_1,\hat{l}_2)$.
This makes the spectator anti-quark be slightly off-shell $\hat{l}_B^2\simeq \xi^2 m_b^2\propto \Lambda^2/E^2$.
However, for $(V-A)(V\pm A)$ and $-2(S-P)(S+P)$ operators, the final expression for 
propagator of the spectator anti-quark appears proportional to $1/(\bar{u}-\xi)$ after we have cancelled some common
factors of the numerator and denominator of the propagator.
This looks like that the spectator anti-quark carries a collinear momentum $\xi E n^{\mu}$.
The subscript $j$ in $A_{j}^{i,f}$ mean that $j=1$ for $(V-A)(V-A)$ operators, 
$j=2$ for $(V-A)(V+A)$ operators, and $j=3$ for $-2(S-P)(S+P)$ operators.
The superscript $i$ and $f$ in $A_{j}^{i,f}$ mean that gluon emissions start from the initial or final state.
The convention used here is that the $M_1$ contains an anti-quark from the weak vertex.
The anti-quark carries longitudinal momentum fraction as $\bar{u}$.
The $M_2$ contains a quark from the weak vertex with momentum fraction $v$. 
Because the weak annihilation terms are power suppressed than the leading hard spectator interactions by a
relative factor $1/m_b$,
we identify the weak annihilation correction as subleading corrections.
Within our precisions, only terms from twist-2 DAs for the light mesons are considered.
Since $M_1$ and $M_2$ are assumed to move in opposite directions and carry energetic momenta,
the gluon momentum $k$ can be hard or soft. 
For soft $k$,
the following contraction vanishes 
\bee
\text{Tr}[(\frac{P_{b,\alpha}}{P_b\cdot k}\bar{\Gamma}_i-
\bar{\Gamma}_i\frac{P_{b,\alpha}}{P_b\cdot k})\phi_{B}(l_B)]\simeq 0 + O(1/E^2)
\eee

The hard $k$ contributions should give finite results.
However, this requires some cares.
The second term in the first trace term of $A^{i}_{j}(\hat{l}_B,\hat{l}_1,\hat{l}_2)$ is proportional to 
$(\bar{u}-\xi)^{-1}$. 
The conventional approach is to neglect the $\xi$ as the authors of \cite{Beneke:2000ry} has done.
This then results in an end-point divergence $X_{A}$ as $\bar{u}\to 0$.
To solve this divergent problem, 
we propose to retain the $\xi$ dependence in $A^{i}_{j}(\hat{l}_B,\hat{l}_1,\hat{l}_2)$.
By introducing a model for $\phi_{B}(\xi)$,
the divergent term $X_A$ appears as
\bee
X_A\to\int_0^{1}d\xi \phi_{B}(\xi)\int_{0}^{1} \frac{du}{(\bar{u}-\xi)\bar{u}}\phi_{P}^{M_1}(u)\;.
\eee

%%%%%%%%%%%%%%%%%%%%%%%%%%%%%%%%%%%%%%%%%%%%%%%%%%%%%%%%%%%%%%%%%%%%%%%%%%%%%%%%%%%%%%%%%%%%%%%%%%%%%%%%%%%%%%%%%%%%
\section{comparisons with other expansion schemes}\label{sec:sec4}
In this section, a comparison between different calculation schemes for the contributions
involving the pseudotensor LCDA of the final state light mesons will be given.

\subsection{BN scheme}
The calculation scheme proposed by Beneke and Neubert in \cite{Beneke:2000ry, Beneke:2003zv} 
is reviewed below for comparison.
We denote this calculation scheme as the BN scheme. 
Let $H(u,k_{\perp},\cdots)$ represent any hard scattering function in the amplitude
\bee\label{BN-eq-1}
\int_{0}^{1} du \cdots\Tr[H(u,k_{\perp},\cdots)\Phi^{M}(u,k_{\perp})]\;,
\eee
such as the one loop vertex function $V$, 
the one loop penguin function $P$, 
the hard spectator function $H$.
The twist order of the annihilation contributions involving the twist-3 pseudotensor LCDA of 
one of final state light mesons in the $B\to M_1 M_2$ decays are identified as twist-4, 
they will be neglected in this comparison.
The variable $u$ and $k_{\perp}$ in $H(u,k_{\perp},\cdots)$ denote 
the momentum fraction and transverse momenta carried by the partons, respectively. 
The momentum carried by the quark $q$ is denoted by $k_{q}$ 
and the momentum carried by the anti-quark $\bar{q}$ is denoted as $k_{\bar{q}}$.  
We write $k_{q}$ and $k_{\bar{q}}$ as
\begin{eqnarray}\label{q-momenta}
k_{q}^{\mu} &=& u E n^{\mu} + \frac{k_{\perp}^2}{2u E} \bar{n}^{\mu} + k_{\perp}^{\mu}\nonumber\\
k_{\bar{q}}^{\mu} &=& u E \bar{n}^{\mu} + \frac{k_{\perp}^2}{2\bar{u}E} \bar{n}^{\mu} - k_{\perp}^{\mu}\;.
\end{eqnarray}
The momentum fraction $u$ is defined as $u=n\cdot k_{q}/n\cdot P$ 
with $P$ being the meson's momentum as defined in Eq.~(\ref{meson_momentum}).
The $\bar{u}=1-u$ is defined as $\bar{u}=n\cdot k_{\bar{q}}/n\cdot P$.
The meson state $\Phi^{M}(u,k_{\perp})$ is defined as
\bee\label{meson-state-1}
\Phi^{M}(u,k_{\perp})=-\frac{if_{P}}{4}\left[\mu_{P}\gamma_5\left(\phi_{p}(u)
-i\sigma_{\alpha\beta}\bar{n}^{\alpha}n^{\beta}\frac{\phi_{\sigma}^{\prime}(u)}{6}
+i\sigma_{\alpha\beta}E\bar{n}^{\alpha}\frac{\phi_{\sigma}(u)}{6}
\frac{\partial}{\partial k_{\perp\beta}}\right)\right]
\eee
where $\phi_{P}^{M}(u)$, $\phi_{p}^{M}(u)$, 
and $\phi_{\sigma}^{M}(u)$ are the twist-2 and twist-3 LCDAs for the pseudoscalar meson $M$,
and $\phi^{M\prime}_{\sigma}(u)=d\phi_{\sigma}(u)/du$.
The $\partial/\partial k_{\perp,\beta}$ in last term of $\Phi^{M}(u,k_{\perp})$ is defined to be applied on 
the $H(u,k_{\perp},\cdots)$.
After the $\partial/\partial k_{\perp,\beta}$ has been applied on $H(u,k_{\perp},\cdots)$, 
a collinear limit $k_{\perp}\to 0$ is followed to be applied \cite{Beneke:2000ry, Beneke:2003zv}.
Because $H(u,k_{\perp},\cdots)$ can be expanded in $k_{\perp}$,
one can use the following transformation  \cite{Beneke:2000ry, Beneke:2003zv}
\begin{eqnarray}\label{k_T_trans-1}
\frac{\partial}{\partial k_{\perp\beta}}\to\frac{2k^\beta_{\perp}}{k^2_{\perp}}
\end{eqnarray}
to simplify the calculation 
\begin{eqnarray}\label{k_T_trans-2}
\frac{\partial}{\partial k_{\perp\beta}}k_{\perp}^{\lambda}\to
\frac{\langle k^{\beta}_{\perp}k^{\lambda}_{\perp}\rangle}{k^2_{\perp}}=d^{\beta\lambda}_{\perp}\;,
\end{eqnarray}
where $d^{\beta\lambda}_{\perp}=g^{\beta\lambda}-\bar{n}^{\beta}n^{\lambda}-\bar{n}^{\lambda}n^{\beta}$ 
and the bracket in $\langle k^{\beta}_{\perp}k^{\lambda}_{\perp}\rangle$ represents an average over the azimuthal angle
in the integration over $k_{\perp}$.
 
The next step is to use the equations of motion, Eq.~(\ref{equation-1}),
to rewrite the twist-3 part of $\Phi^{M}(u,k_{\perp})$ in the following expression \cite{Beneke:2000ry, Beneke:2003zv}
\begin{eqnarray}\label{meson-state-2}
\left.\Phi^{M}(u,k_{\perp},\cdots)\right|_{tw3}=-\frac{if_{P}\mu_{\chi}^M}{4}\gamma_{5}
\frac{\s{k}_{q}\s{k}_{\bar{q}}}{k_{q}\cdot k_{\bar{q}}}\phi_{p}(u)\;.
\end{eqnarray}
We note that some terms of $O(\Lambda/E)$ have been added to arrive at the above compact form.
In addition, the solution based on Eq.~(\ref{equation-1})  
$\phi_{p}^M(u)=1$ and $\phi_{\sigma}^M(u)=6u\bar{u}$
have been used to derive Eq.~(\ref{meson-state-2}). 
Finally, the \eq{meson-state-2} is substituted into \eq{BN-eq-1}
to arrive at 
\bee
-\frac{if_{P}\mu_{P}}{4}\int_{0}^{1} d u \cdots
\text{Tr}[ H(u,k_{\perp},\cdots)\gamma_5 
\left.\frac{\s{k}_{q}\s{k}_{\bar{q}}}{k_{q}\cdot k_{\bar{q}}}]\right|_{k_{\perp}\to 0}
\phi_{p}(u)\;.
\eee
For comparison, we suggest to use the following expression  
\bee\label{BN-scheme}
-\frac{if_{P}\mu_{P}}{4}\int_{0}^{1} d u \cdots
\left[\right. &&\text{Tr}[ H(u,\cdots)\gamma_5 ]\phi_{p}(u)\nn
&&-i\text{Tr}[ H(u,\cdots)\sigma_{\alpha\beta}\gamma_5 ]\bar{n}^{\alpha}n^{\beta}
\frac{\phi_{\sigma}^{\prime}(u)}{6}\nn
&&+i\left.\text{Tr}[ \frac{\partial H(u, k_{q\perp},\cdots)}{\partial k_{q\perp}^{\beta}}
\sigma_{\alpha\beta}\gamma_5 ]\right|_{k_{q\perp}=0}
E\bar{n}^{\alpha}\frac{\phi_{\sigma}(u)}{6}\left.\right ]\;.\nn
\eee
Using this calculation scheme,
the authors of \cite{Beneke:2000ry, Beneke:2003zv} found that the relevant vertex contributions and
penguin contributions are IR finite,
while the hard scattering contributions contain the divergent term $X_{H}$.

\subsection{DYZ scheme}
Du {\it et. al.} \cite{Du:2001ns,Du:2001iu} proposed to perform the calculations (such as the Dirac algebra 
and the integration over the virtual gluon's momentum) for
the contributions involving $\phi^M_{\sigma}(u)$ in the coordinate space.
After the calculations have been completed, the expression is then transformed into the momentum space.
For comparison, we write the resultant expression in the momentum space as
\bee\label{DYZ-scheme}
 -\frac{if_{P}\mu_{P}}{4}\int_{0}^{1} d u\cdots 
&&\left[\right.  \text{Tr}[ H(u,\cdots)\gamma_5 ]\phi_{p}(u)
+i\text{Tr}[ \frac{\partial H(l,\cdots)}{\partial l^{\beta}}\sigma^{\alpha\beta}\gamma_5 ]|_{l=\hat{l}}
E \bar{n}_{\alpha}\frac{\phi_{\sigma}(u)}{6} \left.\right ]\;.
\eee
Using this calculation scheme, Du {\it et. al.} found that
it is necessary to require a symmetric criteria for the $\phi_{\sigma}(u)$ to regularize the infrared divergences 
from the vertex and penguin contributions. 
However, even requiring the symmetric criteria,
the IR divergences still exist in the hard spectator contributions.  
   
\subsection{CE scheme}
Based on the analysis given in Section III,
the relevant amplitudes for one loop contributions with a pseudotensor LCDA calculated under the CE scheme
can be formally written as
\bee\label{CE-eq-1}
 &&-\frac{if_{P}\mu_{P}}{4}\int_{0}^{1} d u\cdots 
\left[\right.  \text{Tr}[ H(u,\cdots)\gamma_5 ]\hat{\phi}_{p}(u)
-\frac{1}{2}
\text{Tr}[ H(u,\cdots)\epsilon_{\perp}^{\alpha\beta}\sigma_{\alpha\beta} ]\hat{\phi}_{\sigma}(u) \left.\right ] \nn
&&+\frac{1}{8}\int_{0}^{1} d u\cdots \text{Tr}[ H_{\mu}(u,u,\cdots)\sigma_{\alpha\beta}\gamma_5]
w^{\mu}_{\mu^\prime }\phi^{\alpha\beta\mu^\prime}_{\partial}(u,u) \;,
\eee
where 
\bee
&&\phi^{\alpha\beta\mu^\prime}_{\partial}(u,u)=
\int_0^{\infty}\frac{d\lambda}{2\pi}e^{-i\bar{u} \lambda }
\langle M|\bar{q}(0)\sigma^{\alpha\beta}\gamma_5 i \partial^{\mu^\prime}(\lambda n/E)
q(\lambda n/E)|0\rangle \;.\nn
\eee
Under the CE scheme, the analysis given in Section III 
and the explicit expressions for the relevant one loop contributions in the next Section,  
show that the vertex, penguin and hard scattering contributions are IR finite up-to twist-3 under the
two particle approximation. The last term in \eq{CE-eq-1} is a of higher twist than three.
We retain it in \eq{CE-eq-1} is only for comparisons with the other schemes.

\subsection{Comparisons}
We now summarize the differences between these three calculation schemes.

\subsubsection{The interpretation of the derivative hard function}
The derivative functions 
$
{\partial H(u,l_{\perp},\cdots)}/{\partial l_{\perp}^{\beta}}
$
in the BN scheme or ${\partial H(u,l,\cdots)}/{\partial l^{\beta}}$  
in the DYZ schemes are interpreted as short distance hard scattering functions related to $\phi_{\sigma}(u)$.
The derivative $\partial/\partial l^{\beta}_\perp$ in the BN scheme, 
or $\partial/\partial l^{\beta}$ in the DYZ scheme 
come from the coordinate variable in the spin projector associated with the $\phi_{\sigma}(u)$. 
In the CE scheme, the derivative function
${\partial H(u,l,\cdots)}/{\partial l^{\mu}}$ arises from a Taylor expansion of the $H$ with respect to the
collinear momentum $\hat{l}$.
By a simple manipulation, we can write the following corresponding relations between the terms calculated in 
the BN and the CE schemes as
\bee
&& \text{Tr}[ H(u,\cdots)\sigma_{\alpha\beta}\gamma_5 \bar{n}^{\alpha}n^{\beta} ]^{BN}
\frac{\phi_{\sigma}^{\prime}(u)}{3}
\leftrightarrow -i\text{Tr}[ H(u,\cdots)\epsilon_{\perp}^{\alpha\beta}\sigma_{\alpha\beta} ]^{CE}
\hat{\phi}_{\sigma}(u)\;,\nn
&& \text{Tr}[ \frac{\partial H(l,\cdots)}{\partial l_{\perp}^{\beta}} \sigma^{\alpha\beta}\gamma_5]^{BN}|_{l=\hat{l}}
E \bar{n}_{\alpha}f_{P}\mu_{P}\frac{\phi_{\sigma}(u)}{3}
\leftrightarrow
-i\text{Tr}[H_{\mu}(u,u,\cdots)\sigma_{\alpha\beta}\gamma_5]^{CE}
d^{\mu}_{\mu^\prime}
\phi^{\alpha\beta\mu^\prime}_{\partial}(u,u)\;.\nn
\eee 
Similarly, we can also obtain the following relationships between terms derived by the DYZ scheme and the CE scheme as
\bee
\text{Tr}[\frac{ H(l,\cdots)}{\partial l^{\beta}}\sigma^{\alpha\beta} ]^{DYZ}_{l=\hat{l}}
E \bar{n}_{\alpha}f_{P}\mu_{P}\frac{\phi_{\sigma}(u)}{3}
\leftrightarrow 
&& i\text{Tr}[ H(u,\cdots)\epsilon_{\perp}^{\alpha\beta}\sigma_{\alpha\beta} ]^{CE}
f_{P}\mu_{P}\hat{\phi}_{\sigma}(u)\nn
&&-i\text{Tr}[H_{\mu}(u,u,\cdots)\sigma_{\alpha\beta}\gamma_5]^{CE}
d^{\mu}_{\mu^\prime}
\phi^{\alpha\beta\mu^\prime}_{\partial}(u,u)\;.\nn
\eee 
In the above expressions, 
we have assumed that $\phi^{\prime}_{\sigma}(u)$ and $\phi_{\sigma}(u)$ are scheme independent.
The BN, DYZ, and CE superscripts in the hard scattering functions 
represent the functions having been calculated under the BN scheme, the DYZ scheme, and the CE scheme, respectively.
Under the two particle approximation, 
$H_{\mu}(u,u,\cdots)$ is identical to $(\partial H/\partial l^{\mu})_{l\to\hat{l}}$. 
And, ${\partial H(l,\cdots)}/{\partial l^{\mu}}$ is $1/E$ suppressed than $H(l,\cdots)$.
The main differences between the CE scheme and the other two schemes are from the spin projector 
$i\epsilon^{\alpha\beta\nu\mu} E n_{\alpha}$ introduced in the BN and DYZ schemes.
The $E$ factor from the spin projector has the effects to increase one order in the dimension of 
${\partial H(l,\cdots)}/{\partial l^{\mu}}$.
Therefore, the product of ${\partial H(l,\cdots)}/{\partial l^{\mu}}$ and 
$i\epsilon^{\alpha\beta\nu\mu} E n_{\alpha}$ becomes of the same order as of order of $H(l,\cdots)$.
On the other hand, in the CE scheme,
the ${\partial H(l,\cdots)}/{\partial l^{\mu}}$ is not assumed to be associated with a large factor $E$.

\subsubsection{The twist identification}
The analysis made in Section II indicates that the $\phi_{\sigma}(u)$ (or $\phi^{\perp}_{\sigma}(u))$) 
should be identified as one twist order higher than the twist order of $\phi^{\prime}_{\sigma}(u)$ 
(or $\phi^{\parallel}_{\sigma}(u))$) under the energetic light meson limit.
Namely, $\phi_{\sigma}(u)$ is of $O(\Lambda/E)$ at $u=O(1)$
and $\phi^{\prime}_{\sigma}(u)=O(1)$ at $u=O(1)$.
The term 
\[
i\text{Tr}[ \left.\frac{\partial H(l,\cdots)}{\partial l_{\perp}^{\beta}}\sigma^{\alpha\beta}]^{(BN)}\right|_{l=\hat{l}}
E \bar{n}_{\alpha}\frac{\phi_{\sigma}(u)}{6}
\]
is $1/E$ suppressed than the term
\[
-i\text{Tr}[ H(u,\cdots)\sigma_{\alpha\beta}\gamma_5 \bar{n}^{\alpha}n^\beta]^{(BN)}
\frac{\phi_{\sigma}^{\prime}(u)}{6}\;.
\]
Thus, the calculations made by the BN scheme under the energetic meson limit 
become consistent with the results obtained by
using the CE scheme.
This fact can be seen in the next Section, where the explicit results for the one loop contributions 
calculated by the CE scheme will be given.
Unlike the BN and CE schemes, 
the identification of different twist order is unclear in the DYZ scheme. 
The contributions of different order of magnitudes are mixing together.

%%%%%%%%%%%%%%%%%%%%%%%%%%%%%%%%%%%%%%%%%%%%%%%%%%%%%%%%%%%%%%%%%%%%%%%%%%%%%
%\section{One Loop Evolution equation}
%In this section, we calculate the one loop radiative corrections to 
%the LCDAs $\phi_{p}(x)$ and $\hat{\phi}_{\sigma}(x)$.

%%%%%%%%%%%%%%%%%%%%%%%%%%%%%%%%%%%%%%%%%%%%%%%%%%%%%%%%%%%%%%%%%%%%%%%%%%%%%
\section{Twist-3 two particle contributions in $B\to\pi K$ decays}\label{sec:sec5}
%%%%%%%%%%%%%%%%%%%%%%%%%%%%%%%%%%%%%%%%%%%%%%%%%%%%%%%%%%%%%%%%%%%%%%%%%%%%%
The predictions for the penguin dominated $B$ decay processes under QCDF approach 
are related to the $X_{H,A}$ terms.
We consider the $B\to\pi K$ decays as an example to illustrate how our results obtained in previous sections
can be used to improve our understandings for these decays.
The matrix elements of the effective weak Hamiltonian can be written as, up to $O(1/m_b)$,
in the convention of \cite{Beneke:2001ev}
\bee
\langle \pi K|H_{eff}|\bar{B}\rangle
=\frac{G_F}{\sqrt{2}}\sum_{p=u,c}\lambda_{p}\langle \pi K|T_{p} + T_{p}^{a n n}|\bar{B}\rangle
\eee
where
\bee
T_{p}&=&a_1(\pi K)\delta_{p u}(\bar{u}b)_{V-A}\otimes(\bar{s}u)_{V-A}
+a_2(\pi K)\delta_{p u}(\bar{s}b)_{V-A}\otimes (\bar{u}u)_{V-A}\nn
&& + a_3(\pi K)\sum_{q}(\bar{s}b)_{V-A}\otimes (\bar{q}q)_{V-A}
+a_4^{p}(\pi K)\sum_{q}(\bar{q}b)_{V-A}\otimes (\bar{s}q)_{V-A}\nn
&& + a_5(\pi K)\sum_{q}(\bar{s}b)_{V-A}\otimes (\bar{q}q)_{V+A}
+ a_6(\pi K)\sum_{q}(-2)(\bar{q}b)_{S-P}\otimes (\bar{s}q)_{S+P}\nn
&& + a_7(\pi K)\sum_{q}(\bar{s}b)_{V-A}\otimes \frac{3}{2}e_q (\bar{q}q)_{V+A}\nn
&& + a_8(\pi K)\sum_{q}(-2)(\bar{q}b)_{S-P}\otimes \frac{3}{2}e_q(\bar{s}q)_{S+P}\nn
&& + a_9(\pi K)\sum_{q}(\bar{s}b)_{V-A}\otimes \frac{3}{2}e_q (\bar{q}q)_{V-A}\nn
&& + a_{10}(\pi K)\sum_{q}(\bar{q}b)_{V-A}\otimes \frac{3}{2}e_q (\bar{s}q)_{V-A}\;,
\eee
where $(\bar{q}_1 q_2 )_{V\pm A}=\bar{q}_1 \gamma^{\mu}(1\pm \gamma_5)q_2$
and $(\bar{q}_1 q_2 )_{S\pm P}=\bar{q}_1 (1\pm \gamma_5)q_2$.
The symbol $\otimes$ in $T_p$ implies that the matrix elements of the operators in $T_{p}$ are evaluated
according to the factorized form 
$\langle \pi K|j_1\otimes j_2|\bar{B}\rangle\equiv \langle \pi|j_1|\bar{B}\rangle\langle K|j_2|0\rangle $
or $\langle K|j_1|\bar{B}\rangle\langle \pi|j_2|0\rangle$.
The contributions relate to $T^{a n n}_p$ arise from the weak annihilation contributions
with a set of coefficients $b_i(\pi K)$.

The expressions for $a_{i}(\pi K)$ are written as \cite{Beneke:2001ev, Beneke:2003zv}
\begin{eqnarray}
a_{i}^{p}(M_1 M_2)&=&(C_{i} + \frac{C_{i\pm 1}}{N_c})N_{i}(M_2)\nn
&& +\frac{C_{i\pm 1}}{N_c}\frac{\alpha_s C_F}{4\pi }[V_{i}(M_2)+\frac{4\pi^2}{N_c}H_{i}(M_1 M_2)]+P_{i}^{p}(M_2)\;,
\end{eqnarray}
where the upper or lower signs correspond to the odd $i$ or even $i$.
The superscript $p$ is only used for $i\ge 3$.
%The function $V_{i}$ denotes the one loop vertex corrections, 
%$H_{i}$ the hard spectator corrections,
%and $P_{i}^{p}$ the penguin corrections.
The leading order coefficients $N_{i}(M_2)$ represent the normalization integral of the distribution amplitude
$\phi_{P}^{M_2}$ or $\hat{\phi}_{p}^{M_2}$
\begin{eqnarray}
N_{i}(M_2)=\left\{
\begin{array}{ll}
0; & i=6,8\\
1; & i=1-5,7,9,10
\end{array}
\right.
\end{eqnarray}
The vertex corrections are written as
\begin{eqnarray}\label{vertex-function}
V_{i}(M_2)=\left\{
\begin{array}{ll}
{\displaystyle \int_{0}^{1}du \phi_{P}^{M_2}(u)\left[12\ln\left(\frac{\mu}{m_b}\right)-18+g(u)\right]} ; & i=1-4,9,10\\
{\displaystyle \int_{0}^{1} du \phi_{P}^{M_2}(u)\left[-12\ln(\frac{\mu}{m_b})+6-g(1-u)\right]}; & i=5,7\\
{\displaystyle \int_{0}^{1} du \hat{\phi}_{p}^{M_2}(u)\left[-6+h(u)\right]}; & i=6,8
\end{array}
\right.
\end{eqnarray}
where
\begin{eqnarray}
g(v)&=&3\left(\frac{1-2u}{1-u}\ln u -i\pi\right)\nn
    && +\left[ 2Li_{2}(u)-\ln^2 u +\frac{2\ln u}{1-u}-(3+2i\pi)\ln u -(u\leftrightarrow (1-u)) \right]\nn
h(v)&=&2\left[Li_{2}(u)-(1+i\pi)\ln u -(u\leftrightarrow (1-u)\right]\;,    
\end{eqnarray}
where we have employed the naive dimensional regularization (NDR) scheme with an anti-commuting $\gamma_5$ 
for the regularization of the UV or IR divergences arising from the loop corrections.
The calculations for $V_{i}(M_2)$, $i=1-5,7,9,10$ have been checked by using the CE scheme.
The results have been found to be consistent with the results derived by using the BN scheme.
This is reasonable because CE and BN schemes are equivalent at the leading twist order.
We also note that the $V_{6,8}$ calculated under the CE scheme are also consistent with those obtained by
the BN scheme.
This is due to the fact that the main contributions in $V_{6,8}(M_2)$ are from the projection onto the 
$\phi_{p}^{M_2}$.
The detailed calculations  for $V_{6}(M_2)$ under the CE scheme will be given in the Appendix A.
The integrations
\bee
\int_{0}^{1} du g(u)\phi^{M_2}_{P}(u)&=&-\frac{1}{2}-3i\pi\;,\nn
\int_{0}^{1} du h(u)\hat{\phi}^{M_2}_{p}(u)&=&0\;,
\eee
lead to the values of the vertex corrections at the scale $\mu$ as
\bee
V_{i}(M_2)=
\left\{
\begin{array}{ll}
{\displaystyle \left[12\ln\left(\frac{\mu}{m_b}\right)-\frac{37}{2}-3i\pi\right]}; & i=1-4,9,10\\
{\displaystyle \left[-12\ln(\frac{\mu}{m_b})+\frac{13}{2}+3i\pi\right]}; & i=5,7\\
{\displaystyle -6 }; & i=6,8\\
\end{array}
\right.
\eee

The penguin contributions are given by
\bee
P_4^p(M_2)
=&&\frac{\alpha_s C_F}{4\pi N_c}
\left\{
C_1 \left[\frac{4}{3}\ln\frac{m_b}{\mu}+\frac{2}{3}-G_{M_2}(s_p)\right]
+C_3 \left[\frac{8}{3}\ln\frac{m_b}{\mu}+\frac{4}{3}-G_{M_2}(0)-G_{M_2}(1)\right]
\right.\nn
&&
\hspace{1.6cm}+(C_4+C_6) \left[\frac{4n_f}{3}\ln\frac{m_b}{\mu}
-(n_f-2)G_{M_2}(0)-G_{M_2}(s_c)-G_{M_2}(1)\right]\nn
&&
\hspace{1.6cm}
-\left.2C_{8g}^{eff}\int_{0}^{1} \frac{du}{1-u}\phi^{M_2}_{P}(u)\right\}\;,\nn
P_6^p(M_2)
=&&\frac{\alpha_s C_F}{4\pi N_c}
\left\{
C_1 \left[\frac{4}{3}\ln\frac{m_b}{\mu}+\frac{2}{3}-\hat{G}_{M_2}(s_p)\right]
+C_3 \left[\frac{8}{3}\ln\frac{m_b}{\mu}+\frac{4}{3}-\hat{G}_{M_2}(0)-\hat{G}_{M_2}(1)\right]
\right.\nn
&&
\hspace{1.6cm}+(C_4+C_6) \left[\frac{4n_f}{3}\ln\frac{m_b}{\mu}+\frac{2}{3}
-(n_f-2)\hat{G}_{M_2}(0)-\hat{G}_{M_2}(s_c)-\hat{G}_{M_2}(1)\right]\nn
&&
\hspace{1.6cm}
-\left.2C_{8g}^{eff}\int_{0}^{1} du \hat{\phi}^{M_2}_{p}(u)\right\}\;,\nn
P_{10}^p(M_2)
=&&\frac{\alpha }{9\pi N_c}
\left\{
(C_1 +N_c C_2) \left[\frac{4}{3}\ln\frac{m_b}{\mu}+\frac{2}{3}-G_{M_2}(s_p)\right]
-3C_{7\gamma}^{eff}\int_{0}^{1} \frac{du}{1-u}\phi^{M_2}_{P}(u)\right\}\;,\nn
P_{8}^p(M_2)
=&&\frac{\alpha }{9\pi N_c}
\left\{
(C_1 +N_c C_2) \left[\frac{4}{3}\ln\frac{m_b}{\mu}+\frac{2}{3}-\hat{G}_{M_2}(s_p)\right]
-3C_{7\gamma}^{eff}\int_{0}^{1} du \hat{\phi}^{M_2}_{p}(u)\right\}\;,\nn
\eee 
where $n_f=5$ denotes the number of light quark flavors.
The mass ratios $s_u=0$ and $s_c=m_c^2/m_b^2$ are introduced.
According to the conventions used in \cite{Beneke:2001ev, Beneke:2003zv}, 
the electroweak corrections from $C_{7-10}$ are neglected.
The functions $G_{M_2}(s)$and $\hat{G}_{M_2}(s)$ are defined as \cite{Beneke:2001ev, Beneke:2003zv}
\bee
G_{M_2}(s)&=&\int_0^1 du G(s-i\epsilon, 1-u)\phi^{M_2}_{P}(u)\;,\nn
\hat{G}_{M_2}(s)&=&\int_0^1 du G(s-i\epsilon, 1-u)\hat{\phi}^{M_2}_{p}(u)\;,\nn
G(s, u)&=&-4\int_{0}^{1} dx x(1-x)\ln[s-x(1-x)u]\;.
\eee
The effective Wilson coefficients are given by 
$C_{7\gamma}^{eff}=C_{7\gamma}-\frac{1}{3}C_5-C_6$ and $C^{eff}_{8g}=C_8 +C_5$.
Under the energetic limit,  we choose the asymptotic form for 
both $\phi^{M_2}_{P}(u)$ and $\hat{\phi}^{M_2}_{p}(u)$ as  $\phi^{M_2}_{P}(u)=\hat{\phi}^{M_2}_{p}(u)=6u\bar{u}$.
The asymptotic models for $\phi^{M_2}_{P}(u,\mu)$ and $\hat{\phi}^{M_2}_{p}(u,\mu)$ 
are defined as the asymptotic limit $\mu\to\infty$ for the distribution amplitudes.
At finite renormalization scale, the distribution amplitudes are expanded into Gegenbauer polynomials
\bee
\phi^{M_2}_{P}(u,\mu)=6u\bar{u}\left[1+\sum_{n=1}^{\infty}\alpha_{n}^{M_2}(\mu)C^{3/2}_{n}(2u-1)\right]\;,
\eee
where the Gegenbauer moments $\alpha_{n}^{M_2}(\mu)$ are multiplicatively renormalized.
The scale dependence of the Gegenbauer moments $\alpha_{n}^{M_2}(\mu)$ enters 
the vertex and penguin corrections at order $\alpha_s^2$ \cite{Beneke:2001ev, Beneke:2003zv}. 
In the next-to-leading calculation as we have done in this paper,
the Gegenbauer moments can be neglected.
In this approximation, we then arrive at a further simplification $G_{M_2}(s)=\hat{G}_{M_2}(s)$.
This results in the following identities
\bee
P_6^p(M_2)&=&P_4^p(M_2)+\frac{\alpha_s C_F}{\pi N_c}C^{eff}_{8g}\;,\\
P_{8}^p(M_2)&=& P_{10}^p(M_2)+\frac{2\alpha }{3\pi N_c}C_{7\gamma}^{eff}\;.
\eee

For practical applications, the $G_{M_{2}}(s)$ are evaluated under the previously mentioned approximations as 
\bee
G_{M_2}(0)&=&\frac{5}{3}+\frac{2i\pi}{3}\;,\\
G_{M_2}(1)&=&\frac{85}{3}-6\sqrt{3}+\frac{4\pi}{9}\;,\\
G_{M_2}(s_c)&=&\frac{5}{3}-\frac{2}{3}\ln s_c +\frac{32}{3}s_c + 16 s_c^2\nn
&&-\frac{2}{3}\sqrt{1-4s_c}\left[1+2s_c+24 s_c^2\right](2 \text{arctanh}\sqrt{1-4s_c}-i\pi)\nn
&& +12s_c^2\left[1-\frac{4}{3}s_c\right]
  (2 \text{arctanh}\sqrt{1-4s_c}-i\pi)^2 +\cdots\;.
\eee
The complete expressions for $G_{M_2}(0)$,  $G_{M_2}(1)$ and $G_{M_2}(s_c)$ with Gegenbauer moments are
referred to \cite{Beneke:2001ev, Beneke:2003zv}.

The scale for the vertex and penguin corrections refers to the parton off-shellness in the loop
diagrams.
The typical setting of the scale is chosen to be $\mu\sim m_b$,
which are the scales substituted in the Wilson coefficients $C_i$
and in the hard scattering kernel $T^{I}$.
The combination of the scale dependences and the renormalization scheme-dependent constants in 
$C_i$, $V_{i}(M_2)$ and $P_{i}^p(M_2)$ give renormalization-group invariant results.
   
The hard spectator corrections are given by
\bee\label{eq:HS-1}
H_{i}(M_1 M_2)&=&\frac{B_{M_1 M_2}}{A_{M_1 M_2}}m_B
\int_{0}^{1} d\xi\frac{\phi_{B}(\xi)}{\xi}
\int_{0}^{1} du \int_{0}^{1} dv
\left[\frac{\phi^{M_1}_P(u)\phi^{M_2}_{P}(v)}{\bar{u}(\bar{v}-\xi)}
+r_{\chi}^{M_1}\frac{\hat{\phi}^{M_1}_p(u)\phi^{M_2}_{P}(v)}{u(\bar{v}-\xi)}
\right]\;,\nn
\eee
for $i=1-4,9,10$, and
\bee\label{eq:HS-2}
H_{i}(M_1 M_2)&=&\frac{B_{M_1 M_2}}{A_{M_1 M_2}}m_B
\int_{0}^{1} d\xi\frac{\phi_{B}(\xi)}{\xi}
\int_{0}^{1} du \int_{0}^{1} dv
\left[\frac{\phi^{M_1}_P(u)\phi^{M_2}_{P}(v)}{u(\bar{v}-\xi)}
+r_{\chi}^{M_1}\frac{\hat{\phi}^{M_1}_p(u)\phi^{M_2}_{P}(v)}{\bar{u}(\bar{v}-\xi)}
\right]\;,\nn
\eee
for $i=5,7$, and $H_{6,8}(M_1 M_2)=0$.
Different from the BN scheme,
we have introduced the distribution amplitude for the $B$ meson and retained the $\xi$ 
dependence in the spectator propagator.
Since the $\hat{\phi}^{M_1}_{p}(u)=6u\bar{u}$ is no longer a constant,
there are no end-point divergences as in those results derived in the BN \cite{Beneke:2001ev, Beneke:2003zv} 
or DYZ schemes \cite{Du:2001ns, Du:2001iu}.
Due to the soft scale associated with the spectator quark,
the introduction of the $B$ meson's DA $\phi_{B}(\xi)$ may need to consider the effects of double logarithms 
from a overlap of soft and collinear divergences associated with the spectator quark.
This may need to introduce a Sudakov form factor to account for the effects of double logarithms. 
A two loop analysis for the hard spectator made by Beneke and Yang \cite{Beneke:2005gs}
indicates that there exists no double logarithms up-to $O(\alpha_s)$.
Therefore, at next-to-leading order, 
we can neglect the Sudakov form factor completely.
By using the model for $\phi_{B}(\xi)$ and the fact that $\phi_P^{M_{1}}(u)=\hat{\phi}_p^{M_{1}}(u)$,
we arrive at a simple expression for $H_{i}(M_1 M_2)$, which is useful for numeric calculations,
\bee
H_{i}(M_1 M_2)=\frac{B_{M_1 M_2}}{A_{M_1 M_2}}m_B(1+r_{\chi}^{M_1})N_{H}\;,
\eee
where $N_{H}=18/\lambda_B$.
The model for $\phi_{B}(\xi)$ is assumed to have the form
\bee\label{phiB-model}
\phi_{B}(\xi)=\frac{N_B\xi^2 \bar{\xi}^2}{[\xi^2+\epsilon_B \bar{\xi}]^2}
\eee 
with $N_B=0.133$, $\epsilon_B=0.005$ for $\lambda_B=350$ MeV.
The $N_B$ and $\epsilon_B$ are determined according to the conditions
\bee\label{phiB-condition}
\int_0^1 d\xi\phi_{B}(\xi) &=& 1\;,\nn
\int_0^1 d\xi\frac{\phi_{B}(\xi)}{\xi} &=& \frac{m_B}{\lambda_B}\;,
\eee
where the errors are controlled within $1\%$.
This is consistent with the conventional assumption $\lambda_B\leq 600$ MeV 
under the condition $3\lambda_B\leq 4\bar{\Lambda}$ with $\bar{\Lambda}=m_B -m_b$.
The scale dependence in $r_{\chi}^{M_1}(\mu_h)$ is chosen as $\mu_h=\sqrt{\Lambda_h \mu}$ with $\Lambda_h=0.5$ GeV
\cite{Beneke:2001ev, Beneke:2003zv}.

The annihilation corrections are expressed in terms of the following $A_{j}^{i,f}$, $j=1,\cdots,3$ functions
\bee\label{eq:ann-1}
A_{1}^{i}&=&\pi\alpha_s \int_{0}^{1} d\xi \phi_{B}(\xi)\int_{0}^{1} du \phi^{M_2}_{P}(u)\int_{0}^{1} dv 
\phi^{M_1}_{P}(v)\left[
\frac{1}{(1-(u-\xi)(\bar{\xi}-v))v}
+\frac{1}{(\bar{u}-\xi )\bar{u}v}
\right]\;,\nn
A_{2}^{i}&=&\pi\alpha_s \int_{0}^{1} d\xi \phi_{B}(\xi)\int_{0}^{1} du \phi^{M_2}_{P}(u)\int_{0}^{1} dv 
\phi^{M_1}_{P}(v)\left[
\frac{1}{(1-(u-\xi)(\bar{\xi}-v))\bar{u}}
+\frac{1}{(v-\xi)\bar{u}v}
\right]\;,\nn
A_{3}^{i}&=&A_{1}^{f}=A_{2}^{f}=A_{3}^{f}=0\;.
\eee
The subscript $j$ in $A_{j}^{i,f}$ mean that $j=1$ for $(V-A)(V-A)$ operators, 
$j=2$ for $(V-A)(V+A)$ operators, and $j=3$ for $-2(S-P)(S+P)$ operators.
The superscript $i$ and $f$ in $A_{j}^{i,f}$ mean that gluon emissions start from the initial or final state.
The convention used here is that the $M_1$ contains an anti-quark from the weak vertex.
The anti-quark carries longitudinal momentum fraction as $\bar{u}$.
The $M_2$ contains a quark from the weak vertex with momentum fraction $v$. 
Because the weak annihilation terms are power suppressed than the leading hard spectator interactions by a
relative factor $1/m_b$,
we identify the weak annihilation correction as subleading corrections.
Within our precisions, only terms from twist-2 DAs for the light mesons are considered.
The non-singlet annihilation coefficients are given by
\bee
b_1 &=& \frac{C_F}{N_c^2}C_1 A_{1}^{i}\;,\hspace{2cm}
b_2 = \frac{C_F}{N_c^2}C_2 A_{1}^{i}\;,\nn
b_3^p &=& \frac{C_F}{N_c^2}C_3 A_{1}^{i}\;,\hspace{2cm}
b_4^p = \frac{C_F}{N_c^2}[C_4 A_{1}^{i}+ C_6 A_{2}^{i}]\;,\nn
b_{3,EW}^p &=& \frac{C_F}{N_c^2}C_9 A_{1}^{i}\;,\hspace{2cm}
b_{4,EW}^p = \frac{C_F}{N_c^2}[C_{10} A_{1}^{i}+ C_{8} A_{2}^{i}]\;.
\eee   
We note that the coefficients $(b_1,b_2)$ corresponds to the current-current annihilation,
$(b_3,b_4)$ corresponds to the penguin annihilation,
and $(b_{3,EW},b_{4,EW})$ to the electroweak penguin annihilation.
The end-point divergences as the spectator quark approaching its on-shell are regularized by
the momentum fraction $\xi$, which is defined as $\hat{l}_B=\xi P_b$.
Such a definition for $\xi$ would make the spectator quark slightly off-shell 
since $\hat{l}_B^2=\xi^2 P_b^2\sim \lambda_B^2$,
which is of order $\Lambda_{QCD}^2$.
The errors from this off-shell-ness for the amplitude is of order $O(\Lambda^2/E^2)$, which is of twist-4.
Within our precision, the errors can be neglected. 

Using the same approximations for the hard spectator corrections, we arrive at the following simplified expressions
for the annihilation corrections
\bee
A_{1}^{i}=A_{2}^{i}\simeq 18\pi\alpha_s \left(\frac{\pi^2-9}{3} + N_A\right)\;,
\eee
where $N_{A}=-12.1-1.7 i$ is related to the model for $\phi_{B}(\xi)$.
The convolution integrations appear to be overlapped integrals over $\xi$, $u$ and $v$ with
$\phi_{B}(\xi)$, $\phi^{M_1}_P(u)$ and $\phi^{M_2}_P(v)$.
The traditional end point divergences as $u\to 1$ or $v\to 1$ are then regularized by the $B$ meson distribution
amplitude $\phi_{B}(\xi)$.
Although the averaged value of the $\xi$ is much smaller than one,
however, we argue that it still can not be neglected.
The scale dependence in $\alpha_s(\mu)$ is chosen as $\mu_h=\sqrt{\Lambda_h \mu}$ with $\Lambda_h=0.5$ GeV.

For penguin dominant $B\to\pi K$ decays,
the relevant decay amplitudes under QCD factorization are parametrized as the following \cite{Beneke:2001ev}
\bee
&&A(B^{-}\to \pi^{-}\bar{K}^0)
=\lambda_{p}\left[(a_4^p -\frac{1}{2} a_{10}^p)+r_{\chi}^{K}(a_6^p -\frac{1}{2} a_8^p)\right] A_{\pi K}\nn
&&\hspace{3.5 cm}+ (\lambda_u b_2 + (\lambda_u +\lambda_c)(b_3 + b_3^{EW}))B_{\pi K}\;,\nn
&& - \sqrt{2}A(B^-\to \pi^0 K^-)=[\lambda_u a_1 +\lambda_p ( a_4^p + a_{10}^p)
+ \lambda_p r_{\chi}^{K}(a_6^p + a_8^p) ]A_{\pi K} \nn
&&\hspace{4.5 cm}+ [\lambda_u a_2 + (\lambda_u +\lambda_c)\frac{3}{2}(-a_7 + a_9)]A_{K\pi}\nn
&&\hspace{4.5 cm}+ [\lambda_u b_2 + (\lambda_u +\lambda_c)(b_3 + b_3^{EW})]B_{\pi K}\;,\nn
&& -A(\bar{B}^{0}\to\pi^+ K^{-})=[\lambda_u a_1 + \lambda_p (a_4^p + a_{10}^p) 
+\lambda_p r_{\chi}^K (a_6^p + a_8^p)]A_{\pi K}\nn
&&\hspace{3.5 cm}+[(\lambda_u +\lambda_c )( b_3 - \frac{1}{2}b_3^{EW})]B_{\pi K}\;\nn
&&\sqrt{2}A(\bar{B}^0 \to \pi \bar{K}^0)=A(B^- \to \pi^- \bar{K}^0) + \sqrt{2}A(B^-\to\pi^0 K^-)
-A(\bar{B}^0\to \pi^+ K^-)
\eee
where $\lambda_p =V_{pb}V_{ps}^*$, $a_i\equiv a_i (\pi K)$, 
and $\lambda_p a_i^p=\lambda_u a^u_i + \lambda_c a^c_i$.
The $CP$ conjugation of decay amplitudes are obtained by replacing $\lambda_p \to \lambda_p^*$ for the above amplitudes.
The $A_{\pi K }$, $A_{K \pi }$, and $B_{\pi K }$ are defined as
\bee
A_{\pi K}&=&i\frac{G_F}{\sqrt{2}}(m_B^2-m_{\pi}^2)F^{B\to\pi}_0 (m_K^2 )f_K\;,\nn
A_{K \pi }&=&i\frac{G_F}{\sqrt{2}}(m_B^2-m_{K}^2)F^{B\to K}_0 (m_{\pi}^2 )f_{\pi}\;,\nn
B_{\pi K} &=&i\frac{G_F}{\sqrt{2}} f_B f_K f_{\pi}\;.
\eee
%The form factors are defined
%\bee
%\langle P(p)|\bar{q}\gamma^{\mu}b|\bar{B}\rangle
%=F_+^{B\to P}(q^2)(P_{B}^{\mu} + p^{\mu})
%+[F_0^{B\to P}(q^2)-F_+^{B\to P}(q^2)]\frac{m_B^2-m_P^2}{q^2}q^{\mu}
%\eee
%The form factors coincide as $q^2=0$, $F^{B\to P}_+ (0)=F_0^{B\to P}(0)$.
%The expressions for the parameters $a_i$ are referred to \cite{Beneke:2000ry,Beneke:2001ev}.
For numerical calculations, we will use the following input parameters
\bee
\begin{array}{cccc}
\Lambda_{\bar{MS}}^{(5)}=0.225 \text{GeV}\ , & m_b(m_b)=4.2 \text{GeV} & m_c(m_b)=1.3 \text{GeV}\ , & m_s(2\text{GeV})=0.090\text{GeV}\ , \\
|V_{cb}|=0.41 \ , & |V_{ub}/V_{cb}|=0.09 \ ,& \gamma=70^{\circ} \ ,& \tau(B^{-})=1.67(\text{ps}) \ , \\
\tau(B_d)=1.54(\text{ps}) \ , & f_{\pi}=131\text{MeV} \ ,& f_K = 160\text{MeV}\ , & f_B=200\text{MeV}\ , \\
F_0^{B\to\pi}=0.28 \ ,& F_0^{B\to K}=0.34 \ .& &\\
\end{array}
\eee
For $\lambda_u$ and $\lambda_c$, we take the following convention for their parametrization
\bee
\frac{\lambda_u}{\lambda_c}=\tan^2\theta_c R_b e^{-i\gamma}
\eee
where
\bee
&& \tan^2\theta_c =\frac{\lambda^2}{1-\lambda^2}\;,\nn
&& R_b =\frac{1-\lambda^2/2}{\lambda}|\frac{V_{ub}}{V_{cb}}|\;,\nn
&& \lambda = |V_{us}|\;.
\eee
The value of $\lambda$ is taken as $0.22$.

The values of the NLO Wilson coefficients $C_{i}$, $i=1,\cdots, 10$
and the LO $C_{7\gamma}^{eff}$ and $C_{8g}^{eff}$, at the scale $m_b=4.2 \text{GeV}$ 
( $\sqrt{\Lambda_h m_b}=1.45 \text{GeV}$)
are given by \cite{Beneke:2001ev}
\bee
\begin{array}{lll}
C_1 =1.081(1.190) \ , & C_2 =-0.191(-0.373) \, & C_3 =0.014(0.027) \ ,\\
C_4 =-0.036(-0.062) \ , & C_5 =0.009(0.012) \, & C_6 =-0.042(-0.086) \ ,\\
C_7/\alpha =-0.001(-0.013) \ , & C_8/\alpha = 0.060(0.133)\, & C_9/\alpha = -1.254(-1.380)\ ,\\
C_{10}/\alpha =0.223(0.432) \ , & C_{7\gamma}^{eff,LO}=-0.318 \ ,&  C_{8g}^{eff,LO}=-0.151\ ,
\end{array}
\eee
where the parameters $\Lambda^{(5)}_{\bar{MS}}=0.225 \text{GeV}$,
$m_t(m_t)=167 \text{GeV}$, $m_{b}(m_b)=4.2 \text{GeV}$, $M_W =80.4 \text{GeV}$,
$\alpha=1/129$, $\sin(\theta_{W})=0.23$, $\alpha_{s}(M_Z)=0.118$ have been used.
By using the input parameters and the Wilson coefficients, 
we list the values of $a_i$, $i=1,\cdots, 10$, 
and $b_j$, $j=1,\cdots, 3$, calculated at the scale $m_b=4.2 \text{GeV}$ as below
\bee 
\begin{array}{lll}
a_1 = 0.938 + 0.014 i\ , & a_2 = 0.351 - 0.081 i\ , & a_3 = -0.011 + 0.003 i \ ,\\
a_4^u =-0.020 - 0.015 i & a_4^c =  -0.025 - 0.006 i\ , & a_5 = 0.017 - 0.003  i \ , \\
r_{\chi}^K a^u_6 = -0.048 - 0.016 i \ , & r_{\chi}^K a^c_6 = -0.054 - 0.006 i \ , & a_7/\alpha = 0.060 + 0.004 i \ ,\\
r_{\chi}^K a^u_8/\alpha = 0.077 -0.014 i \ , & r_{\chi}^K a^c_8/\alpha =0.072-0.07 i \ ,
& a_9/\alpha = -1.149 - 0.017  i \ ,\\
a^u_{10}/\alpha = -0.378 + 0.082 i \ , & a^c_{10}/\alpha = -0.382 + 0.088 i \ , & r_A b_1=-0.154 - 0.022 i\ ,\\
r_A b_2 =0.048 + 0.007 i\ ,& r_A b_3 =-0.003 - 0.0005 i\ , & r_A b_4 = 0.019 + 0.003 i \ ,\\
r_A b_3^{EW}/\alpha = 0.179 + 0.026 i \ , &
r_A b_4^{EW}/\alpha = -0.073 - 0.011 i \ , &  
\end{array}\nn
\eee
in which
\bee
r_A = \frac{B_{\pi K}}{A_{\pi K}} = \frac{f_B f_{\pi}}{m_B^2 F^{B\to\pi}_0 (0)}\simeq 0.004\; ,
\eee
and 
\bee
r_{\chi}^{K}=\frac{2 m_K^2}{\bar{m}_b(\bar{m}_{q}+ \bar{m}_s )}\simeq 1.18 (0.76)\;,
\eee
where $r_{\chi}^{K}=1.18$ is calculated at the scale $m_b=4.2$ GeV,
and $r_{\chi}^{K}=0.76$ is calculated at the scale $\sqrt{\Lambda_h m_b}=1.45$ GeV.
Different contributions to the coefficients $a_i$ are given in Table~I for reference.
The hard scattering contributions are dominant for $a_{3}, a_{5}, a_{7}, a_{10}$.
The penguin contributions are minor for all $a_i$, $i=1,\cdots, 10$.
The vertex contributions are important except of $a_{6,8}$.
\begin{table}[t]\label{Table:ai}
\caption{Different contributions to the coefficients $a_i$, $i=1,\cdots, 10$.}
\begin{tabular}[t]{cccccc}
\hline
$a_i$ & $C_i+\frac{C_{i\pm 1}}{N_c}$ & $\frac{\alpha_s C_{F} C_{i\pm 1}}{4\pi N_c} V_{i}$ &
 $\frac{\pi\alpha_s C_{F}C_{i\pm 1}}{N^2_c} H_{i}$ & $ P^{p}_{i}$ & total\\
\hline \hline
$a_{1}$ & $1.017$ & $0.028 + 0.014 i $  & $-0.107$ & $0$ & $0.938 + 0.014 i$\\ 
\hline 
$a_{2}$ & $0.169$ & $-0.158 -0.081 i $  & $0.340$ & $0$ & $0.351 - 0.081 i$\\ 
\hline 
$a_{3}$ & $0.002$ & $0.005 + 0.003 i $  & $-0.018$ & $0$ & $-0.011 + 0.003 i$\\ 
\hline 
$a^{u}_{4}$ & $-0.029$ & $-0.002 - 0.001 i $  & $0.008$ & $0.003-0.014 i$ & $-0.020 - 0.015 i$\\ 
\hline 
$a^{c}_{4}$ & $-0.029$ & $-0.002 - 0.001 i $  & $0.008$ & $-0.002-0.005 i$ & $-0.025 - 0.006 i$\\ 
\hline 
$a_{5}$ & $-0.005$ & $-0.002 - 0.003 i $  & $0.024$ & $0$ & $0.017 - 0.003 i$\\ 
\hline 
$a^{u}_{6}$ & $-0.039$ & $-4.27\times 10^{-4}$  & $0$ & $-0.002-0.014 i$ & $-0.041 - 0.014 i$\\ 
\hline 
$a^{c}_{6}$ & $-0.039$ & $-4.27\times 10^{-4}$  & $0$ & $-0.007-0.005 i$ & $-0.046 - 0.005 i$\\ 
\hline 
$a_{7}/\alpha$ & $0.019$ & $0.003 + 0.004 i $  & $0.038$ & $0$ & $0.060 + 0.004 i$\\ 
\hline 
$a^{u}_{8}/\alpha$ & $0.060$ & $-4.75\times 10^{-5}$  & $0$ & $0.005-0.012 i$ & $0.065 - 0.012 i$\\ 
\hline 
$a^{c}_{8}/\alpha$ & $0.060$ & $-4.75\times 10^{-5}$  & $0$ & $0.001-0.006 i$ & $0.061 - 0.006 i$\\ 
\hline 
$a_{9}/\alpha$ & $-1.180$ & $-0.033 - 0.017 i $  & $0.064$ & $0$ & $-1.149 - 0.017 i$\\ 
\hline 
$a^{u}_{10}/\alpha$ & $-0.195$ & $0.184 + 0.094 i $  & $-0.395$ & $0.028-0.012 i$ & $-0.378 + 0.082 i$\\ 
\hline
$a^{c}_{10}/\alpha$ & $-0.195$ & $0.184 + 0.094 i $  & $-0.395$ & $0.024-0.006 i$ & $-0.382 + 0.088 i$\\ 
\hline
\end{tabular}
\end{table}

The branching ratio for a $\bar{B}\to \pi K$ decay is given by the expression
\bee
Br(\bar{B}\to \pi K)=\frac{\tau_B}{16\pi m_B}|A(\bar{B}\to\pi K)|^2\;,
\eee
\begin{table}[t]\label{Table:BRi}
\caption{Predictions for $CP$-averaged branching ratios (in unites of $10^{-6}$) for $B\to \pi K$ decays.}
\begin{tabular}[t]{cccccc}
\hline
mode & CE & BN$^{\prime}$ & BN\cite{Beneke:2003zv} & DYZ\cite{Du:2001hr} & data\cite{Barberio:2007cr}\\
\hline \hline
$B^{-}\to\pi^{-} \bar{K}^{0}$   & $10.1$ & $12.1 $  & $19.3$ & $12.6$ & $23.1\pm 1.0$\\ 
\hline 
$B^{-}\to\pi^{0} K^{-}$         & $6.4$ & $5.3 $  & $11.1$ & $6.7$ & $12.8\pm 0.6$\\ 
\hline 
$\bar{B}^{0}\to\pi^{+} K^{-}$   & $9.6$ & $10.2$  & $16.7$ & $9.1$ & $19.4\pm 0.6$\\ 
\hline 
$\bar{B^0}\to\pi^{0} \bar{K^{0}}$ & $3.9$ & $6.0$  & $7.0$ & $4.3$ & $10.0\pm 0.6$\\ 
\hline 
\end{tabular}
\end{table}
The predictions in the CE column of Table~II are calculated according to the results derived in this paper.
The predictions in BN column of Table~II are quoted from the paper \cite{Beneke:2003zv}. 
The predictions in DYZ column of Table~II are quoted from the paper \cite{Du:2001hr}.
We observe that the predictions in the BN column are about $1.7-1.9$ times larger than those in the CE and DYZ columns.
To understand this difference, we employed the theoretic input parameters given in the Table 2 
and the $a_i$ and $b_i$ values given in the Table 3-5 of \cite{Beneke:2001ev} to recalculate the predictions 
for the branching ratios for the $B\to\pi K$ decays.
The predictions are given in the BN$^\prime$ column of Table~II.
To check the consistency between our calculations made in the BN$^{\prime}$ column and those in \cite{Beneke:2001ev},
let's make an example for comparison.
For example, the prediction for $Br(B^{-}\to\pi^{-} \bar{K}^{0})$ decay is calculated to be
$12.1\times 10^{-6}$, 
which is close to the central value given in the Eq.~(93) in the paper \cite{Beneke:2001ev}
\bee
10^{6}\times Br(B^{-}\to\pi^{-} \bar{K}^{0})=[14.1^{+6.4}_{-4.0}(m_s)^{+8.1}_{-3.6}(X_A)]
\times\left[\frac{F^{B\to\pi}_0(0)}{0.28}\right]\;,
\eee
where the first error is due to the parameter variations shown in the Table 2 in that paper,
and the second error is from the uncertainty due to power corrections from weak annihilation
and twist-3 hard spectator contributions.
We find that our calculation is consistent with that made in \cite{Beneke:2001ev}.
We argue that the difference between the predictions given in the BN$^\prime$ and BN columns 
maybe due to different methods used for determining the central values of the predictions
among many sources of theoretical uncertainties. 
In average, the predictions made by the CE, BN$^\prime$ and DYZ schemes 
are only about one half of the experimental data.
The predictions made by the BN scheme given in the paper \cite{Beneke:2003zv} are consistent with 
but lower than the experimental data \cite{Barberio:2007cr}.

\section{discussions and conclusions}
In this paper, we have shown that the factorizability of the QCDF amplitudes for $B\to PP$ decays with twist-3
two parton contributions can be preserved under the energetic meson limit
that the two final state light mesons carry energetic momenta.
Namely, we have extended the factorization theorem \eq{fact-1} to $O(\alpha_s)$ and $O(1/m_b)$ 
under the two parton approximation.
The factorizability is shown by the following facts:
(1)Under the energetic meson limit, 
the pseudoscalar distribution amplitude for a light pseudoscalar meson is allowed to be 
non-constant by the equations of motion for the quark.  
(2)The non-constant $\phi_{p}(x)$ is then used to regularize the end-point divergences 
in the hard spectator corrections at twist-3 order.
(3)By retaining the dependence of the momentum fraction variable of 
the spectator quark of the $B$ meson,
the end-point divergent problem for the annihilation corrections at twist-3 order are solved. 
Based on the factorization for the matrix element at the twist-3 order,
we have constructed a collinear expansion (CE) scheme for calculations of the hard scattering kernels of 
order $O(\alpha_s)$ and $O(\Lambda_{QCD}/m_b)$.
The results were applied to make predictions for the branching ratios of $B\to \pi K$ decays.
The predicted averaged branching ratios of $B\to \pi K$ decays are only about one half of the experimental data.
Because the end point divergences in the hard spectator and annihilation corrections as found in previous studies have
been regularized,
the strong phase related to these two contributions is predicted to be universal as $\phi_{A}=8.2^{\circ}$. 

The predictions for the averaged branching ratios of $B\to \pi K$ decays 
made by three schemes, the CE, BN$^{\prime}$ and DYZ schemes,
all have similar magnitudes.
Although the CE scheme contains no significant improvements in phenomenology than the other two schemes,
however, it has reduced large uncertainties in the $X_{H,A}$ terms, theoretically.
In literature, the $X_{H,A}$ terms are modeled as
\bee
X_{H,A}=\text{ln}{\frac{m_B}{\Lambda_h}}(1+\rho_{H,A}e^{i\phi_{H,A}})\;,
\eee
where $\rho_{H,A}\le 1$ and $\phi_{H,A}$ are unknown parameters and process dependent in general.
In order to reduce the theoretical uncertainties , 
$\rho_{H,A}$ and $\phi_{H,A}$ are further assumed universal and their values are determined phenomenologically \cite{Beneke:2003zv}.
This reduces the prediction power of QCDF.
On the other hand, in our approach, the relevant terms are finite and their values are calculable.
The introduced $B$ meson DA $\phi_{B}(\xi)$ in the weak annihilation and twist-3 hard spectator contributions
results in no additional uncertainties than the other schemes,
because the parameters in the $\phi_{B}(\xi)$ \eq{phiB-model} 
are completely determined by the first two moments of the $\phi_{B}(\xi)$ \eq{phiB-condition}.
Since the most contributions of $A^{i}_{1,2}$ are from the end-point region,
it is also interesting to check whether different models for the $\phi_{B}(\xi)$ can result in
different values for $A^{i}_{1,2}$. 
As found in \cite{Lee:2005gza}, 
the end-point behaviors of the model for the $\phi_{B}(\xi)$ are 
well controlled by the first two moments \eq{phiB-condition} 
and almost independent of the parameterization form of the assumed model.
This implies 
that the $A^{i}_{1,2}$ given by \eq{phiB-model} are almost model independent.
In summary, the predictions under our approach are not only free from the end-point divergences 
but also independent of the model for $\phi_{B}(\xi)$.
In addition, we have unproven the theoretical uncertainties to be of order $O(1/m_b^2)$.

We emphasize that the methods proposed in this work for resolving the end-point divergences is not ad hoc but general.
For example, in the charmless B decays with one scalar meson in the final state,
the chirally enhanced corrections are necessary. 
The twist-3 two parton DAs for the scalar meson can get involved.
Similar to the pseudoscalar meson, 
the equation of motion for the twist-3 two parton DAs for the scalar meson are also used to 
determine the DAs \cite{Cheng:2005nb,Cheng:2007st}.
The physical situation in these decay processes is similar to that in $B\to\pi K$ decays. 
Therefore, the analysis given in Section \ref{sec:sec2} 
and the calculation scheme given in Section \ref{sec:sec3} can be used.
We will show this fact in another place.
Another important contribution of this paper is that the proposed CE scheme not only provides a systematic method for
including higher twist contributions, but also is consistent with the QCD factorization scheme.
This is the first method in the literature that can systematically calculate the higher twist contributions
within the factorization approach.  
Last, the complete twist-3 contributions need to consider the three particle DA.
We plane to study this issue in another preparing work.

\vspace{0.5cm}
\noindent
\begin{center}
{\bf Acknowledgments}
\end{center}
The author would like to appreciate the constant supports from 
the Department of Science Application and Dissemination
and the National Science Council of R.O.C. 
under Grant Numbers NSC92-2112-M-142-001
, NSC93-2112-M-142-001 
and NSC95-2112-M-142-001.
\noindent
%%%%%%%%%%%%%%%%%%%%%%%%%%%%%%%%%%%%%%%%%%%%%%%%%%%%%%%%%%%%%%%%%%%%%%%%%%%%%%%%%%%%%%%%%%%%%%%%%%%%%%%%
\appendix
\section{calculations of twist-3 vertex corrections} 
In this Appendix,
the calculations for the one loop vertex contributions to the matrix element of operators $Q_{i}$, $i=6$ 
will be present.
The amplitude for the Feynman diagram as depicted in \fig{fig:fig3}(a) is written as 
\bee\label{Q6-vertex-1}
&&\langle Q_{5}^{(8)(V-A)(V+A)}\rangle^{vertex, 1-loop}_{(a)}\nn
&=&
if_{M_{2}}\mu_{M_{2}}
\frac{\pi\alpha_{s} C_{F}\mu^{2D}}{N_{c}}
\int_{0}^{1} du 
\Gamma(3)
\int_{0}^{1} dx 
\int_{0}^{1-x} dy 
\int\frac{d^D k}{(2\pi )^D}
\frac{1}{[k^2-2(x P_{b}+ y u q)\cdot k ]^3}\nn
&&\times
\langle M_{1}|\bar{q}
\gamma^{\mu}(1+\gamma_{5})
(\gamma_{5}\hat{\phi}^{M_{2}}_{p}(u)-\frac{1}{2}\epsilon_{\perp}\cdot\sigma \hat{\phi}^{M_{2}}_{\sigma}(u))
\gamma^{\alpha}
(u\s{q}-\s{k})
\gamma_{\mu}
(1-\gamma_{5})
(\s{P}_{b}-\s{k}+ m_{b})\gamma_{\alpha}
b|\bar{B}\rangle\;.\nn
\eee
where NDR has been used. 
In the above expression, we have used the following spin state expansion for the matrix element 
$\langle M_2|\bar{q}(0)q(\lambda n/E)|0\rangle$  
\bee\label{spin-expand-M2}
\int_0^{infty} \frac{d\lambda}{2\pi} e^{-i \bar{u}\lambda }\langle M_2|\bar{q}(0)q(\lambda n/E)|0\rangle
= -\frac{if_{M_2}}{4N_c}[\gamma_5 \s{q}\phi^{M_2}_{P}(u)
+\mu_{\chi}^{M_2}
\left(\gamma_5\hat{\phi}^{M_2}_{p}(u)-\frac{1}{2}\epsilon_{\perp}\cdot\sigma\hat{\phi}^{M_2}_{\sigma}(u)\right)]\;.\nn
\eee
Perform a substitution of $k\to k+ xP_b + yu q$ for $k$ to arrive at
\bee\label{Q6-vertex-2}
&&\langle Q_{5}^{(V-A)(V+A)}\rangle^{vertex, 1-loop}_{(a)}\nn
&=&
if_{M_{2}}\mu_{M_{2}}
\frac{\pi\alpha_{s} C_{F}\mu^{2D}}{N_{c}}
\int_{0}^{1} du \hat{\phi}^{M_{2}}_{p}(u)
\Gamma(3)
\int_{0}^{1} dx 
\int_{0}^{1-x} dy 
\int\frac{d^D k}{(2\pi )^D}
\frac{1}{[k^2-(x P_{b}+ y u q)^2]^3}\nn
&&\times
\langle M_{1}|\bar{q}
\gamma^{\mu}(1+\gamma_{5})
(\gamma_{5}-\frac{1}{2}\epsilon_{\perp}\cdot\sigma )
\gamma^{\alpha}
(u\bar{y}\s{q}-x\s{P}_b -\s{k})
\gamma_{\mu}
(1-\gamma_{5})
(\bar{x}\s{P}_{b}-yu\s{q}-\s{k}+ m_{b})\gamma_{\alpha}
b|\bar{B}\rangle\;,\nn
\eee
where we have used $\hat{\phi}^{M_2}_{p}(u)=\hat{\phi}^{M_2}_{\sigma}(u)$.
The contributions with odd number of $k$ give vanishing results. 
Completing the loop integration over $k$ by NDR gives
\bee\label{Q6-vertex-3}
&&\langle Q_{5}^{(8)(V-A)(V+A)}\rangle^{vertex, 1-loop}_{(a)}\nn
&=&
if_{M_{2}}\mu_{M_{2}}
\frac{\pi\alpha_{s} C_{F}\mu^{2D}}{N_{c}}
\int_{0}^{1} du \hat{\phi}^{M_{2}}_{p}(u)
\Gamma(3)
\int_{0}^{1} dx 
\int_{0}^{1-x} dy 
\frac{(-1)^3}{(4\pi )^{D/2}}\frac{\Gamma(3-D/2)}{\Gamma(3)}
\left( \frac{1}{(x P_{b}+ y u q)^2}\right )^{3-D/2}\nn
&&\times
\left\{\right.
\langle M_{1}|\bar{q}
\gamma^{\mu}(1+\gamma_{5})
(\gamma_{5}-\frac{1}{2}\epsilon_{\perp}\cdot\sigma )
\gamma^{\alpha}
(u\bar{y}\s{q}-x\s{P}_b)
\gamma_{\mu}
(1-\gamma_{5})
(\bar{x}\s{P}_{b}-yu\s{q}+ m_{b})\gamma_{\alpha}
b|\bar{B}\rangle\nn
&&
-\frac{g^{\eta\lambda}}{2}\frac{\Gamma(2-D/2)}{\Gamma(3-D/2)}
\left(\frac{1}{(x P_{b}+ y u q)^2} \right )^{-1}
\langle M_{1}|\bar{q}
\gamma^{\mu}(1+\gamma_{5})
(\gamma_{5}-\frac{1}{2}\epsilon_{\perp}\cdot\sigma )
\gamma^{\alpha}\gamma_{\eta}
\gamma_{\mu}
(1-\gamma_{5})\gamma_{\lambda}\gamma_{\alpha}
b|\bar{B}\rangle
\left.\right\}\;.
\nn
\eee
We note that the first term in the bracket of \eq{Q6-vertex-3} is finite under $x,y\to 1$ 
while becomes divergent as $x,y\to 0$,
and the second term in the bracket of \eq{Q6-vertex-3} is finite under $x,y\to 0$ while becomes divergent as $x,y\to 1$.
This implies that the first term is infrared divergent while the second term is ultra-violate divergent.
We apply $D=4+2 a$ for the first term and $D=4-2\epsilon$ for the second term.

The Dirac identities in $D$-dimension
\bee
\gamma^{\mu}\gamma_{\mu} &=& D\;,\nn
\gamma^{\mu}\gamma^\nu\gamma_{\mu}&=&-(D-2)\gamma^{\nu}\;,\nn
\gamma^{\mu}\gamma^{\alpha}\gamma^{\beta}\gamma_{\mu}&=&4 g^{\alpha\beta}-(4-D)\gamma^{\alpha\beta}\;,\nn
\gamma^{\mu}\gamma^{\alpha}\gamma^{\beta}\gamma^{\gamma}\gamma_{\mu}
&=&-2\gamma^{\gamma}\gamma^{\alpha}\gamma^{\beta} +(4-D)\gamma^{\alpha}\gamma^{\beta}\gamma^{\gamma}\;,
\eee
can be used to simplify the Dirac matrix.
After completing the spin algebra, we found that only $\gamma_5 \phi^{M_2}_{p}(u)$ can contribute.
The UV part of $\langle Q_{5}^{(8)(V-A)(V+A)}\rangle^{vertex, 1-loop}_{(a)}$ is 
\bee\label{Q6-vertex-UV-1}
&&\langle Q_{5}^{(8)(V-A)(V+A)}\rangle^{vertex, 1-loop}_{(a),UV}\nn
&=&
-2if_{M_{2}}\mu_{M_{2}}\langle M_{1}|\bar{q}(1-\gamma_5)b|\bar{B}\rangle
\frac{\alpha_{s} C_{F}}{4\pi N_{c}}
\int_{0}^{1} du \hat{\phi}^{M_{2}}_{p}(u)\nn
&&\times
(1-\frac{5}{2}\epsilon)
\left(\frac{4\pi\mu^2}{m_b^2}\right)^{\epsilon}
\Gamma(\epsilon)
\int_{0}^{1} dx 
\int_{0}^{1-x} dy 
\frac{2}{[x(x+y u)]^{\epsilon}}\;.
\eee
The integrals over $x$ and $y$ are
\bee
\int_{0}^{1} dx 
\int_{0}^{1-x} dy 
\frac{2}{[x(x+y u)]^{\epsilon}}=1+3\epsilon+\frac{u\ln u}{\bar{u}}\;.
\eee
By substituting the above identity, the UV contribution becomes
\bee\label{Q6-vertex-UV-2}
&&\langle Q_{5}^{(8)(V-A)(V+A)}\rangle^{vertex, 1-loop}_{(a),UV}\nn
&=&
-2if_{M_{2}}\mu_{M_{2}}\langle M_{1}|\bar{q}(1-\gamma_5)b|\bar{B}\rangle
\frac{\alpha_{s} C_{F}}{4\pi N_{c}}
\int_{0}^{1} du \hat{\phi}^{M_{2}}_{p}(u)\nn
&&\times
(\frac{1}{\epsilon}-\gamma_E + 4\pi +\ln\frac{\mu^2}{m_b^2}+\frac{1}{2}+\frac{u\ln u}{\bar{u}})\;.
\eee
The calculation for the IR part of $\langle Q_{5}^{(8)(V-A)(V+A)}\rangle^{vertex, 1-loop}_{(a)}$ 
requires some cares.
As we have shown in Sec.III,
the different components of the vertexes $\gamma^{\mu}$, $\gamma^{\alpha}$, $\gamma_{\alpha}$
lead to contributions of different magnitudes.
The IR part of $\langle Q_{5}^{(V-A)(V+A)}\rangle^{vertex, 1-loop}_{(a)}$ appears as
\bee\label{Q6-vertex-IR-1}
&&\langle Q_{5}^{(8)(V-A)(V+A)}\rangle^{vertex, 1-loop}_{(a)}\nn
&=&
if_{M_{2}}\mu_{M_{2}}
\frac{\pi\alpha_{s} C_{F}\mu^{2D}}{N_{c}}
\int_{0}^{1} du \hat{\phi}^{M_{2}}_{p}(u)
\Gamma(3)
\int_{0}^{1} dx 
\int_{0}^{1-x} dy 
\int\frac{d^D k}{(2\pi )^D}
\frac{1}{[k^2-2(x P_{b}+ y u q)\cdot k ]^3}\nn
&&\times
\langle M_{1}|\bar{q}
\gamma^{\mu}(1+\gamma_{5})
\gamma_{5}
(2(\bar{y}u-x)q^{\alpha}-x\gamma^{\alpha}\s{p})
\gamma_{\mu}
(1-\gamma_{5})
(2\bar{x}P^{\alpha}_{b}+(xm_{b}-yu\s{q})\gamma_{\alpha})
b|\bar{B}\rangle\;.\nn
\eee
By some manipulations, the expression becomes
\bee\label{Q6-vertex-IR-2}
&&\langle Q_{5}^{(V-A)(V+A)}\rangle^{vertex, 1-loop}_{(a),IR}\nn
&=&
2if_{M_{2}}\mu_{M_{2}}\langle M_{1}|\bar{q}(1-\gamma_5)b|\bar{B}\rangle
\frac{\alpha_{s} C_{F}}{16\pi N_{c}}
\int_{0}^{1} du \hat{\phi}^{M_{2}}_{p}(u)\nn
&&\times
\left(\frac{4\pi\mu^2}{m_b^2}\right)^{-a}
\Gamma(1-a)
\int_{0}^{1} dx 
\int_{0}^{1-x} dy 
\frac{N_{IR}}{[x(x+y u)]^{1-a}}\;,
\eee
where
\[
N_{IR}=[2Du\bar{x}\bar{y}+4uxy-(4+2D)x\bar{x}]\;.
\]
After completing the integrations over $x$ and $y$,
the $\langle Q_{5}^{(8)(V-A)(V+A)}\rangle^{vertex, 1-loop}_{(a),IR}$ becomes
\bee
&&\langle Q_{5}^{(V-A)(V+A)}\rangle^{vertex, 1-loop}_{(a),IR}\nn
&=&
2if_{M_{2}}\mu_{M_{2}}\langle M_{1}|\bar{q}(1-\gamma_5)b|\bar{B}\rangle
\frac{\alpha_{s} C_{F}}{4\pi N_{c}}
\left(\frac{4\pi\mu^2}{m_b^2}\right)^{-a}
\Gamma(1-a)
\int_{0}^{1} du \hat{\phi}^{M_{2}}_{p}(u)\nn
&&\times
[\frac{1}{a^2}+\frac{2\ln u}{a}+\frac{1}{2a}-\ln u+\ln^2 u-2Li_{2}(1-\frac{1}{u})
+1+\frac{3\ln u}{\bar{u}}]\;.
\eee
As a result, $\langle Q_{5}^{(V-A)(V+A)}\rangle^{vertex, 1-loop}_{(a)}$ is written as
\bee
&&\langle Q_{5}^{(V-A)(V+A)}\rangle^{vertex, 1-loop}_{(a)}\nn
&=&
-2if_{M_{2}}\mu_{M_{2}}\langle M_{1}|\bar{q}(1-\gamma_5)b|\bar{B}\rangle
\frac{\alpha_{s} C_{F}}{4\pi N_{c}}\int_{0}^{1} du \hat{\phi}^{M_{2}}_{p}(u)
\left\{\right.
(\frac{1}{\epsilon}-\gamma_{E}+4\pi+\ln\frac{\mu^2}{m_b^2}+\frac{1}{2}+\frac{u\ln u}{\bar{u}})\nn
&&-\left(\frac{4\pi\mu^2}{m_b^2}\right)^{-a}\Gamma(1-a)
[\frac{1}{a^2}+\frac{2\ln u}{a}+\frac{1}{2a}-\ln u + \ln^2 u
-2 Li_{2}(1-\frac{1}{u})+1+\frac{3\ln u}{\bar{u}}]
\left.\right\}\;.\nn
\eee

The calculations for the Feynman diagrams depicted in \fig{fig:fig3}(b-d) can be done in a similar way.
The amplitudes for these three diagrams are written as
\bee
&&\langle Q_{5}^{(V-A)(V+A)}\rangle^{vertex, 1-loop}_{(b)}\nn
&=&
2if_{M_{2}}\mu_{M_{2}}\langle M_{1}|\bar{q}(1-\gamma_5)b|\bar{B}\rangle
\frac{\alpha_{s} C_{F}}{4\pi N_{c}}\int_{0}^{1} du \hat{\phi}^{M_{2}}_{p}(u)
\left\{\right.
(\frac{1}{\epsilon}-\gamma_{E}+4\pi+\ln\frac{\mu^2}{m_b^2}+\frac{7}{2}+\frac{\bar{u}\ln \bar{u}}{u})\nn
&&-\left(\frac{4\pi\mu^2}{m_b^2}\right)^{-a}\Gamma(1-a)
[\frac{1}{a^2}+\frac{2\ln \bar{u}}{a}+\frac{1}{2a}-\ln \bar{u}+\ln^2 \bar{u}
-2 Li_{2}(1-\frac{1}{\bar{u}})+1+\frac{3\ln \bar{u}}{u}]
\left.\right\}\;.\nn
\eee
\bee
&&\langle Q_{5}^{(V-A)(V+A)}\rangle^{vertex, 1-loop}_{(c)}\nn
&=&
2if_{M_{2}}\mu_{M_{2}}\langle M_{1}|\bar{q}(1-\gamma_5)b|\bar{B}\rangle
\frac{\alpha_{s} C_{F}}{4\pi N_{c}}\int_{0}^{1} du \hat{\phi}^{M_{2}}_{p}(u)
\left\{\right.
(\frac{1}{\epsilon}-\gamma_{E}+4\pi+\ln\frac{\mu^2}{m_b^2}+\frac{7}{2}-\ln u + i\pi)\nn
&&-\left(\frac{4\pi\mu^2}{m_b^2}\right)^{-a}\Gamma(1-a)
[\frac{2}{a^2}+\frac{2\ln u}{a}-\frac{3+2i\pi}{2a}-(4+2i\pi)\ln u+4i\pi + \ln^2 u +\frac{27+2\pi}{6}]
\left.\right\}\;.\nn
\eee
\bee
&&\langle Q_{5}^{(V-A)(V+A)}\rangle^{vertex, 1-loop}_{(d)}\nn
&=&
-2if_{M_{2}}\mu_{M_{2}}\langle M_{1}|\bar{q}(1-\gamma_5)b|\bar{B}\rangle
\frac{\alpha_{s} C_{F}}{4\pi N_{c}}\int_{0}^{1} du \hat{\phi}^{M_{2}}_{p}(u)
\left\{\right.
(\frac{1}{\epsilon}-\gamma_{E}+4\pi+\ln\frac{\mu^2}{m_b^2}+\frac{1}{2}-\ln \bar{u}+i\pi)\nn
&&-\left(\frac{4\pi\mu^2}{m_b^2}\right)^{-a}\Gamma(1-a)
[\frac{2}{a^2}+\frac{2\ln \bar{u}}{a}-\frac{3+2i\pi}{2a}-(4+2i\pi)\ln \bar{u}+4i\pi + \ln^2 \bar{u} +\frac{27+2\pi}{6}]
\left.\right\}\;.\nn
\eee
The summation of the above contributions from \fig{fig:fig3}~(a)-(d) gives the $V_{6,8}$ in Eq.~(\ref{vertex}).
It is obvious that the IR divergences are cancelled. 
The UV divergences are regularized by means of the dimensional regularization
and the remaining scheme constants are subtracted by the $\overline{MS}$ subtraction scheme. 
Our calculation is consistent with the result derived by using the BN scheme \cite{Beneke:2003zv}.
We also note that, for the vertex contributions, 
the pseudotensor LCDA does not contribute in both CE and BN schemes. 
In the DYZ scheme, the pseudotensor LCDA  can involve in the vertex contributions
and there are associated IR divergences .
The authors in \cite{Du:2001ns} argued that a symmetric parameterization for 
pseudotensor LCDA is necessary for eliminating the associated IR divergences.
Although both CE and BN schemes obtain similar results for the vertex contribution $V_{6}(M_2)$,
the meanings for the pseudoscalar LCDA ($\hat{\phi}^{M_2}_{p}(u)$ in the CE scheme 
and the $\phi^{M_2}_p(u)$ in the BN scheme) are different.  
In the CE scheme, the $\hat{\phi}^{M_2}_{p}(u)=6u\bar{u}$ is used.
On the other hand, in the BN scheme, $\phi^{M_2}_{p}(u)=1$ was used.

\section{calculations of twist-3 penguin corrections}
The calculations for twist-3 penguin contributions from the penguin contractions as depicted in the diagrams in 
\fig{fig:fig3}~(e) and (f) are given in this Appendix. 
We denote the penguin topology in \fig{fig:fig3}~(e) as type-$I$ and  penguin topology in  
\fig{fig:fig3}~(f) as type-$II$.
The contributions from $Q_{1,3}$ appear in the type-$I$ penguin topology
and the contributions from $Q_{4,6}$ are identified as the type-$II$ penguin topology.
For the type-$I$ penguin topology, 
the amplitude from operators $Q_{1,3}$ are written as
\bee\label{penguin-I-1}
&&\langle Q_{1,3}^{(V-A)(V-A)}\rangle^{penguin,1-loop}\nn
&=& \frac{C_F\pi\alpha_s \mu^{4-D}}{N_c}f_{M_2}\mu_{\chi}^{M_2}
\int_{0}^{1} du \int_{0}^{1} dx\int\frac{d^D k}{(2\pi)^D}\frac{1}{[k^2-R^2]^2}\nn
&&
\times
%\left\{\right.
\frac{1}{(p-uq)^2}\langle M_1|\bar{q}\gamma^{\alpha}
(\gamma_5 \hat{\phi}^{M_2}_{p}(u)-\frac{1}{2}\epsilon_{\perp}\cdot\sigma \hat{\phi}^{M_2}_{\sigma}(u))
\gamma^{\mu}(1-\gamma_5) \nn
&&
\hspace{0.8cm}(\s{k}-\bar{x}(\s{p}-u\s{q})+m_q)
\gamma_{\alpha}
(\s{k}+x(\s{p}-u\s{q})+m_q)
\gamma_{\mu}(1-\gamma_5)b|B\rangle
%\left.\right\}
\;,
\eee
where $R^2=m_b^2(s_q -x\bar{x}u)$ with $s_q = m_q^2/m_b^2$.
For the type-$II$ penguin topology, 
the relevant amplitude from operators $Q_{4,6}$ is expressed as
\bee\label{penguin-II-1}
&&\langle Q_{4,6}^{(V-A)(V-A)}\rangle^{penguin,1-loop}\nn
&=& -\frac{C_F\pi\alpha_s \mu^{4-D}}{N_c}f_{M_2}\mu_{\chi}^{M_2}
\int_{0}^{1} du \int_{0}^{1} dx\int\frac{d^D k}{(2\pi)^D}\frac{1}{[k^2-R^2]^2}\nn
&&
\times\frac{1}{(p-uq)^2}
\langle M_1|\bar{q}\gamma^{\alpha}(\gamma_5 \hat{\phi}^{M_2}_{p}(u)
-\frac{1}{2}\epsilon_{\perp}\cdot\sigma \hat{\phi}^{M_2}_{\sigma}(u))
 \gamma^{\mu}(1-\gamma_5)b|B\rangle\nn
&&
\times
\sum_{q=u,c,b}
\left\{\right.\text{Tr}[
\gamma_{\alpha}
(\s{k}-\bar{x}(\s{p}-u\s{q})+m_q)
\gamma_{\mu}(1-\gamma_5)
(\s{k}+x(\s{p}-u\s{q})+m_q)
]
\left.\right\}
\;.
\eee
where $R^2=m_b^2(s_q -x\bar{x}u)$ with $s_q = m_q^2/m_b^2$.
We first complete the $k$ integration to obtain
\bee\label{penguin-I-2}
&&\langle Q_{1,3}^{(V-A)(V-A)}\rangle^{penguin,1-loop}\nn
&=& -if_{M_2}\mu_{\chi}^{M_2}\frac{C_F\alpha_s}{16\pi N_c}
\int_{0}^{1} du \int_{0}^{1} dx
\nn
&&
\times\left\{\right.
\frac{g_{\eta\lambda}}{2}R^2
(
\langle M_1|\bar{q}\gamma^{\alpha}
(\gamma_5 \hat{\phi}^{M_2}_{p}(u)-\frac{1}{2}\epsilon_{\perp}\cdot\sigma \hat{\phi}^{M_2}_{\sigma}(u))
\gamma^{\mu}(1-\gamma_5)
\gamma^{\lambda}
\gamma_{\alpha}
\gamma^{\eta}
\gamma_{\mu}(1-\gamma_5)b|B\rangle \nn
&&
\hspace{0.5cm}
+x\bar{x}\left(\frac{4\pi\mu^2}{m_b^2}\right)^{\epsilon}
\frac{\Gamma(\epsilon)}{\bar{R}^{2\epsilon}}
\langle M_1|\bar{q}\gamma^{\alpha}
(\gamma_5 \hat{\phi}^{M_2}_{p}(u)-\frac{1}{2}\epsilon_{\perp}\cdot\sigma \hat{\phi}^{M_2}_{\sigma}(u))
\gamma^{\mu}(1-\gamma_5)
((\s{p}-u\s{q})+m_q)\nn
&&\hspace{4.5cm}\times\gamma_{\alpha}
((\s{p}-u\s{q})+m_q)
\gamma_{\mu}(1-\gamma_5)b|B\rangle
\left.\right\}\frac{1}{(p-uq)^2}
\eee
and
\bee\label{penguin-II-2}
&&\langle Q_{4,6}^{(V-A)(V\pm A)}\rangle^{penguin,1-loop}\nn
&=& if_{M_2}\mu_{\chi}^{M_2}\frac{C_F\alpha_s}{16\pi N_c}
\int_{0}^{1} du \int_{0}^{1} dx
\nn
&&
\times\left\{\right.
\frac{g_{\eta\lambda}}{2}R^2
\langle M_1|\bar{q}\gamma^{\alpha}
(\gamma_5 \hat{\phi}^{M_2}_{p}(u)-\frac{1}{2}\epsilon_{\perp}\cdot\sigma \hat{\phi}^{M_2}_{\sigma}(u))
\gamma^{\mu}(1-\gamma_5)b|B\rangle
\text{Tr}[\gamma_{\alpha}\gamma^{\lambda}\gamma_{\mu}(1-\gamma_5)\gamma^{\eta}]
\nn
&&
+x\bar{x}\left(\frac{4\pi\mu^2}{m_b^2}\right)^{\epsilon}
\frac{\Gamma(\epsilon)}{\bar{R}^{2\epsilon}}
\langle M_1|\bar{q}\gamma^{\alpha}
(\gamma_5 \hat{\phi}^{M_2}_{p}(u)-\frac{1}{2}\epsilon\cdot\sigma \hat{\phi}^{M_2}_{\sigma}(u))
\gamma^{\mu}(1-\gamma_5)b|B\rangle\nn
&&
\hspace{1.5cm}
\times\left(\right.
\sum_{q=u,c,b}\text{Tr}[\gamma_{\alpha}
((\s{p}-u\s{q})+m_q)
\gamma_{\mu}(1-\gamma_5)
((\s{p}-u\s{q})+m_q)]
\left.\right)
\left.\right\}
\frac{1}{(p-uq)^2}\;.
\eee
where $\bar{R}=\sqrt{s_q -x\bar{x}u}$.
After completing the spin algebra, we arrive at
\bee
\langle Q_{1,3}^{(V-A)(V-A)}\rangle^{penguin,1-loop}
&=& 4if_{M_2}\mu_{\chi}^{M_2}\langle M_1|\bar{q}(1-\gamma_5)b|B\rangle
\frac{C_F\alpha_s}{4\pi N_c}
\int_{0}^{1} du \hat{\phi}^{M_2}_{p}(u)
\int_{0}^{1} dx x\bar{x}\nn
&&[\frac{1}{\epsilon}-\gamma_{E}+\ln(4\pi)+\ln\frac{\mu^2}{m_b^2}-2-2\ln\sqrt{s_q-x\bar{x} u}]
\eee
and
\bee
\langle Q_{4,6}^{(V-A)(V-A)}\rangle^{penguin,1-loop}
&=& 4if_{M_2}\mu_{\chi}^{M_2}\langle M_1|\bar{q}(1-\gamma_5)b|B\rangle
\frac{C_F\alpha_s}{4\pi N_c}
\int_{0}^{1} du \hat{\phi}^{M_2}_{p}(u)
\int_{0}^{1} dx x\bar{x}\nn
&&\sum_{q=u,c,b}[\frac{1}{\epsilon}-\gamma_{E}+\ln(4\pi)+\ln\frac{\mu^2}{m_b^2}-2\ln\sqrt{s_q-x\bar{x} u}]\;.
\eee
After completing the integral over $x$, 
we obtain the contributions from different operators contained in the $P_{6}(M_2)$ function.
Similar to the vertex contributions, 
the pseudotensor LCDA does not get involved in the penguin contributions.
The penguin contributions calculated by the CE scheme are consistent with the results calculated by 
using the BN scheme \cite{Beneke:2003zv},
while are different from the results calculated by the DYZ scheme \cite{Du:2001ns}.

The calculation for the contributions from the chromo-magnetic dipole operator $Q_{8G}$ 
is straightforward in the CE scheme.
The amplitude is written as
\bee
-i\frac{f_{M_2}\mu_{\chi}^{M_2}\alpha_s m_b}{8\pi N_c}\int_{0}^{1} du
\langle M_1|\bar{q}\gamma^{\alpha}(\gamma_5 \hat{\phi}^{M_2}_{p}(u)
-\frac{1}{2}\epsilon\cdot\sigma \hat{\phi}^{M_2}_{\sigma}(u))
[\s{k}\gamma_{\alpha}-\gamma_{\alpha}\s{k}](1+\gamma_5)b|B\rangle\frac{1}{(P_b-uq)^2}\;.\nn
\eee 
After completing the spin algebra, we arrive at
\bee
&&-i\frac{f_{M_2}\mu_{M_2}\alpha_s}{2\pi N_c}\int_{0}^{1} du
\langle M_1|\bar{q}(\frac{3}{2}\hat{\phi}^{M_2}_{p}(u)
+\frac{1}{2}\hat{\phi}^{M_2}_{\sigma}(u))(1-\gamma_5) b|B\rangle\nn
&=&(-4if_{M_2}\mu_{M_2})\frac{\alpha_s C_F}{4\pi N_c}\int_{0}^{1} du \hat{\phi}^{M_2}_{p}(u)
\langle M_1|\bar{q}(1-\gamma_5) b|B\rangle\;,
\eee  
where the second line is obtained by using the equations of motion Eq.~(\ref{equation-2}).
For comparison, we note that the calculation for the contributions from the chromo-magnetic dipole operator $Q_{8G}$ 
by the BN scheme gives the following expression \cite{Beneke:2003zv}
\bee
&&-i\frac{f_{M_2}\mu_{M_2}\alpha_s}{2\pi N_c}\int_{0}^{1} du
\langle M_1|\bar{q}(\frac{3}{2}\phi_{p}(u)+\frac{1}{2}\frac{\phi_{\sigma}^{\prime}(u)}{6}
+\frac{1}{\bar{u}}\frac{\phi_{\sigma}(u)}{6})(1-\gamma_5) b|B\rangle\nn
&=&(-4if_{M_2}\mu_{M_2})\frac{\alpha_s C_F}{4\pi N_c}\int_{0}^{1} du \phi_{p}(u)
\langle M_1|\bar{q}(1-\gamma_5) b|B\rangle\;,
\eee 
where the second line is derived by using the Eq.~(\ref{equation-1}).
We note that the results from the penguin contractions calculated by the DYZ scheme are quite different
from our results. 
The detailed expressions for the penguin contributions by the DYZ scheme 
are referred to their original paper \cite{Du:2001ns}.
\section{Wilson coefficients}
The NLO Wilson coefficients used in this paper are summarized below for reference.
The solution to the renormalization group equation for the Wilson coefficients $C_1,\cdots,C_{10}$ can be written as
\bee
\vec{C}(\mu)=
\left[
{\bf U_0} +\frac{\alpha_s(\mu)}{4\pi}{\bf J_0}{\bf U_0}
-\frac{\alpha_s(M_W)}{4\pi}{\bf U_0}{\bf J_0}
+\frac{\alpha}{4\pi}\left(\frac{4\pi}{\alpha_s(\mu)}{\bf R_0}+{\bf R_1}\right)
\right]\vec{C}(M_W)
\eee
where the $U_0$ and $R_0$ are the LO parts  
and $J_0$ and $R_1$ are the NLO parts of $C_i$, respectively.
The $\alpha_s$ is used its NLO expression.
The expressions for the $U_{0}, J_0, R_0$ and $R_1$ matrices can be found in \cite{Buchalla:1995vs}.
   
%\bibliography{BCPreference}
%\bibliographystyle{plain}

\end{document}